\documentclass[10pt,journal,compsoc]{IEEEtran}

\usepackage[T1]{fontenc}
\ifCLASSOPTIONcompsoc

\usepackage{cite}

\usepackage{bm}
\usepackage{amsmath}
\usepackage{amsthm}
\usepackage{float}
\usepackage{setspace}
\usepackage{amssymb}
\usepackage{stfloats}
\usepackage{cite}
\usepackage{ragged2e}
\usepackage[ruled,vlined,linesnumbered]{algorithm2e}
\usepackage{amsfonts}
\usepackage{mathrsfs}
\usepackage{amsmath,amsthm}
\usepackage{array,booktabs}
\usepackage{subfigure}
\usepackage{multirow}
\usepackage{cuted}
\usepackage{multicol}
\usepackage{graphicx}
\usepackage{subfigure}
\usepackage{graphicx,xcolor,bm}
\usepackage{hyperref}
\usepackage{threeparttable}
\usepackage{dcolumn}
\usepackage{setspace}
\usepackage{makecell}
\usepackage{lipsum}
\usepackage{enumerate}
\usepackage{bbm}

\newtheorem{Defn}{Definition}

\usepackage{cite,bm}
\graphicspath{{figures/}}
\begin{document}

	\title{Matching-based Hybrid Service Trading \\ for Task Assignment over \\ Dynamic Mobile Crowdsensing Networks}

	\author{\vspace{-0.2cm}Houyi Qi*, Minghui Liwang*, \IEEEmembership{Member}, \IEEEmembership{IEEE}, Seyyedali Hosseinalipour, \IEEEmembership{Member}, \IEEEmembership{IEEE}, Xiaoyu Xia, \IEEEmembership{Member}, \IEEEmembership{IEEE}, Zhipeng Cheng, \IEEEmembership{Student Member}, \IEEEmembership{IEEE}, Xianbin Wang,~\IEEEmembership{Fellow}, \IEEEmembership{IEEE}, Zhenzhen Jiao\vspace{-0.2cm}
\thanks{This work is supported in part by the National Natural Science Foundation of China under Grant 62271424, the Natural Science Foundation of Xiamen City under Grants 3502Z20227002 and 3502Z20227007, the Fundamental Research Funds for the Central Universities under Grant 20720230035, and the Basic and Applied Basic Research Foundation of Guangdong Province under Grant no. 2022A1515110042. (Corresponding author: Minghui Liwang; *H. Qi and M. Liwang contributed equally to this work)

H. Qi (qihouyi@stu.xmu.edu.cn), M. Liwang (minghuilw@xmu.edu.cn) and are with the School of Informatics, Xiamen University, Fujian, China. 
S. Hosseinalipour (alipour@buffalo.edu) is with Department of Electrical Engineering, University at Buffalo-SUNY, NY, USA. X. Xia (xiaoyu.xia@rmit.edu.au) is with the School of Computing Technologies, RMIT University, Melbourne, Victoria, Australia. Z. Cheng (chengzp\_x@163.com) is with the School of Future Science and Engineering, Soochow University, Jiangsu, China. X. Wang (xianbin.wang@uwo.ca) is with the Department of Electrical and Computer Engineering, Western University, Ontario, Canada. Z. Jiao (jiaozhenzhen@teleinfo.cn) is with the iF-Labs, Beijing Teleinfo Technology Co., Ltd., CAICT, China.

	}}

	\IEEEtitleabstractindextext{
		\begin{abstract}
			\justifying
	By opportunistically engaging mobile users (workers), mobile crowdsensing (MCS) networks have emerged as important approach to facilitate sharing of sensed/gathered data of heterogeneous mobile devices. To assign tasks among workers and ensure low overheads, we introduce a series of stable matching mechanisms, which are integrated into a novel hybrid service trading paradigm consisting of \textit{futures trading} and \textit{spot trading} modes, to ensure seamless MCS service provisioning. In futures trading, we determine a set of long-term workers for each task through an \textbf{o}verbooking-enabled \textbf{i}n-\textbf{a}dvance \textbf{m}any-to-\textbf{m}any \textbf{m}atching (OIA3M) mechanism, while characterizing the associated risks under statistical analysis. In spot trading, we investigate the impact of fluctuations in long-term workers' resources on the violation of service quality requirements of tasks, and formalize a spot trading mode for tasks with violated service quality requirements under practical budget constraints, where the task-worker mapping is carried out via \textbf{o}nsite \textbf{m}any-to-\textbf{m}any \textbf{m}atching (O3M) and \textbf{o}nsite \textbf{m}any-to-\textbf{o}ne \textbf{m}atching (OMOM). We theoretically show that our proposed matching mechanisms satisfy stability, individual rationality, fairness, and computational efficiency. Comprehensive evaluations confirm the satisfaction of these properties in practical network settings and demonstrate our commendable performance in terms of service quality, running time, and decision-making overheads, e.g., delay and energy consumption.
		\end{abstract}

		\begin{IEEEkeywords}
			Mobile crowdsensing, matching theory, futures and spot trading, overbooking, risk analysis\vspace{-0.1cm}
		\end{IEEEkeywords}
	
}

\maketitle

\IEEEdisplaynontitleabstractindextext

\IEEEpeerreviewmaketitle

\section{Introduction }
\IEEEPARstart{T}{he} past decade has witnessed a leap in proliferation of smart devices and their sensing/computing capabilities. Meanwhile, new applications such as environmental monitoring and traffic forecasting have emerged, bringing the need for responsive and cost-effective data collection/sensing through smart mobile devices. To meet such demands, mobile crowdsensing (MCS) networks have been studied \cite{1,2,3,4,5} to facilitate data sharing among mobile devices (users) and MCS platforms, which host MCS tasks with various quality of service (QoS) requirements \cite{6,7,8,9,10,11,12}. 

One of the main goals of MCS networks is the efficient recruitment of heterogeneous mobile devices/users (referred to as \textit{workers}) to participate in MCS tasks. However, worker recruitment is a non-trivial procedure due to the selfish behaviors of workers \cite{13,14,15,16}, preventing them from sharing their limited communication and computation resources. This is due to the fact that supporting MCS tasks often consume a large portion of workers' limited resources and could also compromise their privacy \cite{17,18,19,20,21}. As a result, proper design of incentive mechanisms is of great importance to encourage the participation of workers in MCS networks, which draws a resemblance between MCS networks and service trading markets, where MCS tasks recruit workers to acquire services while providing payments.

\vspace{-0.2cm}
\subsection{Motivations}
In general, the existing literature on MCS networks tends to focus on one of two service trading strategies: \textit{spot trading} and \textit{futures trading}. Spot trading involves task owners and workers coming to a real-time agreement for service trading, usually on-site and based on current market conditions. Futures trading, on the other hand, allows task owners and workers to participate in secure futures trading by establishing forward contracts for future resource provisioning. 
While each strategy has its own benefits and drawbacks, it is important to also manage and configure resource utilization according to the uncertain and dynamic nature of the network. One approach to cope with this is to apply overbooking. In the following, we discuss these methods and detail the notion of overbooking.
\vspace{-0.1cm}
\subsubsection{Spot and futures service trading} 
Spot trading is a widely adopted technique for resource and service trading, which is commonly formulated as finding a proper matching between \textit{sellers} (e.g., workers that generate or collect data) and \textit{buyers} (e.g., MCS task owners) in real-time \cite{22}. 
Although spot trading is an intuitive service trading procedure, it suffers from the following limitations.

\noindent
$\bullet$ \textit{Excessive latency:} The onsite decision-making procedure in spot trading (i.e., looking for a many-to-many matching between task owners and workers) can lead to an excessive delay/latency, especially upon having a large number of participants in the service trading market. This negatively impacts the task completion and resource utilization.

\noindent
$\bullet$ \textit{Undesired energy consumption:} Obtaining a proper many-to-many matching in spot trading requires forming complex patterns of interactions among participants, the realization of which can lead to excessive energy consumption \cite{23}. Such an overhead can negatively impact the quality-of-experience (QoE) of participants, especially for sellers with limited battery and power supply (e.g., smart phones, electrical vehicles) \cite{24}.

\noindent
$\bullet$ \textit{Potential trading failures:} Following the many-to-many matching process in spot trading, a select few winners - those who have been matched with task owners - will be compensated, while the remaining workers who were not matched to any task, despite investing their time and energy in the trading decision, will not receive any payment.

To address the aforementioned drawbacks of spot trading, futures trading promotes the participants to establish trading contracts prior to future practical transactions, based on historical data/statistical analysis. With these contracts in place, contractual participants can engage in trading with full confidence that the agreed-upon terms will be upheld during each practical transaction. Specifically, a \textit{practical transaction} refers to a trading event between task owners and workers, where workers deliver services to tasks while receiving payment from them.
Nevertheless, implementing futures trading can cause a variety of issues, such as improper contract terms and excessive payments, especially when statistical analysis fails to provide an accurate prediction on the future market condition (e.g., in terms of resource supply/demand). In this work, we argue that the main limitations of the existing works are due to the fact that they consider service trading either through spot trading or futures trading mode, leading to the deployment of these trading modes in isolation. We subsequently propose a hybrid service trading market for MCS networks that encompasses both futures and spot trading, and further introduces the notion of \textit{overbooking}.

\vspace{-0.15cm}
\setlength{\skip\footins}{0.15cm}
\subsubsection{Overbooking: from flight tickets to MCS networks}
In the conventional resource pre-selling service buyers face the possibility of "no-show"\footnote{"No-show" describes a situation in which a worker is unable to provide services for MCS tasks due to a variety of factors, such as its location and battery level.} events\cite{25}. Due to this reason, airlines \cite{26}, hotels \cite{27} and telecom companies \cite{28}, \cite{29} actively \textit{overbook} \cite{30} their resources. For example, to maximize their revenue, airlines overbook tickets by letting the number of sold tickets exceed the number of available seats; otherwise, flights will often take off with 15\% vacant seats, causing a waste of resources and economic losses \cite{31}. 
In the case of the service trading market we have considered, overbooking is encouraged. This is because resource supply can fluctuate over time, which encourages each MCS task owner to book more resources than its practical budget, in case some long-term workers fail to show up\footnote{For example, in our considered market, an MCS task with a practical budget that requires 3 workers, is allowed to recruit 4 workers.}, where \textit{long-term workers} refer to the workers who have signed contracts with sellers in the futures market.

Motivated by the above discussions, this paper explores a novel hybrid trading market that blends futures and spot trading modes, to achieve efficient mappings between multiple workers and MCS tasks, facilitating responsive and cost-effective services in dynamic MCS networks. In futures trading, we employ the \textbf{o}verbooking-enabled \textbf{i}n-\textbf{a}dvance \textbf{m}any-to-\textbf{m}any \textbf{m}atching (OIA3M) mechanism to determine long-term workers for each task, with risk analysis, based on the statistics of the ever-changing nature of MCS networks. This allows us to maximize the expected utility of both workers and tasks, which is essential for successfully navigating the dynamic service trading market over the long term. In the spot market, we introduce the \textbf{o}nsite \textbf{m}any-to-\textbf{m}any \textbf{m}atching (O3M) mechanism that enables each task owner with surplus budget to recruit surplus workers to improve its received service quality during practical transactions. We also develop \textbf{o}nsite \textbf{m}any-to-\textbf{o}ne \textbf{m}atching (OMOM) to help free up some of the long-term workers if the overall payment of a task exceeds its practical budget. 

\vspace{-0.2cm}
\subsection{A Toy Example}
To better illustrate our proposed hybrid service trading paradigm, we consider an example of taxi booking.

In taxi booking, taxi companies can offer in-advance reservations based on customer demand. In order to increase its revenue, e.g., a company $ \mathbb{A} $ is permitted to overbook its cars in case some customers do not show up or cancel their reservations. For instance, $ x $ taxis (which represents the practical resource supply) can be reserved for $ y $ customers ($ x<y $) assuming that $ z $ customers ($ z<y $) may cancel their bookings. By doing this, the taxi company should assess the risks via analyzing past data of historical transactions, such as if $ y-z>x $ or if $ y-z $ falls significantly below $ x $. To ensure a seamless travel experience, customers can also reserve taxis from other companies while evaluating potential risks, such as making a deposit. Let us refer to customers who book reservations with company $ \mathbb{A} $ and do not cancel them as "$ \mathbb{A}$-customers". During each actual transaction (e.g., during actual travels), there are three potential scenarios to consider:

\noindent
$\bullet$ \textit{Case 1.} When the number of $ \mathbb{A} $-customers equals to the number of taxis (i.e., $ y-z=x $), company $ \mathbb{A} $ can provide services according to its in-advance reservations.

\noindent
$\bullet$ \textit{Case 2.} When the number of taxis exceeds the number of $ \mathbb{A} $-customers (i.e., $ y-z<x $), $ x-y+z $ taxis are available to serve other customers under an onsite/spot manner.

\noindent
$\bullet$ \textit{Case 3.} When the number of $ \mathbb{A} $-customers exceeds the number of taxis (i.e., $ y-z>x $), taxi company $ \mathbb{A} $ needs to cancel the reservation of $ y-z-x $ $ \mathbb{A} $-customers.

Apparently, although overbooking can help taxi company $ \mathbb{A} $ cope with the absence of some $ \mathbb{A} $-customers during actual travels, there still exist nonnegligible risks. For example, company $ \mathbb{A} $ may need to cancel the reservation of some $ \mathbb{A} $-customers (i.e., Case 3), which can leave bad impressions. It is thus essential to consider appropriate overbooking rate, to alleviate financial losses.

\vspace{-0.2cm}
\subsection{Novelty and Contribution}
This paper investigates MCS service provisioning through a novel hybrid trading market that integrates futures and spot trading modes, consisting of multiple workers and tasks. To achieve responsive and cost-effective MCS services, we analyze potential risks that participants may encounter and introduce the notion of overbooking. For the futures trading mode, the OIA3M mechanism is investigated, allowing each MCS task owner to employ multiple workers via signing long-term contracts, according to the overbooked budget, to cope with uncertain resource supply. For the spot trading mode, we introduce O3M and OMOM to modify/conduct resource trading given the current network/market conditions. To the best of our knowledge, this paper is a first of its kind in designing a series of matching mechanisms for hybrid service trading markets. Our main contributions are summarized below:

\noindent 
$\bullet$ This paper introduces a hybrid service trading market over dynamic MCS networks via integrating both futures and spot trading modes, while obtaining proper matchings between multiple MCS tasks and workers. We further introduce overbooking to the considered market to cope with the dynamics of the resources offered by workers.

\noindent 
$\bullet$ For the futures market, we develop OIA3M to determine a set of workers that can offer services to MCS tasks in future transactions. For each MCS task, we also characterize the risks of receiving unsatisfying service quality mainly due to the uncertainties in workers' participation. We then exploit overbooking to manage such risks. We show that OIA3M ensures key properties of matching stability, individual rationality, fairness, and non-wastefulness.

\noindent
$\bullet$ We consider the spot market as a backup plan to ensure the satisfaction of service quality requirements of MCS tasks. In the spot market, we consider two scenarios during each practical transaction: \textit{i)} the overall payment of a task owner exceeds its budget (e.g., due to overbooking), for which we design OMOM matching mechanism; and \textit{ii)} task owners have surplus budgets, for which we design O3M matching mechanism. Both of these mechanisms satisfy the above-mentioned key matching properties.

\noindent 
$\bullet$ We conduct simulations using a real-world dataset to demonstrate that our proposed mechanism achieves a commendable performance in terms of service quality, while outperforming baseline methods in terms of running time and the overhead of interactions among participants.
	
\vspace{-0.3cm}
\section{Related Work} 
\begin{table}[b!] 
	\vspace{-0.6cm}
	{\footnotesize
		\caption{\footnotesize{A summary of related studies (Sta.: Stability, Ind.: Individual rationality, Fai.: Fairness, NonW.: Non-wastefulness, Ove.: Overbooking, Ris.: Risk evaluation)}} \vspace{-0.6cm} 
		\begin{center}
			\setlength{\tabcolsep}{0.5mm}{
				\begin{tabular}{|c|c|c|c|c|c|c|c|c|}
					\hline
					\multirow{2}{*}{\textbf{Reference}} & \multicolumn{2}{c|}{\textbf{Trading mode}}&\multicolumn{4}{c|}{\textbf{Key property}} & \multicolumn{2}{c|}{\makecell[c]{\textbf{Innovative}\\ \textbf{attributes}}}\\  \cline{2-9} 
					&\makecell[c]{Spot}&\makecell[c]{Futures}&Sta.&\makecell[c]{Ind.}&Fai.&\makecell[c]{NonW.}&Ove.&\makecell[c]{Ris.}\\ \hline
					\makecell[l]{\cite{22,36,38}} &$\surd$& &$\surd$&$\surd$&$\surd$&$\surd$& &\\ \hline
					\makecell[l]{\cite{32,34}} &$\surd$& &$\surd$&$\surd$&$\surd$& & &\\ \hline
					\makecell[l]{\cite{33,35,37},\\\cite{39,40,41}} &$\surd$& & & & & & &\\ \hline
					\makecell[l]{\cite{42,43,45}} & &$\surd$& & & & & &\\ \hline
					\makecell[l]{\cite{44}} & &$\surd$& & & & &$\surd$&$\surd$ \\ \hline
					our work &$\surd$&$\surd$&$\surd$&$\surd$&$\surd$&$\surd$&$\surd$&$\surd$\\ \hline
			\end{tabular}}
	\end{center}}
\end{table}
This section delves into existing works related to resource trading in MCS networks.

\textbf{Spot trading.} Efficient worker recruitment in MCS networks has been mostly studied in spot trading mode \cite{22,32,33,34,35,36,37,38,39,40,41}. 
In \cite{22}, \textit{Dai et al.} considered MCS task owners' budgets and built a distributed many-to-many matchings between MCS tasks and smart phone users.
In \cite{32}, \textit{Yucel et al.} proposed a task allocation algorithm based on dynamic programming.
In \cite{33}, \textit{Nie et al.} developed a two-stage Stackelberg game for MCS networks to incentivize mobile users to participate in resource sharing.
In \cite{34}, \textit{Yucel et al.} obtained stability conditions for task assignment in the context of semi-opportunistic mobile crowdsensing.
In \cite{35}, \textit{Tan et al.} studied cooperative crowdsensing to tackle cooperative task allocation problems in social mobile crowdsensing. 
In \cite{36}, \textit{Yucel et al.} proposed three stable task assignment algorithms for MCS networks.
In \cite{37}, \textit{Wu et al.} proposed a personalized task recommender system for MCS, which recommends tasks to users.
In \cite{38}, \textit{Yucel et al.} found task assignments that can fulfill both preferences of service requesters and workers in a budget-constrained MCS system. 
In \cite{39}, \textit{Xu et al.} investigated collaborations among multiple edge nodes to match a set of edge nodes to a worker.
In \cite{40}, \textit{Peng et al.} studied the online task assignment in MCS, where each task has a desired time window for data collection.
In \cite{41}, \textit{Zhang et al.} proposed a privacy-preserving multitask allocation for MCS, which selects high-quality users for task execution.

\textbf{Futures trading.} Since spot trading suffers from excessive delay, energy consumption, and potential trading failures, researchers have looked into the futures trading mode for not only MCS but also other wireless edge applications.
In \cite{42}, \textit{Liwang et al.} proposed a futures resource trading mechanism for edge computing-assisted UAV networks.
In \cite{43}, \textit{Sheng et al.} studied futures-based spectrum trading in wireless networks.
In \cite{44}, \textit{Liwang et al.} investigated futures-enabled worker recruitment in MCS networks while considering uncertainties in workers' participation and their local workloads. 
In \cite{45}, \textit{Sexton et al.} introduced a futures-enabled resource slicing scheme for wireless edge networks.
Nevertheless, futures trading also carries certain risks that call for careful consideration, which brings out our key motivations in investigating hybrid service trading mechanisms. A summary of related studies are provided in Table 1.

\textbf{Hybrid service trading.} In this work, we introduce a hybrid service trading mode over dynamic MCS networks by integrating both futures and spot markets. Our methodology offers several key advantages over existing approaches as summarized below.

	\noindent
	$\bullet$ \textbf{Hybrid matching mechanisms:} We explore a new approach to service trading that combines both futures and spot trading modes. By adopting futures trading (i.e., OIA3M), we address some of the weaknesses of spot trading, such as delays, excessive energy usage, and potential trading errors. We further recognize and acknowledge that network uncertainties can pose risks if we solely rely on futures trading. To cope with this issue, we use spot trading (i.e., O3M and OMOM) as a backup plan to ensure reliable and timely service trading. These two modes complement each other, while enhancing the quality of MCS services and user experience.
	
	\noindent
	$\bullet$ \textbf{Overbooking and risk evaluation:} We consider the uncertainties that in MCS networks, such as infrequent worker participation, which can cause fluctuations in resource supply. We thus adopt the concept of overbooking, allowing task owners to purchase more resources than their practical budget to better cope with the dynamics of service provisioning. Our analysis also includes an estimation of the risks that MCS task owners may encounter during practical transactions, which are then kept within an acceptable range.

\vspace{-0.18cm}
\section{Overview and System Model}
This paper examines a dynamic MCS network with multiple workers and tasks (owners) with service quality requirements and budget constraints. Our primary objective is to establish practicable/feasible mappings, along with service prices, between the two parties. To achieve efficient and cost-effective MCS service provisioning, we explore a novel hybrid trading market that blends futures and spot trading modes. A series of matching mechanisms will be devised for this market as follows.
	
	\noindent
	$\bullet$ Prior to practical service transactions, MCS tasks can enlist workers through long-term contracts in futures trading mode. To this end, we propose OIA3M mechanism, which selects a group of long-term workers for each task. Additionally, we design corresponding long-term contracts (e.g., service prices), taking into account the risks involved in futures trading, between tasks and workers. These contracts will be executed during future practical transactions (detailed in Section 4).
	
	\noindent
	$\bullet$ During practical transactions, the spot trading mode will activate, allowing participants with pre-signed contracts to complete transactions. This mode further ensures high service quality for MCS tasks with two backup plans in place. Firstly, if a task owner exceeds its practical budget (e.g., due to overbooking), this mode uses the OMOM mechanism to free up some of the long-term workers. Secondly, if task owners have surplus budgets, this mode utilizes the O3M mechanism to recruit temporary workers in the spot market to improve service quality (detailed in Section 5).
	
	In summary, in this work, we regard the service trading between MCS tasks and workers as a series of matching problems. Our objective is \textit{to achieve efficient matching outcomes between participants, thereby enabling cost-effective and responsive MCS services.}

\begin{figure*} \centering    
	\vspace{-0.2cm}
	\subfigbottomskip=-1pt
	\subfigcapskip=-15cm
	\setlength{\abovecaptionskip}{-0cm}
	\setlength{\belowcaptionskip}{-0.5cm}
	\subfigure[] {    
		\includegraphics[width=2\columnwidth]{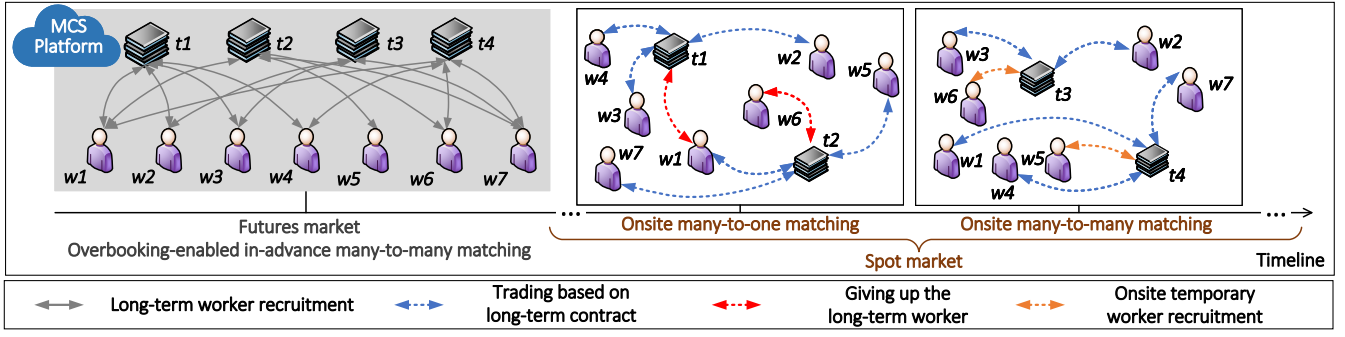}  
	}  

	\caption{Framework and timeline associated with our proposed matching-based hybrid service trading market. }     
	\label{fig} 
	\vspace{-0.52cm}    
\end{figure*}

\begin{table}[]
	\vspace{-0.1cm}
	{\footnotesize
		\caption{\footnotesize{Key notations used in the paper.}}\vspace{-0.6cm} 
\setlength{\tabcolsep}{1mm}{ 
		\begin{center}
			\begin{tabular}{|l|l|}
				\hline
				\multicolumn{1}{|l|}{\textbf{Notation}} & \multicolumn{1}{l|}{\textbf{Explanation}} \\ \hline
				$ \bm{T} $, $ \bm{T^{\prime}} $, $ \bm{T^{\prime\prime}} $ & MCS task set in OIA3M, OMOM, O3M \\ \hline
				$ \bm{W} $, $ \bm{W^{\prime}_i} $, $ \bm{W^{\prime\prime}} $ & Worker set in OIA3M, OMOM, O3M\\\hline
				$ t_i $ & $ i^\text{th} $ MCS task in $ \bm{T} $, $ \bm{T^{\prime}} $, $ \bm{T^{\prime\prime}} $ \\\hline
				$ w_j $ & $ j^\text{th} $ worker in $ \bm{W} $, $ \bm{W^{\prime}_i} $, $ \bm{W^{\prime\prime}} $ \\\hline
				$ c_{i,j} $	&\makecell[l]{Service cost of worker $ w_j $ participated in task $ t_i $} \\\hline
				$ p_{i,j}^F $, $ p_{i,j}^S $ & \makecell[l]{Payment of task $ t_i $ offered to worker $ w_j $ in the \\futures market and the spot market} \\\hline
				$ p^{Desire}_{i,j} $ & \makecell[l]{Desire payment that worker $ w_j $ wants from task $ t_i $} \\\hline
				$ q_{i,j} $&\makecell[l]{Service quality provided by worker $ w_j $ for task $ t_i $} \\\hline
				$\alpha_{j}$ &\makecell[l]{Random variable describes the participation of\\worker $ w_j  $}\\\hline
				$\tau$& Overbooking rate \\\hline
				\makecell[l]{$ B_{i} $, $ (1+\tau)B_{i} $,\\$ B_i^{\prime} $} & \makecell[l]{Budget, overbooked budget and remaining budget\\ for task $ t_i $} \\\hline
				$ Q_i $ &Desired service quality for task $ t_i $\\\hline
				\makecell[l]{$\gamma\left(t_i\right)$, $\mu \left(t_i\right)$,\\$\varphi\left(t_i\right)$} & \makecell[l]{Set of workers recruited for processing task $ t_i $ in\\ OIA3M, OMOM, O3M} \\\hline
				\makecell[l]{$\gamma\left(w_j\right)$, $\mu \left(w_j\right)$,\\$\varphi\left(w_j\right)$}&\makecell[l]{Set of tasks assigned to worker $ w_j $ in OIA3M, \\OMOM, O3M} \\\hline
				$ P_j^F $, $ P_j^{S1} $, $ P_j^{S2} $& \makecell[l]{Payment profile of worker $ w_j $ in OIA3M, OMOM,\\ O3M} \\\hline
				$ \mathbb{P}^F $, $ \mathbb{P}^{S1} $, $ \mathbb{P}^{S2} $ &\makecell[l]{Payment profile of all workers in OIA3M,OMOM,\\ O3M} \\\hline
				$ \mathbb{P}_{-j}^F $, $ \mathbb{P}_{-j}^{S1} $, $ \mathbb{P}_{-j}^{S2} $ & \makecell[l]{Payment profile of all workers excluding worker $ w_j $\\ in OIA3M, OMOM, O3M} \\\hline
				$ {U_{i}^{T}} $, $ \overline{U_{i}^{T}} $& Utility and expected utility of task $ t_i  $                               \\\hline
				$ {U_{j}^{W}} $, $ \overline{U_{j}^{W}} $& Utility and expected utility of worker $ w_j  $                                 \\\hline
				$ {R_{i}^{T}} $& Risk associated with task $ t_i $ \\ \hline
				$ \mathrm{\Delta}p$ & Payment reduction rate of workers 
				\\
				\hline
			\end{tabular}
		\end{center}}
	}
	\vspace{-0.6cm}   
\end{table}

\vspace{-0.2cm}
\subsection{Overview}
We consider an MCS network consisting of two parties: \textit{i)} multiple MCS tasks (owners) collected in set $ \bm{T} = \left\{ \begin{matrix} 	\left. {}t_{1},t_{2},\ldots,t_{|\bm{T}|} \right\} \\ \end{matrix} \right. $, and \textit{ii)} multiple workers gathered via set $ \bm{W} = \left\{ \begin{matrix}
	\left. {}w_{1},w_{2},\ldots,w_{|\bm{W}|} \right\} \\ \end{matrix} \right. $. Different workers may offer a task with different service quality, due to \textit{i)} different qualities, reliability, and accuracy of data; \textit{ii)} various sensing/computing capability; \textit{iii)} different network conditions (e.g., wireless channel quality)\cite{22,32}. To model such heterogeneities, we let $ q_{i,j} $ denote the service quality provided by worker $ w_j $ for task $ t_i $. Also, performing the same task on different workers may incur different costs due to having different hardware and software settings \cite{32,36}. Subsequently, we use $ c_{i,j} $ to represent the incurred cost on worker $ w_j $ when executing task $ t_i $. Also, let $p_{i,j}$ denote the asked payment/compensation of worker $ w_j $ for executing task $ t_i $, and $ p^{Desire}_{i,j} $ indicate the desired payment that worker $ w_j $ wants from task $ t_i $ (i.e., $ p^{Desire}_{i,j} $ is the highest asked payment that worker $ w_j$ can get from task $ t_i $ according to the market). The overall payment from a task to the recruited workers is limited by the task owner's budget. Moreover, a worker can offer services for multiple tasks.

Fig. 1 depicts a schematic of our proposed matching-based MCS hybrid service trading market. In the futures market, MCS tasks can recruit workers via reaching long-term contracts; while in the spot market, in each practical transaction\footnote{A practical transaction denotes a service trading event between tasks and workers.}, long-term workers and tasks fulfill their obligations as stipulated in long-term contracts, while some tasks may further employ temporary workers upon experiencing unsatisfying service quality. 
Examples of transactions are illustrated in Fig. 1. In particular, considering the existence of two tasks $t_1$ and $t_2$ in the system, in the designed futures market, MCS task owners and workers sign long-term contracts through an overbooking-enabled many-to-many matching mechanism. Then, in each practical transaction (spot market), onsite many-to-one matching between $ t_1 $ and $ t_2 $ is conducted when the overall payment that needs to be made for tasks $ t_1 $ and $ t_2 $ exceeds their budgets, in which case they should give up some workers (e.g., in the figure, task $ t_1 $ gives up worker $ w_1 $ and task $ t_2 $ gives up worker $ w_6 $). Also, in each practical transaction, due to an insufficient number of existing long-term workers, tasks may fail to reach their required service quality and thus aim to use their surplus budgets to recruit temporary workers to improve their service quality (e.g., in the figure, tasks $ t_3 $ and $ t_4 $ recruit workers $ w_6 $ and $ w_5 $, respectively). 

\vspace{-0.13cm}
\subsection{Modeling of Tasks and Workers}
Each MCS task $ t_i \in \bm{T} $ is associated with a desired service quality denoted by $ Q_i $ \cite{22,36}, which can be achieved via recruiting multiple workers. Also, each task $ t_i\in \bm{T} $ is associated with a budget\footnote{In this paper, "budget" and "practical budget" are used interchangeably (i.e., $ B_i $), while "overbooked budget" refers to $(1+\tau)B_i$.} $ B_i $, which limits the number of recruited workers. In the futures market, to cope with the dynamics of resource supply, we allow the tasks to overbook their limited budgets through introducing a scaling factor $\tau$ and refer to $ (1 + \tau )B_i $ as the \textit{overbooked budget}.

Motivated by the dynamics of workers (e.g., their locations local tasks, and limited battery) that can lead to the \textit{no-show} phenomenon, we model the uncertainties in the participation of each worker through a random variable $ \alpha_j $ drawn from a Bernoulli distribution $ \alpha_{j} \sim \textbf{\text{B}} \left( \left\{ 1,0 \right\},\left\{ a_{j},1 - a_{j} \right\} \right) $ with the probability mass function (PMF) shown below
\begin{equation}\label{key}	
	\setlength{\abovedisplayskip}{2pt} 
		\setlength{\belowdisplayskip}{2pt}
{\small
	\begin{aligned} 
	\text{Pr}\left( \alpha_{j} = n \right) = \left\{ \begin{matrix}
		{~~a_{j}~~~~,n = 1} \\
		{1 - a_{j},n = 0} ,\\ 
	\end{matrix} \right.
\end{aligned} }    
\end{equation}
where $ \alpha_j=1 $ indicates that worker $ w_j $ participates in a transaction; and $ \alpha_j=0 $ otherwise. We summarize the key notations used in our system model in Table 2.

\vspace{-0.2cm}
\section{Proposed OIA3M in The Futures Market}
\noindent
In this section, we formalize the futures market and our proposed OIA3M. We first define the following key parameters:

\noindent
$\bullet$ $\gamma\left(t_i\right)$: the set of workers recruited for processing task $ t_i $ in the futures market, i.e., $ \gamma\left(t_i\right)\subseteq \bm{W} $;

\noindent
$\bullet$	$ \gamma\left(w_j\right) $: the set of tasks that is assigned to worker $ w_j $ in the futures market, i.e., $ \gamma\left(w_j\right)\subseteq \bm{T} $;

In addition, we change the notation for $ p_{i,j} $ in the futures market to $ p_{i,j}^F $ to avoid confusion with spot trading payment. Without loss of generality, let $ P_j^F=\left\{\left.p_{i,j}^F \right|\forall t_i\in \bm{T}\right\} $, $ \mathbb{P}^F=\bigcup_{\ w_j\in \bm{W}} P_j^F $ and $ \mathbb{P}_{- j}^F = \mathbb{P}^F\backslash P_{j}^F $, denote the payment profile, the payment profile of all workers, and the payment profile of all the workers excluding worker $ w_j $, respectively. 

\vspace{-0.2cm}
\subsection{Utility, Expected Utility and Risk}
We define the utility of task $ t_i\in \bm{T} $ as its overall experienced service quality defined as
\begin{equation}\label{key}
		\setlength{\abovedisplayskip}{3pt} 
{\small
\begin{aligned}  
	U^T(t_i,\gamma\left(t_i\right))=\sum_{w_j\in\gamma\left(t_i\right)}{\alpha_jq_{i,j}}.
\end{aligned} }
\end{equation}
\vspace{-3mm}

Due to the uncertain participation of workers captured via $ \{ \alpha_j\}$, it is challenging to maximize $ U ^T(t_i,\gamma\left(t_i\right)) $ in the futures market. Thus, we opt to consider the expected value of $ U^T(t_i,\gamma\left(t_i\right) ) $ as the utility of each task, which is given by
\begin{equation}\label{key}
	\setlength{\abovedisplayskip}{2pt}
	\setlength{\belowdisplayskip}{2pt}
		{\small
		\begin{aligned} 
	\overline{U^{T}}\left(t_{i},\gamma\left( t_{i} \right) \right) =E[U^T(t_i,\gamma(t_i))]= {\sum\limits_{w_{j} \in \gamma{(t_{i})}}{a_{j}q_{i,j}}}.
			\end{aligned} }
\end{equation}

Also, we define the utility of worker $ w_j\in \bm{W} $ as the difference between its total received payments and cost of task execution given by
\begin{equation}\label{key}
\setlength{\abovedisplayskip}{2pt}  
	\setlength{\belowdisplayskip}{2pt}
{\small
\begin{aligned} 
	U^W(w_j,\gamma\left(w_j\right))=\sum_{t_i\in\gamma\left(w_j\right)}{\alpha_j(p_{i,j}^F-}c_{i,j}),
\end{aligned} }
\end{equation}
based on which, the expected utility can be calculated as
\begin{equation}\label{key}
\setlength{\abovedisplayskip}{2pt}  
	\setlength{\belowdisplayskip}{2pt}
	{\small
		\begin{aligned} 
	\overline{U^W}(w_j,\gamma\left(w_j\right))=E[U^W(w_j,\gamma\left(w_j\right))]=\sum_{t_i\in\gamma\left(w_j\right)}{a_j(p_{i,j}^F-}c_{i,j}).
		\end{aligned} }
\end{equation}

We further define $ P_j^E= a_jP_j^F=\left\{\left.a_jp_{i,j}^F \right|\forall t_i\in \bm{T}\right\} $, $ \mathbb{P}^E=\bigcup_{\ w_j\in \bm{W}} P_j^E $ and $ \mathbb{P}_{- j}^E = \mathbb{P}^E\backslash P_{j}^E $ as the profile of the expected payment of worker $ w_j $, the profile of the expected payment of all workers, and the profile of the expected payment of all workers excluding $ w_j $, respectively.

Our goal in the futures market is to facilitate signing long-term contracts between tasks and workers via analyzing historical statistics of the existing uncertainties in the market captured via no-show events $ \alpha_j $, $ \forall w_j $. To this end, we define a \textit{risk factor} for each task $ t_i\in \bm{T} $, which is formalized as the probability of receiving an unsatisfying service quality due to the dynamics of workers
\begin{equation}\label{key}
\setlength{\abovedisplayskip}{2pt} 
	\setlength{\belowdisplayskip}{2pt}
		{\small
		\begin{aligned} 
	R^{T}\left(t_{i},\gamma\left( t_{i} \right)\right) = \text{Pr}\left\{ \frac{\sum\limits_{w_{j} \in \gamma{(t_{i})}}{\alpha_{j}q_{i,j}}}{Q_{i}} \leq \lambda_{1}^{T} \right\},
			\end{aligned} }
\end{equation}
where $ \lambda_1^T $ is a positive tunable threshold and $ Q_i $ represents the desired service quality. For each task $ t_i $, a larger value of $ 	R^{T}\left(t_{i},\gamma\left( t_{i} \right) \right) $ implies a higher risk of receiving an unsatisfying service quality. Thus, if the risk is unaccepted (e.g., under high no-show chance described by $ \alpha_j=0 $, $ \forall w_j $), the task will not sign long-term contracts with workers in the futures market, and will later engage in spot trading.

\vspace{-0.2cm}
\subsection{Problem Formulation}
Resource provisioning in the futures market can be described as an overbooking-enabled in-advance many-to-many matching, which is prior to future transactions. In this matching, we aim to achieve feasible mappings characterized by $\gamma(.)$ between tasks and workers, in facilitating long-term contracts.

\textbf{Service quality optimization of tasks under budget and risk constraints.} In the futures market, each task $ t_i\in \bm{T} $ aims to maximize its overall expected service quality, which can be formulated as the following optimization problem:
\begin{equation}\label{key}
\setlength{\abovedisplayskip}{2pt} 
\setlength{\belowdisplayskip}{2pt}
			{\small
		\begin{aligned} 
	\underset{\gamma{(t_{i})}}{\max}~\overline{U^{T}}\left( t_{i},\gamma\left( t_{i} \right) \right) 
				\end{aligned} }   
\end{equation}
\vspace{-1.6mm} 
			{\small\setlength{\abovedisplayskip}{2pt} 
				\setlength{\belowdisplayskip}{2pt}
		\begin{flalign} 
&\ \text{s.t.}~~~~~~~~~~~~~~~~~~~~~~	\sum_{w_j\in\gamma\left(t_i\right)}p_{i,j}^F\le(1+\tau)B_i ,&
					\end{flalign} } 
\vspace{-1.6mm}                
\begin{equation}\label{key}
\setlength{\abovedisplayskip}{2pt} 
\setlength{\belowdisplayskip}{2pt}
			{\small
		\begin{aligned} 
	\gamma\left(t_i\right)\subseteq \bm{W},
					\end{aligned} } 
\end{equation}                        
\begin{equation}\label{key}
\setlength{\abovedisplayskip}{2pt} 
\setlength{\belowdisplayskip}{2pt}
			{\small
		\begin{aligned} 
		R^{T}\left(t_{i},\gamma\left( t_{i} \right) \right)\le\lambda_2^T ,
						\end{aligned} } 
\end{equation}
where $ \lambda_2^T $ is a threshold constant within interval $ (0,1] $. In the above formulation, constraint (8) ensures that the expenses of task $ t_i $ devoted to recruiting workers $ \gamma\left(t_i\right) $ do not exceed the overbooked budget $ (1+\tau)B_i $, constraint (9) forces recruited workers $ \gamma\left(t_i\right) $ to belong to set $ \bm{W} $, and constraint (10) dictates the tolerance of each MCS task on receiving an undesired service quality.

In the above optimization problem, considering the definition of risk factor in (6), constraint (10) is a probabilistic constraint, which makes obtaining the solution non-trivial. We thus transform this constraint to a tractable deterministic constraint through exploiting a set of bounding techniques. Particularly, (10) can be rewritten as
\begin{equation}\label{key}
\setlength{\abovedisplayskip}{2pt} 
\setlength{\belowdisplayskip}{2pt}
	{\small
			\begin{aligned}
 R^{T}\left( t_{i},\gamma\left( t_{i} \right) \right) \leq  \lambda_{2}^{T}\Rightarrow \text{Pr}\left\{ {\sum\limits_{w_{j} \in \gamma{(t_{i})}}{\alpha_{j}q_{i,j}}} \geq \lambda_{1}^{T}Q_{i} \right\} \geq 1 - \lambda_{2}^{T}.  
	\end{aligned} }
\end{equation}
To obtain a tractable form for (11), we first obtain an upperbound on its left-hand side using Markov inequality \cite{46} as follows:
\begin{equation}
\setlength{\abovedisplayskip}{2pt} 
\setlength{\belowdisplayskip}{2pt}
	{\small
	\begin{aligned}
			&\text{Pr}\left\{ {\sum\limits_{w_{j} \in \gamma{(t_{i})}}{\alpha_{j}q_{i,j}}} \geq \lambda_{1}^{T}Q_{i} \right\}\leq\\ & \frac{\text{E}\left\lbrack {\sum\limits_{w_{j} \in \gamma{(t_{i})}}{\alpha_{j}q_{i,j}}} \right\rbrack}{\lambda_{1}^{T}Q_{i}} = \frac{\sum\limits_{w_{j} \in \gamma{(t_{i})}}{a_{j}q_{i,j}}}{\lambda_{1}^{T}Q_{i}}.\\
	\end{aligned} }
\end{equation}   
Combining (11) and (12), we get the following tractable form for (10):              
\begin{equation}\label{key}
\setlength{\abovedisplayskip}{2pt} 
\setlength{\belowdisplayskip}{2pt}
	{\small
	\begin{aligned}
	\frac{\sum\limits_{w_{j} \in \gamma{(t_{i})}}{a_{j}q_{i,j}}}{\lambda_1^TQ_i}\geq1-\lambda_2^T.
\end{aligned} }       
\end{equation}           

\textbf{Workers' utility optimization with payment constraints.} In our market of interest, each worker $ w_j\in \bm{W} $ aims to maximize its expected utility, which can be formulated as the following optimization problem:

~

\begin{equation}\label{key}
\setlength{\abovedisplayskip}{2pt} 
\setlength{\belowdisplayskip}{2pt}
	{\small
	\begin{aligned}
	\underset{\gamma{(w_{j})}}{\max}\overset{¯}~{\overline{U^{W}}}\left(w_{j}, \gamma\left( w_{j} \right) \right)
	\end{aligned} }
\end{equation}        \vspace{-2.9mm}          
	{\setlength{\abovedisplayskip}{2pt} 
		\setlength{\belowdisplayskip}{2pt}
		\small
	\begin{flalign}
&\ \text{s.t.}~~~~~~~~~~~~~~~~~~~~~~~~~~~~~~~~~~\gamma\left( w_{j} \right) \subseteq \bm{T},&
	\end{flalign} }
\vspace{-1.8mm}
\begin{equation}\label{key}
	\setlength{\abovedisplayskip}{2pt} 
	\setlength{\belowdisplayskip}{2pt}
	{\small
	\begin{aligned}
	c_{i,j}\le p_{i,j}^F\le p^{Desire}_{i,j}, \forall t_i \in \gamma\left( w_{j} \right),
	\end{aligned} }
\end{equation}                   
where constraint (15) ensures that task set $ \gamma\left(w_j\right) $ belongs to set $ \bm{T} $, and constraint (16) ensures the payments asked by $ w_j $ can guarantee its non-negative expected utility.

It can be construed that in the futures market, we are interested in a many-to-many matching $\gamma(.)$ between the tasks and workers that holds in the conditions of the above two optimization problems, which is obtained in the following.

\vspace{-0.2cm}
\subsection{Overbooking-enabled In-Advance Many-to-Many Matching (OIA3M)}
We next introduce OIA3M, which is a unique many-to-many matching tailored to the characteristics of the futures market. This matching is designed in such a way that can cope with the existing uncertainties and risks in MCS networks, which differentiates it from conventional matching mechanisms \cite{22,32,34}, in which matching decisions are made based on the current network/market conditions.

OIA3M is formalized through a set of definitions. We first define the basic properties of a many-to-many matching below.
\begin{Defn}(Many-to-many matching in the futures market)
	A many-to-many matching $ \gamma $ in the futures market constitutes a mapping between task set $ \bm{T} $ and worker set $ \bm{W} $, which satisfies the following properties:
	
	\noindent
	$\bullet$ for each task $  t_{i} \in \bm{T},\gamma\left( t_{i} \right) \subseteq \bm{W} $,
	
	\noindent
	$\bullet$ for each worker $ w_{j} \in \bm{W}, \gamma\left( w_{j} \right) \subseteq \bm{T} $,
	
	\noindent
	$\bullet$ for each task $ t_i $ and worker $ w_j $, $  t_i\in\gamma(w_j)$ if and only if $ w_j\in\gamma\left(t_i\right) $.
\end{Defn}

We next define a concept called \textit{blocking coalition}, which is a significant factor that may make the matching unstable.

\begin{Defn}(Blocking coalition in OIA3M)
Under a given matching $ \gamma $ and an expected payment profile $ \mathbb{P}^E $, worker $ w_j $ and task set $ \mathbb{T} \subseteq \bm{T}$ may form one of the following two types of blocking coalition $ (w_j; \mathbb{T}) $ under an expected payment $ \widetilde{ P_j^E} $.

\textbf{Type 1 blocking coalition:} Type 1 blocking coalition satisfies the following two conditions:

\noindent
$\bullet$ Worker $ w_j $ prefers execution of task set $ \mathbb{T} \subseteq \bm{T} $ to its currently matched task set $ \gamma(w_j) $, i.e., 
\begin{equation}\label{key}
\setlength{\abovedisplayskip}{2pt} 
\setlength{\belowdisplayskip}{2pt} 
	{\small
	\begin{aligned}
	\overline{U^W}(w_j,\mathbb{T})>\overline{U^W}(w_j,\gamma(w_j)). 
\end{aligned} }        
\end{equation}

\noindent
$\bullet$ Every task in $ \mathbb{T} $ prefers to recruit workers rather than being matched to its currently matched/assigned worker set. That is, for any task $ t_i\in \mathbb{T} $, there exists worker set $ \gamma^\prime(t_i) $ that constitutes the workers that need to be evicted, satisfying
\begin{equation}\label{key}
\setlength{\abovedisplayskip}{2pt} 
\setlength{\belowdisplayskip}{2pt}
	{\small
	\begin{aligned}
\overline{U^{T}}\left(t_{i},\left\{\gamma\left( t_{i} \right)\backslash\gamma^{\prime}\left( t_{i} \right)\right\} \cup \left\{ w_{j} \right\} \right) > \overline{U^{T}}\left(t_{i},\gamma\left( t_{i} \right)  \right).\\
	\end{aligned} }
\end{equation} 

\textbf{Type 2 blocking coalition:} Type 2 blocking coalition satisfies the following two conditions:

\noindent
$\bullet$ Worker $ w_j $ prefers executing task set $ \mathbb{T} \subseteq \bm{T} $ to its currently matched task set $ \gamma(w_j) $, i.e.,
\begin{equation}\label{key}
\setlength{\abovedisplayskip}{2pt} 
\setlength{\belowdisplayskip}{2pt}
		{\small
		\begin{aligned}
	\overline{U^W}(w_j,\mathbb{T} )>\overline{U^W}(w_j,\gamma(w_j) ).
		\end{aligned} }
\end{equation} 

\noindent
$\bullet$ Every task in $ \mathbb{T} $ prefers to further recruit worker $ w_j $ under expected payment profile $ \widetilde{\ P_j^E}\cup\mathbb{P}^E $ in conjunction to its currently matched/assigned worker set. That is, for any task $ t_i\in \mathbb{T} $, we have
\begin{equation}\label{key}
\setlength{\abovedisplayskip}{2pt} 
\setlength{\belowdisplayskip}{2pt}
{\small
\begin{aligned}
	\overline{U^T}(t_i,\gamma(t_i)\cup\left\{ w_{j} \right\})>\overline{U^T}(t_i,\gamma(t_i) ) .
\end{aligned} }
\end{equation}
\end{Defn}

Considering the above definition, Type 1 blocking coalition leads to unstability of matching since the task is incentivized to recruit a different set of workers as compared to what is dictated by the matching to acquire a higher service quality. Similarly, Type 2 blocking coalition makes the matching unstable since the task has a left-over budget that can be used to recruit more workers to increase its service quality.

\vspace{-0.2cm}
\subsection{Design Targets}
Key design targets of our interest for OIA3M in the futures market are described below.

\begin{Defn}(Individual rationality of OIA3M) A matching $ \gamma $ ensures the individual rationality of participants if the following conditions are jointly satisfied:

\noindent
$\bullet$ For tasks, the risk of each task receiving an undesired service quality is controlled to be within a certain range, i.e., constraint (10) is satisfied.

\noindent
$\bullet$ For tasks, the overall payment of a task $ t_i $ matched to workers $ \gamma\left(t_i\right) $ under expected payment profile $ \mathbb{P}^E $ does not exceed $ (1+\tau)B_i $, i.e., constraint (8) is met.

\noindent
$\bullet$ For workers, each	worker $ w_j $ matched to tasks $ \gamma\left(w_j\right) $ under expected payment profile $ P_j^E $ obtains a non-negative expected revenue, i.e.,
\begin{equation}\label{key}
\setlength{\abovedisplayskip}{2pt} 
\setlength{\belowdisplayskip}{2pt}
{\small
\begin{aligned}
	\overline{U^W}(w_j,\gamma\left(w_j\right) ) \geq 0 . 
\end{aligned} } 
\end{equation}
\end{Defn}

\begin{Defn}(Fairness of OIA3M) A matching $ \gamma $ is fair if and only if it imposes no type 1 blocking coalition.
\end{Defn}

\begin{Defn}(Non-wastefulness of OIA3M) A matching $ \gamma $ is non-wasteful if and only if it imposes no type 2 blocking coalition.
\end{Defn}

\noindent
\begin{Defn}(Strong stability of OIA3M) A matching $ \gamma $ is strongly stable if it is individual rationality, fair, and non-wasteful.
\end{Defn}

We next formalize OAI3M and then demonstrate that it conforms to all the aforementioned design targets.

\vspace{-0.2cm}
\subsection{Algorithm Design}
The interaction between MCS task owners and workers is carried out through multiple rounds in the futures market, where the details of each round are described below:

\noindent
$\bullet$ At the beginning of each round, workers announce their asked payments and offered service qualities to MCS tasks.

\noindent
$\bullet$ After collecting workers' reports, each task temporarily selects some workers who can provide the highest expected service quality as candidates, while satisfying its budget constraints (8). It then informs workers whether they have been selected or not.

\noindent
$\bullet$ After obtaining the decisions from tasks, if a worker is rejected by a task, it can reduce its asked payments to elevate its competitiveness in the next round. If a worker is selected by a task, its asked payment will remain unchanged in the next round.

\noindent
$\bullet$ Repeat the above steps until either of the following conditions is met: \textit{i)} all workers are recruited by tasks; \textit{ii)} no worker can further reduce its asked payments (e.g., to avoid negative utility).

Based on the above procedure, a task signs long-term contracts with a subset of workers as described later, let $ p_{i,j}^{F}\left\langle k \right\rangle $ denote the payment asked by worker $ w_j $ for task $ t_i $ in the $ k^\text{th} $ round. Also, let $ \Delta{p}_j $ denote the reduction in the asked payment that worker $ w_j $ is willing to tolerate every time being rejected by a task, as long as the worker's expected utility (5) remains to be non-negative. 

We next describe the steps of OIA3M, the pseudo-code of which is given in Algorithm 1, which takes into account the overbooked budgets and risk analysis in the futures market.

\noindent
\textbf{Step 1. Initialization:} At the beginning of each transaction, each worker $ w_j $'s asked payment is set to $ p_{i,j}^{F}\left\langle 1 \right\rangle = p^{Desire}_{i,j} $ (line 1, Algorithm 1). We also initialize $ \gamma^k (w_j ) $ that contains the tasks that the worker $ w_j $ is interested in and $ \gamma^k(t_i) $ that contains the workers temporarily selected by task $ t_i $ in the $ k^\text{th} $ round.

\noindent
\textbf{Step 2. Proposal of workers:} At each round $ k $, each worker $ w_j $ first chooses tasks that generate non-negative
revenue (16), and records them in $ \gamma^k(w_j ) $, i.e., $ 
\gamma^{k}\left( w_{j} \right) \leftarrow \left\{ t_{i} \middle| {p_{i,j}^{F}\left\langle k \right\rangle \geq c_{i,j},\forall t_{i} \in \bm{T}} \right\} $. Then, $ w_j $ sends a proposal to each task $ t_i $ in $ \gamma^k(w_j) $, including its asked payments $ p_{i,j}^{F}\left\langle k \right\rangle $ and offered service quality $ q_{i,j} $ (lines 5-8, Algorithm 1).

\noindent
\textbf{Step 3. Worker selection on tasks' side:} After collecting the information from workers in set $ {\widetilde{\gamma}}^k\left(t_i\right) $, each task $ t_i $ solves a knapsack problem to determine a collection of temporary workers, e.g., set $ \gamma^k (t_i)$, where $\gamma^k (t_i)\subseteq {\widetilde{\gamma}}^k\left(t_i\right)  $ that can bring the maximum expected service quality under budget $ (1+\tau)B_i $, as given in (22).

\noindent
\textbf{Step 4. Solving the 0-1 knapsack problem:} Considering the overbooked budget $ (1+\tau)B_i $, each task $ t_i $ obtains a set of workers to ensure the highest expected service quality, through solving a 0-1 knapsack problem,
\begin{equation}\label{key}
\setlength{\abovedisplayskip}{2pt} 
\setlength{\belowdisplayskip}{2pt}
	{\small
		\begin{aligned}
		\underset{\gamma^k{(t_{i})}}{\max} \sum_{w_j\in\gamma^k\left(t_i\right)}{a_j(p_{i,j}^F-}c_{i,j})
	\end{aligned} }
\end{equation}
\vspace{-1.5mm}
{
\setlength{\abovedisplayskip}{2pt} 
\setlength{\belowdisplayskip}{2pt}
	\small
	\begin{flalign}
&\ \text{s.t.}~~~~~~~~~\sum_{w_j\in\gamma^k\left(t_i\right)}{a_jp_{i,j}^F} \leq (1+\tau)B_i,~ \forall w_j\in {\widetilde{\gamma}}^k\left(t_i\right),&
\end{flalign} }
which can generally be solved via dynamic programming (DP) \cite{22,47,48,49} (lines 19-22, Algorithm 1). Then, each $ t_i $ reports the selected result of the current round to workers. 

\noindent
\textbf{Step 5. Decision-making on workers' side:} After obtaining the decisions from each task $ t_i\in\gamma^k (w_j) $, worker $ w_j $ inspects the following conditions:

\textbf{Condition 1.} If $ w_j $ is accepted by task $ t_i $ or its current asked payment $ p_{i,j}^{F}\left\langle k \right\rangle $ equals to its cost $ {\ c}_{i,j} $, its payment asked for $ t_i $ remains unchanged, i.e., $ p_{i,j}^{F}\left\langle {k + 1} \right\rangle \leftarrow p_{i,j}^{F}\left\langle k \right\rangle $ (line 14, Algorithm 1);

\textbf{Condition 2.} If $ w_j $ is rejected by a task $ t_i $ and its asked payment $ p_{i,j}^{F}\left\langle k \right\rangle $ is still above its cost $ {c}_{i,j} $, it decreases its asked payment for $ t_i $ in the next round, while avoiding a negative utility, as follows (line 12, Algorithm 1):
\begin{equation}\label{key}	
\setlength{\abovedisplayskip}{2pt} 
\setlength{\belowdisplayskip}{2pt}
{\small
	\begin{aligned}
	p_{i,j}^{F}\left\langle {k + 1} \right\rangle = \max\left\{ p_{i,j}^{F}\left\langle k \right\rangle - \mathrm{\Delta}p_j~,{~c}_{i,j} \right\},
\end{aligned} }
\end{equation}

\noindent
\textbf{Step 6. Repeat:} If all the asked payments stay unchanged from $ (k-1)^{\text{th}} $ round to $ k^{\text{th}} $ round, the matching will be terminated at round $ k $. We use $ \Sigma_{w_j\in \bm{W}}{flag}_j=0 $ to denote this situation (line 4, Algorithm 1).
Otherwise, the algorithm repeats the above steps (e.g., lines 3-16, Algorithm 1) in the next round.

\noindent
\textbf{Step 7. Risk analysis:} When the algorithm is terminated, each task $ t_i $ will choose whether to sign long-term contracts with the matched workers according to the risks of receiving an unsatisfying service quality, given by constraint (10).

Note that the computational complexity of our proposed OIA3M for each task $ t_i \in \bm{T} $ can be described by $ O(k_{Max}|\bm{W}|(1+\tau)B_i) $ (see Appendix A).
\begin{algorithm}[t!]  
	{\small
		\caption{\small{Proposed OIA3M in the futures market}}
		\LinesNumbered 
		\textbf{Initialization: $ k \leftarrow  1 $, $ p_{i,j}^{F}\left\langle 1 \right\rangle \leftarrow p^{Desire}_{i,j}$, for $ \forall i,j $, $ {flag}_{j} \leftarrow  1 $};\ 
		
		\% \textit{\textbf{Action of each worker $ w_j\in \bm{W} $}}
		
		\While{$ {flag}_{j} $}{
			\textbf{$ {flag}_{j} \leftarrow  0 $};
			
			$ \gamma^{k}\left( w_{j} \right) \leftarrow \left\{ t_{i} \middle| {p_{i,j}^{F}\left\langle k \right\rangle \geq c_{i,j},\forall t_{i} \in \bm{T}} \right\} $;
			
			\If{$ \forall\gamma^{k}\left( w_{j} \right) \neq \varnothing $}{
				\For{$ \forall t_{i} \in \gamma^{k}\left( w_{j} \right) $}{$ w_j $ sends a proposal including $ p_{i,j}^{F}\left\langle k \right\rangle $ and $ q_{i,j} $ to $ t_i $;}
				\textbf{Wait decisions from tasks in $ \gamma^k (w_j ) $};\
				
				\For{
					$ \forall t_{i} \in \gamma^{k}\left( w_{j} \right) $
				}{
					\If{$ w_j $ is rejected by $ t_i $ and $ p_{i,j}^{F}\left\langle k \right\rangle>c_{i,j} $}{
						$ p_{i,j}^{F}\left\langle {k + 1} \right\rangle \leftarrow  \max\left\{ p_{i,j}^{F}\left\langle k \right\rangle - \mathrm{\Delta}p~,{ c}_{i,j} \right\} $;}
					\Else{$ p_{i,j}^{F}\left\langle {k + 1} \right\rangle \leftarrow  p_{i,j}^{F}\left\langle k \right\rangle $;}
				}
				
				\If{there exists $p_{i,j}^{F}\left\langle {k + 1} \right\rangle \neq p_{i,j}^{F}\left\langle k \right\rangle\ $, $ \forall t_{i}\in\gamma^{k}\left( w_{j} \right) $}{
					$ {flag}_j\leftarrow 1 $,	$ k\leftarrow k+1 $;\
				}
			}
		}
		
		\% \textit{\textbf{Action of each task $ t_i \in \bm{T} $}}\

		\While{
			$ \Sigma_{w_{j}\in \bm{W}}{flag}_{j} > 0 $}{
			Collect proposals from the workers in $ \bm{W} $, e.g., using $ {\widetilde{\gamma}}^k\left(t_i\right) $ to include the workers that send proposals to $ t_i $;
			
			$ \gamma^k(t_i)\leftarrow $ choose workers from $ {\widetilde{\gamma}}^k\left(t_i\right) $ that can achieve the maximization of the expected utility under budget $ (1+\tau)B_i $;
			
			use DP;
			
			$ t_i $ temporally accepts the workers in $ \gamma^k(t_i) $, and rejects the others;
		}
		\If{
			$ R^{T}\left( t_{i} ,\gamma\left( t_{i} \right) \right) > \lambda_{2}^{T} $}{$ t_i $ gives up trading in the futures market;}
	}\vspace{-1mm}
\end{algorithm}
\vspace{-2mm}

\subsection{Property Analysis}
We next show that OIA3M satisfies the properties of convergence, individual rationality, fairness, non-wastefulness, and strong stability in the futures trading market. 

\noindent
\textbf{Lemma 1.} \textit{(Convergence, individual rationality, fairness, non-wastefulness of OIA3M) Algorithm 1 converges within finite rounds. Furthermore, Algorithm 1 ensures individual rationality of all tasks and workers, fairness, and non-wastefulness.}
\vspace{-0.1cm}
\begin{proof}
	See Appendix B.
\end{proof}
\vspace{-0.05cm}
\noindent
	\textbf{Theorem 1.} \textit{(Strong stability of OIA3M) OIA3M is strongly stable.}
	\vspace{-0.1cm}
\begin{proof}
	See Appendix B.
\end{proof}

\vspace{-0.4cm}
\section{Proposed OMOM and O3M in The Spot Market}
 Since the desired service quality of each task (i.e., $ Q_i $) may not be reached by long-term workers (e.g., due to the no-show phenomenon reflected by the random variable $\alpha_j$), we further consider spot trading mode as a backup plan for service trading. Specifically, the task-worker mapping under spot trading can be considered as either OMOM or O3M. OMOM is deployed when the overall payment exceeds the budget $ B_i $ of task $ t_i $, and $ t_i $ should give up some of the long-term workers. This scenario happens due to recruiting extra workers via overbooked budgets. On the other hand, since the long-term workers that actually participate in a transaction may be unable to provide sufficient services to task $ t_i $, O3M is deployed for $t_i$ to recruit more workers. To avoid confusion with the futures market, we rewrite $ p_{i,j} $ in the spot market as $ p_{i,j}^S $.
 
 \vspace{-0.2cm}
\subsection{Onsite Many-to-One Matching (OMOM)} 
\noindent
We next formalize OMOM mechanism for the spot market, which triggers when a task does not have enough budget to recruit all the long-term workers. We first define the following key parameters:

\noindent
$\bullet$ $ \bm{T^{\prime}} $: $ \{t_i \in \bm{T}\  |  \ \mathrm{\Sigma}_{w_j\in\gamma(t_i)}\alpha_jp_{i,j}^F>B_i\} $;

\noindent
$\bullet$ $ \bm{W_i^{\prime}} $: the set of actual participated workers of $ \gamma\left(t_i\right) $ in the spot market, i.e., $ \bm{W_i^{\prime}}\subseteq \gamma\left(t_i\right) $;

\noindent
$\bullet$ $ \mu\left(t_i\right) $: workers in $ \bm{W_i^{\prime}} $ that are recruited for processing task $ t_i $ in the spot market, i.e., $ \mu\left(t_i\right)\subseteq\ \bm{W_i^{\prime}} $;

\noindent
$\bullet$ $ \mu\left(w_j\right) $: a task $ t_i $ in $ \bm{T^{\prime}} $ that is assigned to worker $ w_j $ in the spot market, i.e., $ \mu\left(w_j\right)=\left\{ t_{i} \right\} $.

Without loss of generality, let $ P_j^{S1}=\left\{ p_{i,j}^S |t_i\in \bm{T^\prime}\right\} $, $ \mathbb{P}^{S1}=\bigcup_{\ w_j\in\bm{W_i^{\prime}}} P_j^{S1} $ and $ 
\mathbb{P}_{- j}^{S1} = \mathbb{P}^{S1}\backslash P_{j}^{S1} $ denote the payment profile, the payment profile of workers $ \ w_j\in \bm{W_i^{\prime}} $, and the payment profile of all the long-term workers who have already been recruited for task $ t_i $ in the futures market excluding worker $ w_j $, respectively.

\vspace{-0.2cm}
\subsubsection{Utilities of tasks and workers}
In the spot market, we define the utility of task $ t_i\in \bm{T^{\prime}} $ as its overall received service quality as
\begin{equation}\label{key}
\setlength{\abovedisplayskip}{2pt} 
\setlength{\belowdisplayskip}{2pt}
{\small
	\begin{aligned}
	U^T(t_i,\mu\left(t_i\right) )=\sum_{w_j\in\mu\left(t_i\right)} q_{i,j}.
\end{aligned} }
\end{equation}

Similarly, we define the utility of worker $ w_j\in\mu\left(t_i\right) $ as the difference between its total received payments and the cost of task execution given by
\begin{equation}\label{key}
\setlength{\abovedisplayskip}{2pt} 
\setlength{\belowdisplayskip}{2pt}
{\small
	\begin{aligned}
	U^W(w_j,\mu\left(w_j\right) )=\sum_{t_i\in\mu\left(w_j\right)}{(p_{i,j}^S-}c_{i,j}).
\end{aligned} }
\end{equation}   

\vspace{-0.2cm}    
\subsubsection{Problem formulation}
It can be construed that in the spot market, we are interested in a many-to-one matching $\mu(.)$ between a task and workers that holds in the below two optimization problems.

\textbf{Service quality optimization of tasks under budget constraints.} In the spot market, each task $ t_i \in \bm{T^\prime} $ aims to maximize its overall service quality, which can be formulated as the following optimization problem:
\begin{equation}\label{key}
\setlength{\abovedisplayskip}{2pt} 
\setlength{\belowdisplayskip}{2pt}
	{\small
		\begin{aligned}
	\underset{\mu{(t_{i})} }{\max}~U^{T}\left(t_{i},\mu\left( t_{i} \right) \right) 
\end{aligned} }
\end{equation} 
\vspace{-1.7mm}
{
\setlength{\abovedisplayskip}{2pt} 
\setlength{\belowdisplayskip}{2pt}
	\small
		\begin{flalign}
&\ \text{s.t.}~~~~~~~~~~~~~~~~~~~~~~~~~~~~~~~~~~~	\mu\left(t_i\right)\subseteq\bm{W_i^{\prime}},&
\end{flalign} }
\vspace{-1.7mm}
\begin{equation}\label{key}
\setlength{\abovedisplayskip}{2pt} 
\setlength{\belowdisplayskip}{2pt}
	{\small
		\begin{aligned}
	\sum_{w_j\in\mu\left(t_i\right)} p_{i,j}^S\le B_i.
\end{aligned} }
\end{equation} 
 
In the above formulation, constraint (28) forces the recruited workers $ \mu\left(t_i\right) $ to be in $ \bm{W_i^{\prime}} $ as stipulated by pre-signed long-term contracts, and (29) ensures that task $ t_i $ recruits workers $ \mu\left(t_i\right) $ under its budget $ B_i $.

\textbf{Workers' utility optimization with payment constraints.} Each worker $ w_j\in\bm{W_i^{\prime}} $ aims to maximize its utility through the following optimization problem:
\begin{equation}\label{key}
\setlength{\abovedisplayskip}{2pt} 
\setlength{\belowdisplayskip}{2pt} 
	{\small
		\begin{aligned}
	\underset{\mu\left( w_{j} \right)}{\max}~U_{j}^{W}\left(w_{j}, \mu\left( w_{j} \right)  \right)
\end{aligned} }
\end{equation}   
\vspace{-1.7mm}
{
	\setlength{\abovedisplayskip}{2pt} 
	\setlength{\belowdisplayskip}{2pt}
	\small
	\begin{flalign}
		&\ \text{s.t.}~~~~~~~~~~~~~~~~~~~~~~~~~~~~~~	\mu(w_j)\in \left\{ \left\{t_{i}\right\},\varnothing \right\} ,&
\end{flalign} }
\vspace{-1.7mm}          
\begin{equation}\label{key}
\setlength{\abovedisplayskip}{2pt} 
\setlength{\belowdisplayskip}{2pt}
	{\small
		\begin{aligned}
	c_{i,j}\le p_{i,j}^S\le p^{Desire}_{i,j} ,~\text{if}~\mu(w_j)\neq \varnothing,~t_i \in \mu(w_j),
\end{aligned} }
\end{equation}                        
where constraint (31) guarantees that the task set $ \mu\left(w_j\right) $ is either empty or includes only one task (e.g., $ t_i $), and constraint (32) ensures that the payments asked by worker $ w_j $ for task $ t_i $ satisfy its non-negative utility.

Service provisioning in the spot market can be described as an OMOM matching, which makes a task $ t_i $ reject/evict some existing long-term workers under its practical budget constraint $ B_i $, which is less than the overbooked budget $ (1+\tau)B_i $. We next define the basic characteristics of OMOM and introduce the notion of blocking pair, which may result in an unstable OMOM.

\noindent
\begin{Defn}(Many-to-one matching in the spot market) A many-to-one matching $ \mu $ in the spot market is a mapping between $ \bm{W_i^{\prime}} $ and a task $ t_{i} $, which satisfies the following conditions:

\noindent
$\bullet$ for a task $ t_i\in \bm{T^{\prime}} $, $ \mu\left(t_i\right)\subseteq\bm{W_i^{\prime}} $,

\noindent
$\bullet$ for any worker $ w_{j} \in \bm{W_i^{\prime}} $, $ \mu\left( w_{j} \right) = \left\{ t_{i} \right\} $,

\noindent
$\bullet$ for a task $ t_i $ and worker $ w_j $, $ \left\{ t_{i} \right\}=\mu(w_j) $ if and only if $ w_j\in\mu\left(t_i\right) $.
\end{Defn}

\noindent
\begin{Defn}(Blocking pair of OMOM)
Under a matching $ \mu $ and payment profile $ \mathbb{P}^{S1} $, worker $ w_j $ and a task $ t_i $ form a blocking pair $ (w_j; t_i) $ under a payment $ \widetilde{ P_j^{S1}} $.

\textbf{Type 1 Blocking pair:} Type 1 blocking pair satisfies the following condition
\begin{equation}\label{key}
\setlength{\abovedisplayskip}{2pt} 
\setlength{\belowdisplayskip}{2pt}
	{\small
	\begin{aligned} 
 U ^{T}\left(t_{i},\left\{\mu\left( t_{i} \right)\backslash\mu^{\prime}\left( t_{i} \right)\right\} \cup \left\{ w_{j} \right\} \right) > U ^{T}\left(t_{i},\mu\left( t_{i} \right)  \right), 
	\end{aligned} }
\end{equation}    
which indicates that task $ t_i $ can increase its service quality by giving up some workers $ \mu^{\prime}\left( t_{i} \right) $ and recruiting worker $ w_j $.

\textbf{Type 2 blocking pair:} Type 2 blocking pair satisfies the following condition
	\begin{equation}\label{key}
\setlength{\abovedisplayskip}{2pt} 
\setlength{\belowdisplayskip}{2pt} 
	{\small
	\begin{aligned}
		U ^T(t_i,\mu(t_i)\cup\left\{ w_{j} \right\} )>U ^T(t_i,\mu(t_i) ) ,
	\end{aligned} }
\end{equation} 
which makes a matching unstable since task $ t_i $ can recruit more workers under its budget constraint to increase its utility.
\end{Defn}

\vspace{-0.25cm}
\subsubsection{Design targets}
The key design targets of our interest for OMOM in the spot market are detailed below.

\vspace{-0.2cm}
\noindent
\begin{Defn}(Individual rationality of OMOM) Under a matching $ \mu $, participants are individually rational if the following conditions are satisfied:

\noindent
$\bullet$ For a task, the practical payment of task $ t_i $ that is matched to a set of workers $ \mu\left(t_i\right) $ under payment profile $ \mathbb{P}^{S1} $ will not exceed $ B_i $, i.e., constraint (28). 

\noindent
$\bullet$ For a task, its overall practical service quality of task $ t_i $ that fails to meet the desired service quality by long-term workers can be maximized, under its remaining budget, i.e., (27) and (28) are satisfied.

\noindent
$\bullet$ Every worker $ w_j $ that is matched to a task $ \mu\left(w_j\right) $ under payment profile $ P_j^{S1} $ obtains a non-negative revenue as
\begin{equation}\label{key}
\setlength{\abovedisplayskip}{2pt} 
\setlength{\belowdisplayskip}{2pt}
	{\small
	\begin{aligned}
	U ^W(w_j,\mu\left(w_j\right) ) \geq0 ~.
	\end{aligned} }
\end{equation}           
\end{Defn}

Also, definitions of fairness, non-wastefulness, and strong stability of OMOM are similar to Definitions 4-6.

\vspace{-0.2cm}
\subsubsection{Algorithm and Property Analysis}
Since many steps in Algorithm 2 are similar to Algorithm 1, the pseudo-code of Algorithm 2 (our proposed OMOM) is detailed in Appendix B due to space limitations. Besides, the computational complexity of OMOM for each task $ t_i \in \bm{T} $ can be described by $ O(k_{Max}^{\prime}|\bm{W_i^{\prime}}|B_i) $ (see Appendix A).
Properties of OMOM are provided below.

\noindent
\textbf{Lemma 2.} \textit{(Convergence, individual rationality, fairness, non-wastefulness of OMOM) Algorithm 2 converges within finite rounds. Furthermore, Algorithm 2 ensures individual rationality of all tasks and workers, fairness, and non-wastefulness.} 
\vspace{-0.15cm}
\begin{proof}
	See Appendix C.
\end{proof}

\vspace{-0.1cm}
\noindent
\textbf{Theorem 2.} \textit{(Strong stability of OMOM) OMOM is strongly stable.} 
\vspace{-0.15cm}
\begin{proof}
	See Appendix C.
\end{proof}

\vspace{-0.5cm}
\subsection{Onsite Many-to-Many Matching (O3M)}
We next discuss our proposed O3M mechanism in the spot market, which triggers when a task has a surplus budget in a practical transaction. We first define a set of parameters.

\noindent
$\bullet$ $ \bm{T^{\prime\prime}} $: $ \{t_i \in \bm{T}\  |  \ \mathrm{\Sigma}_{w_j\in\gamma(t_i)}\alpha_jp_{i,j}^F<B_i\} $;

\noindent
$\bullet$ $ \bm{W^{\prime\prime}} $: certain participation of workers in the spot market, i.e., $ \bm{W^{\prime\prime}}\subseteq \bm{W} $;

\noindent
$\bullet$ $ \varphi\left(t_i\right) $: workers in $ \bm{W^{\prime\prime}} $ that are recruited for processing task $ t_i $ in the spot market, and $ w_j \in \gamma(t_i) $ can't be recruited for $ t_i $, i.e., $ \varphi\left(t_i\right)\subseteq \bm{W^{\prime\prime}} $, $ \varphi\left(t_i\right) \bigcap \gamma(t_i)= \varnothing $;

\noindent
$\bullet$ $ \varphi\left(w_j\right) $: tasks in $ \bm{T^{\prime\prime}} $ that are assigned to workers $ w_j $ in the spot market, and $ t_i \in \gamma(w_j) $ can't be assigned to workers $ w_j $, i.e., $ \varphi\left(w_j\right)\subseteq \bm{T^{\prime\prime}} $, $ \varphi\left(w_j\right) \bigcap \gamma(w_j)= \varnothing $.

Without loss of generality, let $ P_j^{S2}=\left\{\left.p_{i,j}^S\right |\forall t_i\in \bm{T^{\prime\prime}}\right\} $, $ \mathbb{P}^{S2}=\bigcup_{\ w_j\in \bm{W^{\prime\prime}}} P_j^{S2} $ and $ \mathbb{P}_{- j}^{S2} = \mathbb{P}^{S2}\backslash P_{j}^{S2} $ denote payment profile of worker $ w_j $, the payment profile of all workers in $ \bm{W^{\prime\prime}} $, and payment profile of all workers in $ \bm{W^{\prime\prime}} $ excluding worker $ w_j $, respectively.

\vspace{-0.1cm}
\subsubsection{Utilities of tasks and workers}
We define the utility of task $ t_i\in \bm{T^{\prime\prime}} $ as its overall received service quality as
\begin{equation}\label{key}
\setlength{\abovedisplayskip}{2pt} 
\setlength{\belowdisplayskip}{2pt} 
		{\small
		\begin{aligned}
	U^T(t_i,\varphi\left(t_i\right) )=\sum_{w_j\in\varphi\left(t_i\right)} q_{i,j}.
		\end{aligned} }
\end{equation}
             
We also define the utility of worker $ w_j\in \bm{W^{\prime\prime}} $ as the difference between its total received payments and the cost of task execution given by
\begin{equation}\label{key}
\setlength{\abovedisplayskip}{2pt} 
\setlength{\belowdisplayskip}{2pt}
		{\small
		\begin{aligned}
	U ^{W}\left(w_{j},\varphi\left( w_{j} \right) \right) = \sum_{t_i\in\varphi\left(w_j\right)}{(p_{i,j}^S-}c_{i,j}).
		\end{aligned} }
\end{equation}      

\vspace{-0.15cm}      
\subsubsection{Problem formulation}
It can be construed that in the spot market, we are interested in a many-to-many matching $\varphi(.)$ between the tasks and workers that hold in the conditions of the above two optimization problems, which is obtained in the following.

\textbf{Service quality optimization of tasks under budget constraints.}
In the spot market, each task $ t_i\in \bm{T^{\prime\prime}} $ aims to maximize its overall service quality, which can be given by the following optimization problem:
\begin{equation}\label{key}
\setlength{\abovedisplayskip}{2pt} 
\setlength{\belowdisplayskip}{2pt}
		{\small
		\begin{aligned}
	\underset{\varphi{(t_{i})} }{\max}~U ^{T}\left( t_{i},\varphi\left( t_{i} \right) \right)
			\end{aligned} }
\end{equation}   
\vspace{-1.7mm}
{
	\setlength{\abovedisplayskip}{2pt} 
	\setlength{\belowdisplayskip}{2pt}
	\small
	\begin{flalign}
		&\ \text{s.t.}~~~~~~~~~~~~~~~~~~~~~~~~~~~~	{\sum\limits_{w_{j} \in \varphi{(t_{i})}}p_{i,j}^S} \leq B_{i}^{\prime} ~,&
\end{flalign} }
\vspace{-1.7mm}
\begin{equation}\label{key}
\setlength{\abovedisplayskip}{2pt} 
\setlength{\belowdisplayskip}{2pt}
		{\small
		\begin{aligned}
	\varphi\left(t_i\right)\subseteq \bm{W^{\prime\prime}}  ,
			\end{aligned} }
\end{equation}            
where $ B_i^\prime $ denotes the remaining budget of task $ t_i $ after paying for the long-term workers who take part in the transaction, which is given by 
\begin{equation}\label{key}
\setlength{\abovedisplayskip}{2pt} 
\setlength{\belowdisplayskip}{2pt}
		{\small
		\begin{aligned}
	B_i^\prime=B_i-\sum_{w_j\in\gamma\left(t_i\right)}{\alpha_jp_{i,j}^F} .
			\end{aligned} }
\end{equation}      

In the above formulation, constraint (39) ensures that task $ t_i $ recruits workers in set $ \varphi\left(t_i\right) $ under its remaining budget $ B_i^\prime $ and constraint (40) guarantees that the recruited workers $ \varphi\left(t_i\right) $ belong to $ \bm{W^{\prime\prime}} $.

\textbf{Workers' utility optimization with payment constraints.} Each worker $ w_j\in \bm{W^{\prime\prime}} $ aims to maximize its practical net revenue when participating in the spot market, which is modeled as the following optimization problem:
\begin{equation}\label{key}
\setlength{\abovedisplayskip}{2pt} 
\setlength{\belowdisplayskip}{2pt}
{\small
\begin{aligned}
	\underset{\varphi{(w_{j})} }{\max}~U ^{W}\left(w_{j},\varphi\left( w_{j} \right) \right)
\end{aligned} }
\end{equation}    
\vspace{-1.7mm}
{
	\setlength{\abovedisplayskip}{2pt} 
	\setlength{\belowdisplayskip}{2pt}
	\small
	\begin{flalign}
		&\ \text{s.t.}~~~~~~~~~~~~~~~~~~~~~~~~~~~~~~~~~~	\varphi(w_j)\ \subseteq \bm{T^{\prime\prime}} ~,&
\end{flalign} }
\vspace{-1.7mm}         
\begin{equation}\label{key}
\setlength{\abovedisplayskip}{2pt} 
\setlength{\belowdisplayskip}{2pt}
{\small
\begin{aligned}
	c_{i,j}\le p_{i,j}^S\le p^{Desire}_{i,j} ,~\forall t_i\in \varphi{(w_{j})},
\end{aligned} }
\end{equation}                    
where (43) guarantees that tasks in $ \varphi\left(w_j\right) $ belong to set $ \bm{T^{\prime\prime}} $ and (44) ensures that the payment asked by worker $ w_j $ for task $ t_i $ can be limited within a certain range.

When there exist multiple tasks that fail to meet their desired service qualities and still have surplus budgets, service provisioning in the spot market can be considered as onsite many-to-many matching\footnote{Although there may be only one task, we generally call it a many-to-many matching to cope with the common cases with multiple tasks.}, aiming at recruiting more workers to achieve better service qualities. We next describe the basic characteristics of this matching.

\vspace{-0.1cm}
\noindent
\begin{Defn}(Many-to-many matching in the spot market) A many-to-many matching $ \varphi $ in the spot market is a mapping between $ \bm{W^{\prime\prime}} $ and $ \mathbb{T}^{\prime\prime} $, which satisfies the following conditions:

\noindent
$\bullet$ for each task $ t_{i} \in \mathbb{T}^{\prime\prime}, \ \varphi\left( t_{i} \right) \subseteq \bm{W^{\prime\prime}}, $

\noindent
$\bullet$ for each worker $ 
w_{j} \in \bm{W^{\prime\prime}},\ \varphi\left( w_{j} \right)  \subseteq \mathbb{T}^{\prime\prime}, $

\noindent
$\bullet$ for each task $ t_i $ and worker $  w_j,~t_i\in\varphi(w_j) $ if and only if $ w_j\in\varphi\left(t_i\right). $
\end{Defn}

\vspace{-0.1cm}
\noindent
\begin{Defn}(Blocking coalition of O3M) Under a matching $ \varphi $ and the payment profile $ \mathbb{P}^{S2} $, worker $ w_j $ and task set $ \mathbb{T}^{\prime\prime} $ form a blocking coalition ($ w_j; \mathbb{T}^{\prime\prime} $) under a payment$ \widetilde{ P_j^{S2}} $.
	
\textbf{Type 1 blocking coalition:} Type 1 blocking coalition happens when the following conditions are met:
\begin{equation}\label{key}
\setlength{\abovedisplayskip}{2pt} 
\setlength{\belowdisplayskip}{2pt}
{\small
	\begin{aligned}
	U ^W(w_j,\mathbb{T}^{\prime\prime} )>U ^W(w_j,\varphi(w_j) ),
\end{aligned} }
\end{equation}  
\begin{equation}\label{key}
\setlength{\abovedisplayskip}{2pt} 
\setlength{\belowdisplayskip}{2pt} 
{\small
	\begin{aligned} 
 U ^{T}\left( t_{i},\left\{\varphi\left( t_{i} \right)\backslash\varphi^{\prime}\left( t_{i} \right) \right\}\cup \left\{ w_{j} \right\} \right) > U ^{T}\left( t_{i},\varphi\left( t_{i} \right)  \right). 
	\end{aligned} }
\end{equation}  

\textbf{Type 2 blocking coalition:} Type 2 blocking coalition happens when the following conditions are met:
\begin{equation}\label{key}
\setlength{\abovedisplayskip}{2pt} 
\setlength{\belowdisplayskip}{2pt} 
{\small
	\begin{aligned}
	U ^W(w_j,\mathbb{T}^{\prime\prime} )>U ^W(w_j,\varphi(w_j) ),
	\end{aligned}} 
\end{equation}          
\begin{equation}\label{key}
\setlength{\abovedisplayskip}{2pt} 
\setlength{\belowdisplayskip}{2pt}
	{\small
		\begin{aligned}
 U ^T(t_i,\varphi(t_i)\cup\left\{ w_{j} \right\}) >U ^T(t_i,\varphi(t_i) ). 
	\end{aligned}} 
\end{equation}   
\end{Defn}

Considering Definition 11, Type 1 blocking coalition implies that the matching result is unstable because worker $ w_j $ can obtain a higher profit from task set $ \mathbb{T}^{\prime\prime} $, rather than its current matched task set $ \varphi\left( w_{j} \right) $. Also, any task $ t_i $ in set $ \mathbb{T}^{\prime\prime} $ can increase its service quality by evicting some workers $ \varphi^{\prime}\left( t_{i} \right) $ and recruiting worker $ w_j $. Also, Type 2 blocking coalition makes matching unstable because task $ t_i $ can recruit more workers under its budget constraint to improve its utility, and that can benefit worker $ w_j $ as well.

\vspace{-0.1cm}
\subsubsection{Design targets}
The desired properties of O3M are detailed below.

\vspace{-0.1cm}
\noindent
\begin{Defn}(Individual rationality of O3M) Under a given matching $\varphi$, participants are individual rationality if the following conditions are jointly satisfied:

\noindent
$\bullet$ For tasks, the overall practical payment of each task $ t_i $ that is matched to a set of workers $ \varphi\left(t_i\right) $ under payment profile $ \mathbb{P}^{S2} $ will not exceed budget $ B^{\prime}_i $, i.e., constraint (39) is satisfied.

\noindent
$\bullet$ For tasks, the overall practical service quality of tasks that fail to meet the desired service quality by long-term workers can be maximized, under limited service supply and the remaining budget, i.e., (38) and (39).

\noindent
$\bullet$ Each worker $ w_j $ that is matched to a set of tasks $ \varphi\left(w_j\right) $ under payment profile $ P_j^{S2} $ obtains a non-negative revenue as
\begin{equation}\label{key}
\setlength{\abovedisplayskip}{2pt} 
\setlength{\belowdisplayskip}{2pt}
	{\small
	\begin{aligned}
		U ^{W}\left( w_{j},\varphi\left( w_{j} \right) \right)\geq0 ~.
	\end{aligned}} 
\end{equation}                
\end{Defn}

\vspace{-0.1cm}
The definitions of fairness, non-wastefulness, and strong stability for O3M are the same as Definitions 4-6.

\vspace{-0.1cm}
\subsubsection{Algorithm and Property Analysis}
Note that many steps in O3M (shown by Algorithm 3) are similar to Algorithm 1, the pseudo-code of Algorithm 3 (our proposed O3M) is thus moved to Appendix C. Besides, the computational complexity of O3M for each task $ t_i \in \bm{T^{\prime\prime}} $ can be expressed as $ O(k_{Max}^{\prime\prime}|\bm{W^{\prime\prime}}|B_i^{\prime}) $, the relevant analysis is given by Appendix A.
Properties of O3M are given below:

\noindent
\textbf{Lemma 3.} \textit{(Convergence, individual rationality, fairness, non-wastefulness of O3M) Algorithm 3 converges within finite rounds. Furthermore, Algorithm 3 ensures individual rationality of all tasks and workers, fairness, and non-wastefulness.}
\vspace{-0.1cm}
\begin{proof}
	See Appendix D.
\end{proof}
\vspace{-0.10cm}
\noindent
\textbf{Theorem 3.} \textit{(Strong stability of O3M) O3M is strongly stable.}
\vspace{-0.1cm}
\begin{proof}
	See Appendix D.
\end{proof}

\vspace{-0.3cm}
\section{Evaluation}
\noindent
We conduct comprehensive evaluations to verify the effectiveness of our methods. Simulations are carried out via MATLAB R2019a on a Windows machine with 12th Gen Intel Core i5-12400 2.5 GHz and 16 GB RAM. \vspace{-0.1cm}

\vspace{-0.2cm}
\subsection{Simulation Setting}
To emulate an MCS network, we utilize the real-world dataset of Chicago taxi trips \cite{50} which records taxi rides in Chicago from 2013 to 2016 with 77 community areas. We consider the $ 77^{\text{th}} $ community area as our sensing area\cite{51}, with 271259 data points. 

The probability of participation of each worker (taxi) $w_j$ in each transaction, i.e., $ a_j $, is obtained by analyzing the data of 200 taxis in January 2013, i.e., we calculate $ a_j $ by counting the number of days that taxis arrive at $ 77^{\text{th}} $ community area in January 2013 and dividing it by 31 days.
To capture the uncertainties of MCS tasks, the locations of tasks are randomly distributed in the $ 77^{\text{th}} $ community area. Moreover, to better quantize the service cost of workers (i.e., $ c_{i,j} $), we record three key factors from the dataset \cite{22,49}: \textit{i)} the distance traveled by taxi (i.e., the distance between pick-up locations and drop-off locations); \textit{ii)} the distance between the current positions of workers (e.g., pick-up location of taxis) and the location of a task; \textit{iii)} the distance between the position after task completion (e.g., the drop-off location of taxis) and the task location. We choose service costs $ c_{i,j} $, $ \forall i,j $ to be proportional to these factors. Also, service qualities, i.e., $ q_{i,j}, \forall i,j $, are chosen to be inversely proportional to the sum of distances in factors \textit{ii} and \textit{iii}.

Conducting the above procedure leads us to the following parameters:
$ c_{i,j} \in \lbrack 3,6\rbrack $, $p_{i,j} \in \lbrack 6,10\rbrack$, $ q_{i,j} \in \lbrack 1,5\rbrack$, $ B_{i} \in \lbrack 30,50\rbrack$, $\mathrm{\Delta}p = 1 $, $ a_{j} \in \left\lbrack {64.52\% ,96.77\%} \right\rbrack $, $ Q_{i} \in \lbrack 30,35\rbrack$, $ \lambda_{1}^{T} \in \lbrack 1,1.05\rbrack $, $ \lambda_{2}^{T} = 0.2$ \cite{22,44}.
To simplify the presentation, our proposed matching mechanisms are named by "Hybrid\_F\_S". In addition, we adopt Monte Carlo method to generate the results, where presented results are averaged over 1000 independent simulations.

\vspace{-0.2cm}
\subsection{Baseline Methods}
To better evaluate the performance of Hybrid\_F\_S, we consider a set of baseline methods described below.

\noindent
$ \bullet $ \textbf{Many-to-many matching-enabled conventional spot trading (Conventional\_S)\cite{22}:} Conventional\_S presumes a pure spot trading market for MCS network, where many-to-many matching is considered.

\noindent
$ \bullet $ \textbf{Many-to-many matching-enabled conventional futures trading (Conventional\_F)\cite{42}:} Conventional\_F presumes a pure futures trading market for MCS network, where many-to-many matching is considered to determine long-term workers for tasks. 

\noindent
$ \bullet $ \textbf{Service quality-preferred method (Quality\_P)\cite{52}:} Quality\_P considers a spot trading mode where each task selects workers with the best service qualities under its budget constraint. 

\noindent
$ \bullet $ \textbf{Random matching (Random\_M)\cite{52}:} Random\_M considers spot trading mode where each task selects workers randomly under its budget constraint.

\noindent
$ \bullet $ \textbf{Negotiation method (Negotiation)\cite{25}:} Negotiation considers spot trading mode where each task negotiates a unified asked payment with workers, and determines feasible workers under its budget constraint.

\vspace{-0.2cm}
\subsection{Performance Metrics}
To conduct a quantitative performance analysis, we focus on the following performance indicators:

\noindent
$ \bullet $ \textbf{Service quality:} Service quality is one of the most important factors in MCS networks, which is calculated in terms of the overall received service quality of tasks. 

\noindent
$ \bullet $ \textbf{The ratio of service quality of each method to that of Conventional\_S (RoSQ):} RoSQ is the ratio of the service quality of each method to that of Conventional\_S, which captures the performance gains of methods as compared to Conventional\_S.

\noindent
$ \bullet $ \textbf{The fraction of tasks that meet their desired service quality (FoDSQ):} FoDSQ is the ratio of the number of tasks that meet their desired service quality to the total number of tasks.

\noindent
$ \bullet $
\textbf{Utility of workers:} The overall payment received by workers in the market. 

\noindent
$ \bullet $ \textbf{Running time (RT, measured in ms):} The running time is obtained by MATLAB 2019a. 

\noindent
$ \bullet $ \textbf{Number of interactions (NI):} Total number of interactions between tasks (owners) and workers to obtain the matching results, capturing decision-making overhead.

\noindent
$ \bullet $ \textbf{Delay incurred by interactions among participants (DIP):} To evaluate the overall time spent on trading decision-making, we consider two types of latency during decision-making: \textit{i)} For each worker $ w_j \in \bm{W} $, we consider the time spent on informing task $ t_i $ about the asked payment as well as the provided service quality (e.g., line 8, Algorithm 1). We represent this through the uplink latency (denoted by $ \mathbbm{t}_{i,j}^{U} $), within interval $ [0.5, 11] $ milliseconds\cite{53,54,55}; \textit{ii)} For each task owner $ t_i \in \bm{T} $, we consider the time spent on reporting the decision of worker selection, to worker $ w_j $ (e.g., line 22, Algorithm 1). This is modeled by the downlink latency (denoted by $ \mathbbm{t}_{i,j}^{D} $), within interval $ [0.5, 4] $ milliseconds\cite{53,54,55}. 
 $ \mathbbm{t}_{i,j}^{U} $ and $ \mathbbm{t}_{i,j}^{D} $ are chosen to be proportional to the distance between the position of worker $ w_j $ (e.g., pick-up location of taxis associated with Chicago taxi trips introduced by Section 6.1) and the location of task $ t_i $.
Thus, the summation of the time consumed by decision-making is obtained as $ \text{DIP}=\sum_{i=1}^{|\bm{T}|}{\sum_{j=1}^{|\bm{W}|}\mathbb{N}_{i,j}\left(\mathbbm{t}_{i,j}^{D}+\mathbbm{t}_{i,j}^{U}\right)} $, where $ \mathbb{N}_{i,j} $ refers to the number of times that a task $ t_i $ reports the decision of worker selection (to worker $ w_j $), which also equals to the number of times that a worker $ w_j $ announces its asked payment as well as the provided service quality (to task $ t_i $).

\noindent
$ \bullet $ \textbf{Energy consumption incurred by interactions among participants (ECIP):} To evaluate the energy consumed by trading decision-making, we first consider the transmission power of workers denoted by $ e_j^{W} $ ($ \forall w_j\in\bm{W} $), within $ [0.2, 0.4] $ Watt\cite{25,56}, and the transmission power of task owner denoted by $ e_i^{T} $ ($ \forall t_i\in\bm{T} $), within $ [6, 20] $ Watt\cite{57}. Accordingly, the overall energy consumption caused by decision-making can be given by $ \text{ECIP} =\sum_{i=1}^{|\bm{T}|}{\sum_{j=1}^{|\bm{W}|}\mathbb{N}_{i,j}\left(e_i^{T}\mathbbm{t}_{i,j}^{D}+e_j^{W}\mathbbm{t}_{i,j}^{U}\right)} $.

\vspace{-0.2cm}
\subsection{Performance Evaluations}
\subsubsection{Service quality of tasks}
\begin{figure} 
	\centering 
	\vspace{-2.2cm}
	\subfigtopskip=2pt
	\subfigbottomskip=1pt
	\subfigcapskip=-2.0cm
	\setlength{\abovecaptionskip}{-1.6cm}
	\subfigure[] {
		\label{fig:a}     
		\includegraphics[width=0.506\columnwidth]{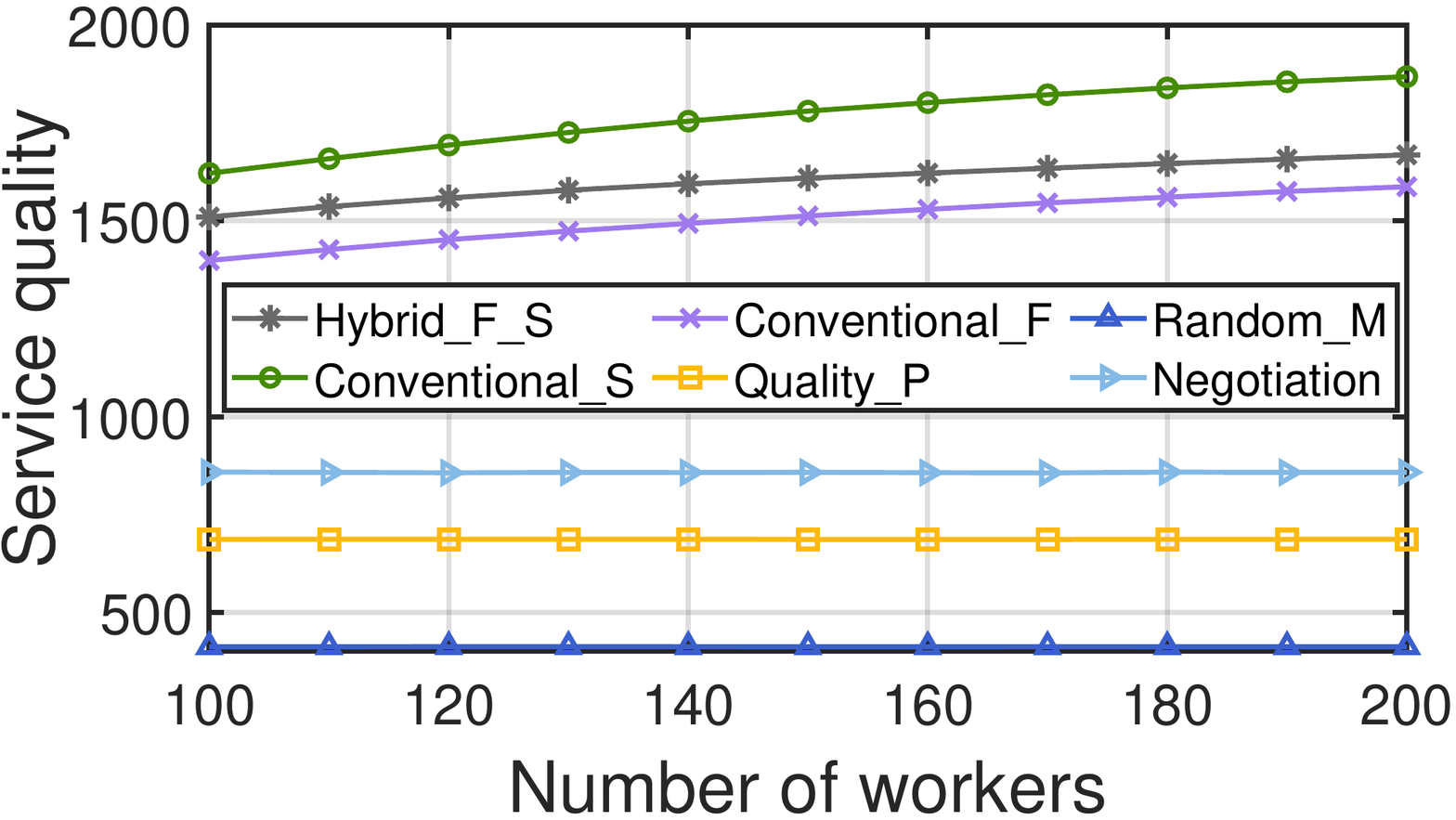}  
	}    \hspace{-6.7mm} \vspace{-35mm}
	\subfigure[] { 
		\label{fig:b}     
		\includegraphics[width=0.506\columnwidth]{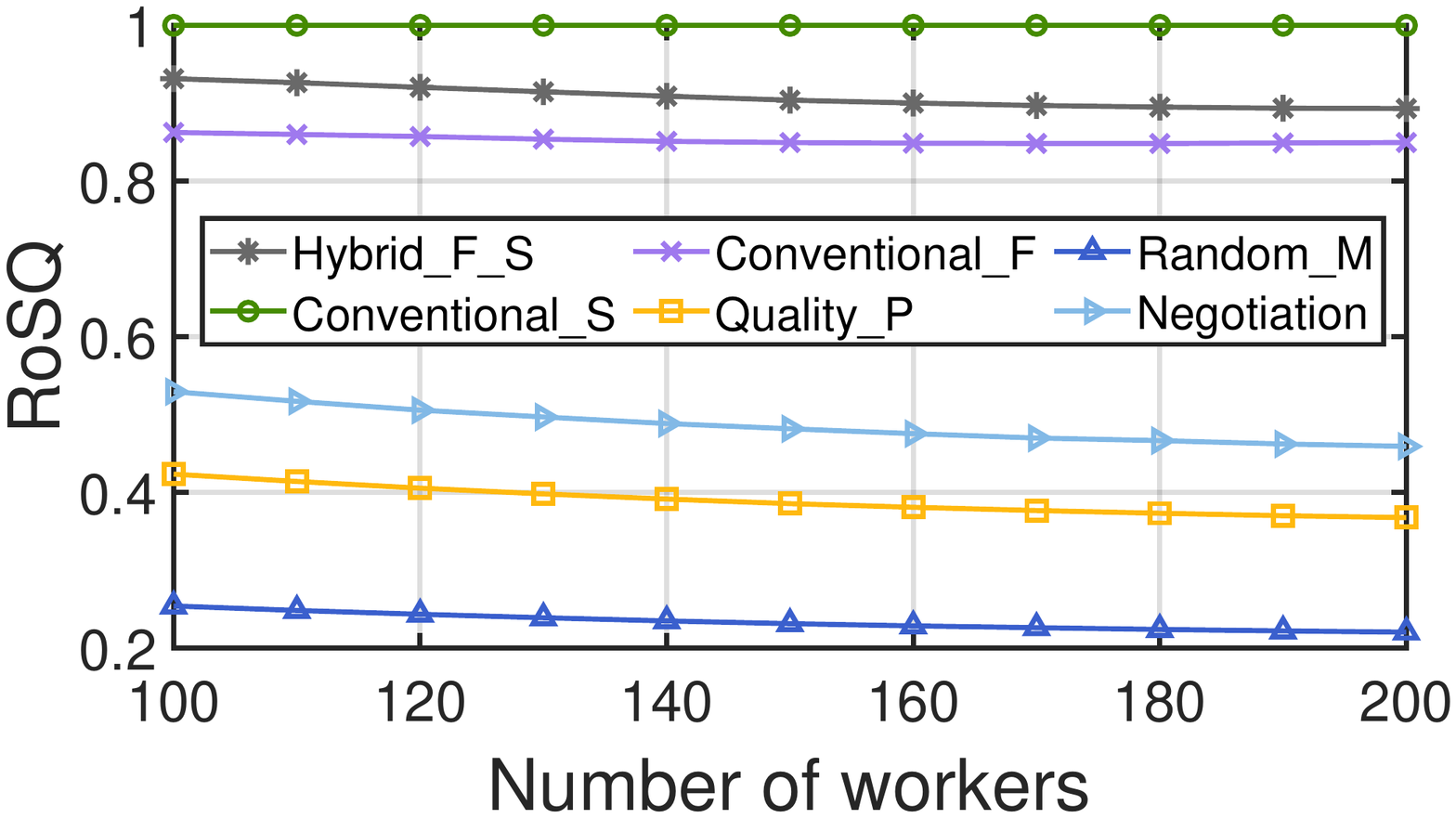}     
	}     
	\subfigure[] {
	\label{fig:c}     
	\includegraphics[width=0.506\columnwidth]{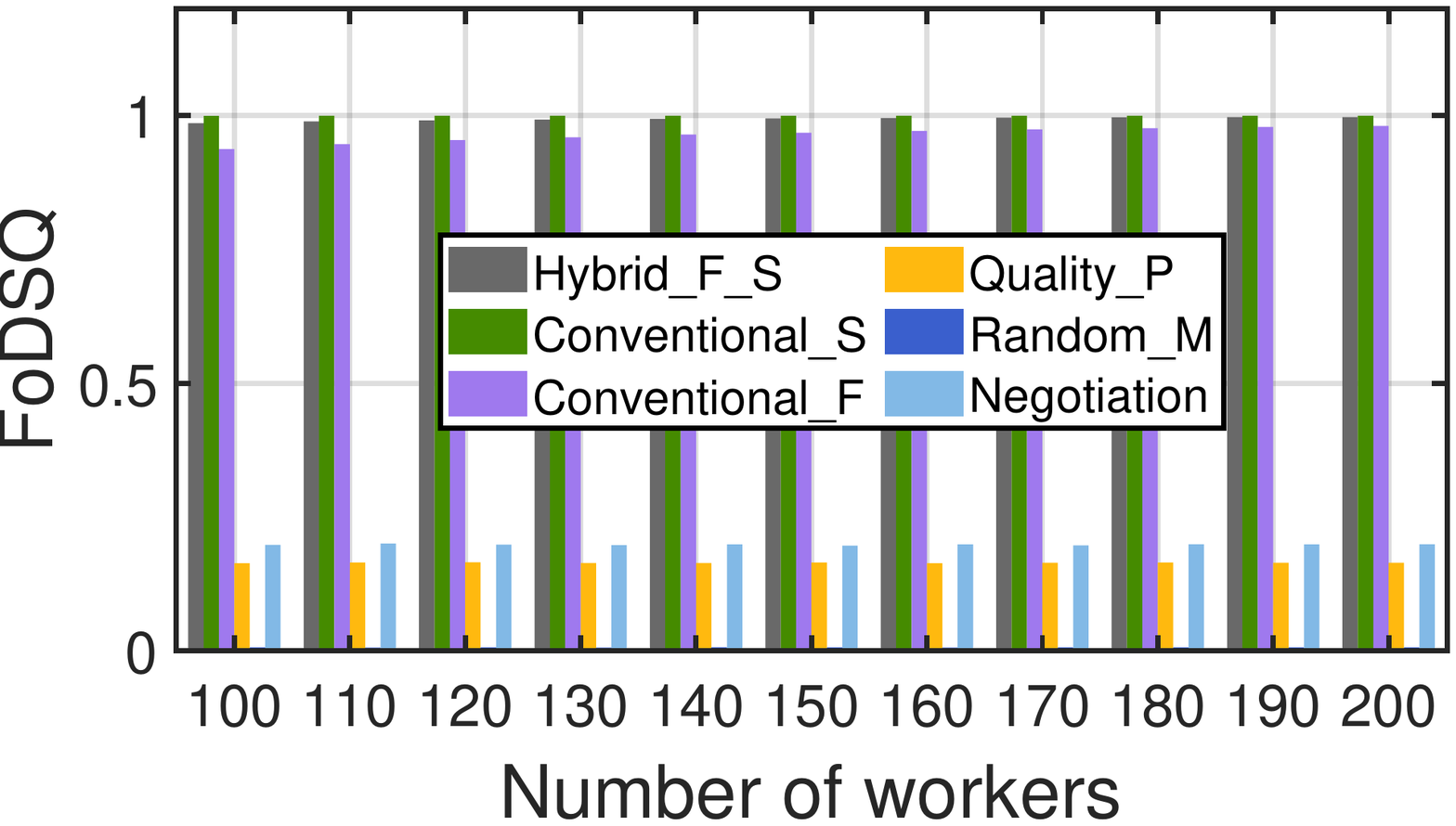}  
}    \hspace{-6.9mm} \vspace{-35mm}
\subfigure[] { 
	\label{fig:d}     
	\includegraphics[width=0.506\columnwidth]{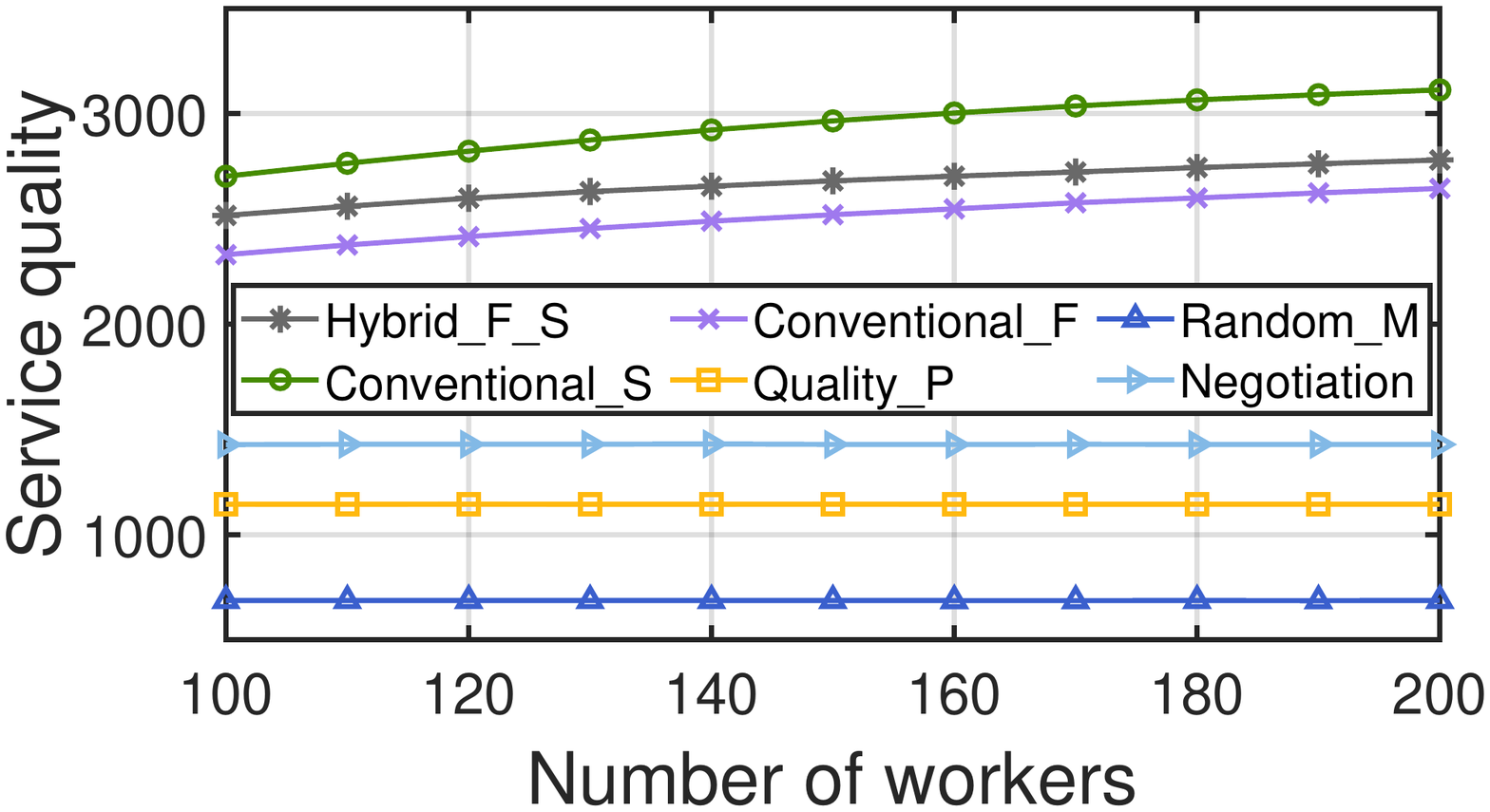}     
}  
	\subfigure[] { 
		\label{fig:e}     
		\includegraphics[width=0.506\columnwidth]{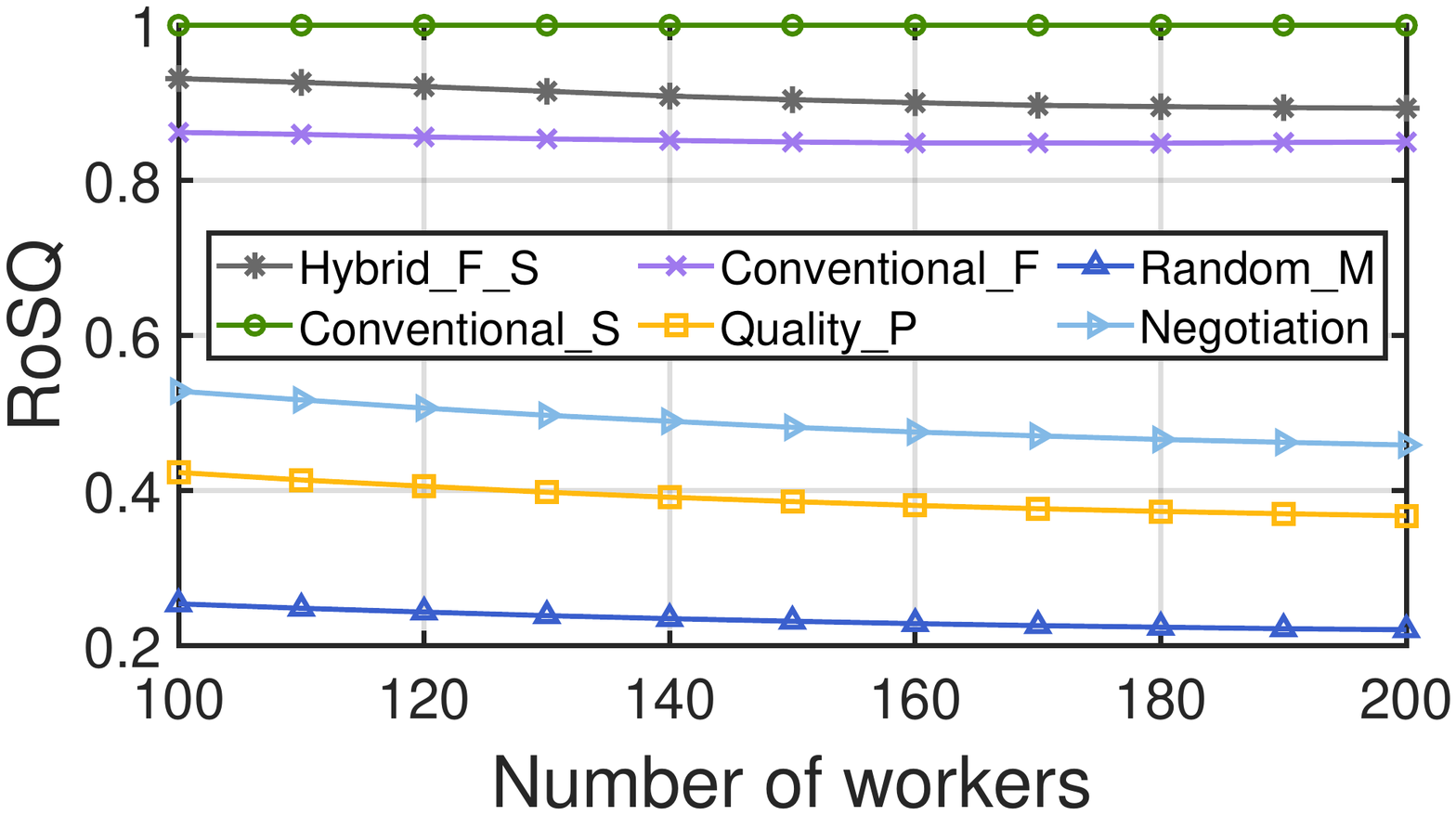}     
	}     \hspace{-6.9mm} 
	\subfigure[] {
		\label{fig:f}     
		\includegraphics[width=0.506\columnwidth]{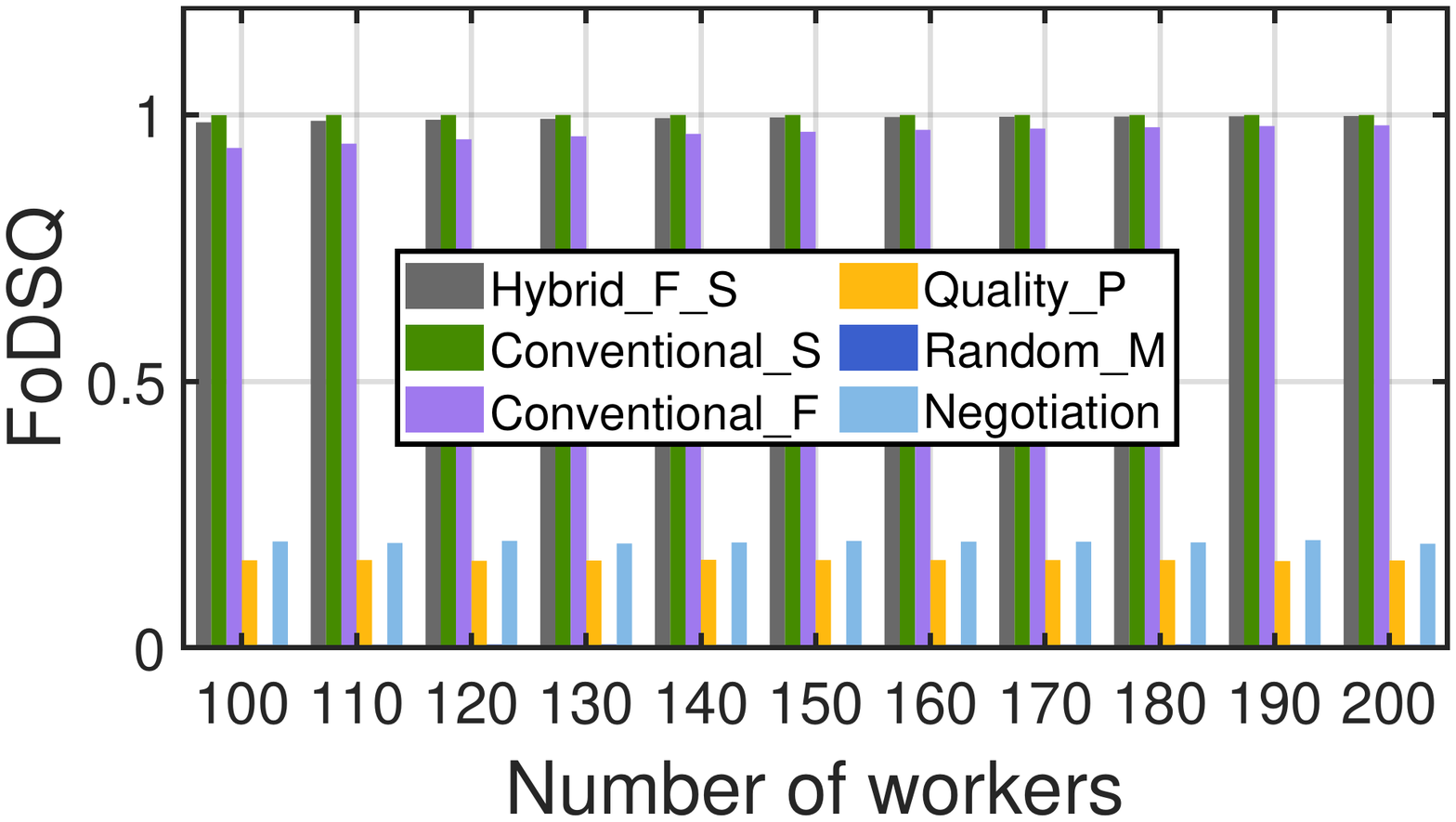}  
	}     
	\caption{Performance comparisons in terms of service quality, RoSQ and FoDSQ under different problem sizes, where (a)-(c) consider 30 tasks, and (d)-(f) consider 50 tasks in the network. }     
	\label{fig}   
	\vspace{-0.45 cm}  
\end{figure}

We study the quality of experience of task owners in terms of the overall service quality, RoSQ, and FoDSQ in Fig. 2. To study various problem sizes, Figs. 2(a)-2(c) consider 30 tasks, while Figs. 2(d)-2(f) consider 50 tasks in the system.

	Fig. 2(a) and Fig. 2(d) show that the service quality of Hybrid\_F\_S, Convention\_S, and Convential\_F increases as the number of workers rises, which is expected since the existence of more workers implies a larger pool of resources. As can be seen from these figures, Conventional\_S achieves the best performance in terms of service quality since matching decisions are made based on the current network conditions during each transaction. Nevertheless, this method suffers from excessive overhead, as will be discussed in Section 6.4.3. Our method, i.e., Hybrid\_F\_S, outperforms Conventional\_F thanks to its carefully crafted hybrid trading mode, where spot trading serves as a backup plan to support resource acquisition. Also, Negotiation, Quality\_P, and Random\_M perform worse than the other three methods due to their design philosophy. For example, the randomness in Random\_M can cause uncertainties in task-worker matching performance, while making it indifferent to the number of workers. Also, Negotiation offers a unified payment to the workers, while Quality\_P only considers the service quality and ignores the payments.

Fig. 2(b) and Fig. 2(e) reveal that RoSQ performance of Hybrid\_F\_S significantly outperforms Conventional\_F, Negotiation, Quality\_P, and Random\_M, owing its hybrid trading strategy. This further reveals that our backup plans (e.g., OMOM and O3M) offer a viable way to utilize the available budget.

As can be seen from Fig. 2(c) and Fig. 2(f), the FoDSQ of Hybrid\_F\_S approaches that of Conventional\_S and surpasses other methods. Such a phenomenon indicates that most tasks can meet their desired service quality while benefiting from time and cost savings (shown in Section 6.4.3), using our method.

\vspace{-0.1cm}
\subsubsection{Utility of workers}
\begin{figure} \centering  
	\vspace{-2.2cm}
	\subfigtopskip=2pt
	\subfigbottomskip=2pt
	\subfigcapskip=-1.95cm
	\setlength{\abovecaptionskip}{-1.5cm}
	\subfigure[] {
		\label{fig:a}     
		\includegraphics[width=0.503\columnwidth]{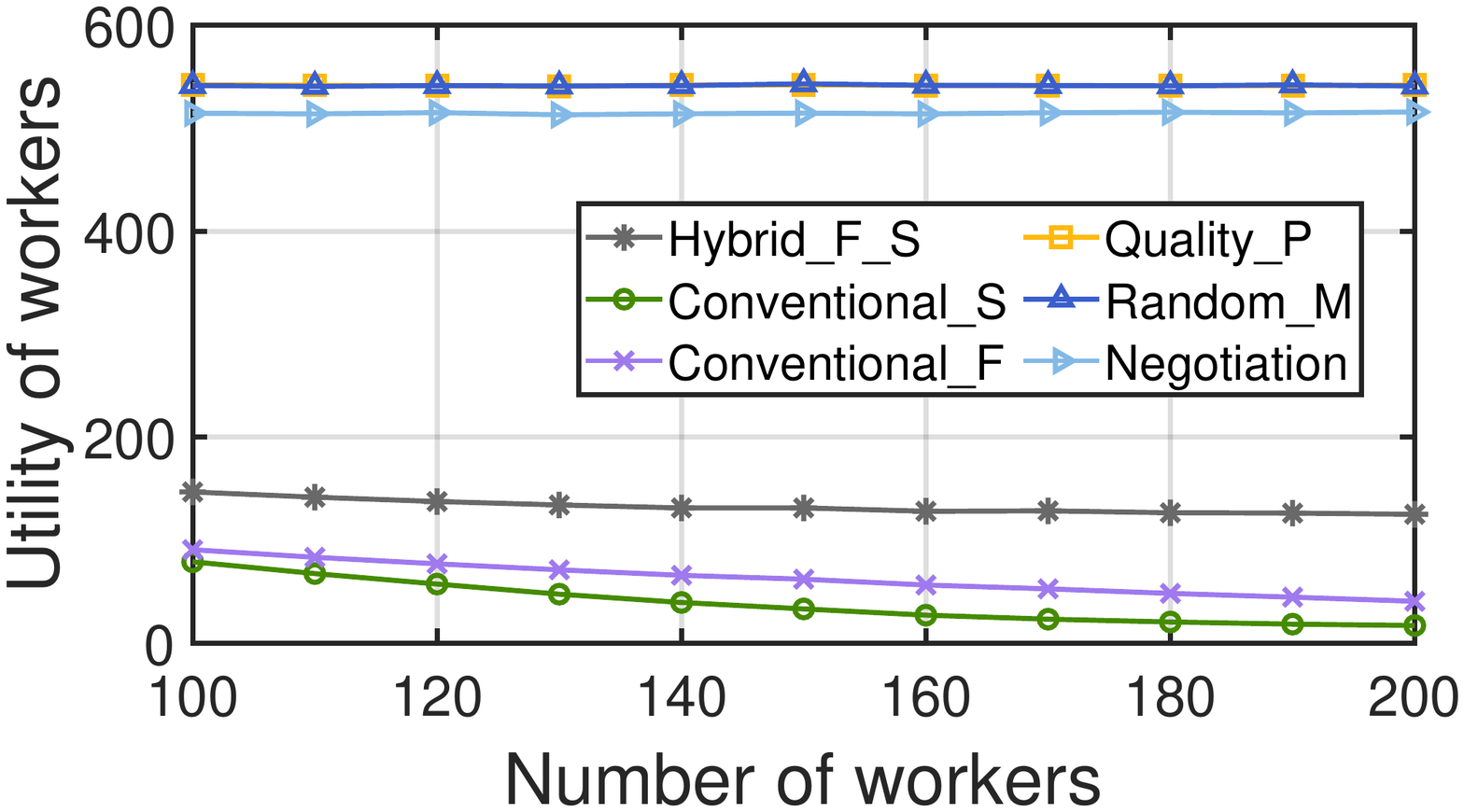}  
	}    \hspace{-6.35mm} 
	\subfigure[] { 
		\label{fig:b}     
		\includegraphics[width=0.503\columnwidth]{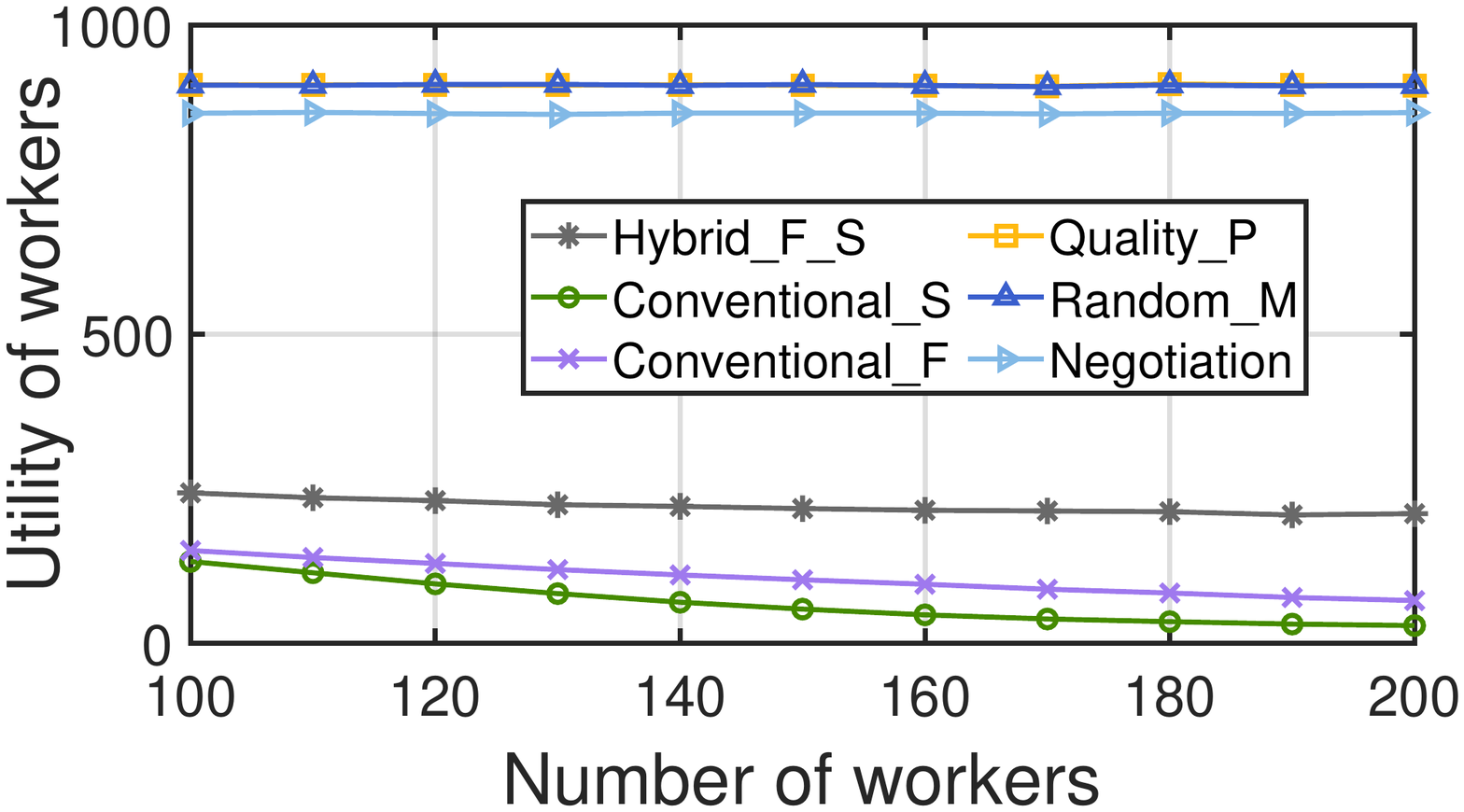}     
	}     
	\caption{Performance comparisons in terms of the utility of workers under different problem sizes, with 30 tasks (a) and 50 tasks (b).}     
	\label{fig}
		\vspace{-0.45cm}      
\end{figure}
We study the utility of workers under Hybrid\_F\_S and Conventional\_S under different numbers of workers in Fig. 3. To study various problem sizes, Fig. 3(a) considers 30 tasks, while Fig. 3(b) presumes 50 tasks in the system.

Fig. 3(a) and Fig. 3(b) show that the utility of workers under Hybrid\_F\_S, Conventional\_F, and Conventional\_S decreases as the number of workers increases. This is because the existence of more workers leads to the presence of more options for tasks. Therefore, to increase their competitiveness, workers have to reduce their asked payments. We can see that Negotiation, Quality\_P, and Random\_M can achieve good performance in terms of the utility of workers, where for Negotiation method. Note that Hybrid\_F\_S outperforms Conventional\_F owing to its hybrid mode. The curve of Conventional\_S falls below that of the other five methods since matching decisions are made based on the current network/market conditions, which impose competition among workers. From Fig. 2 and Fig. 3, we can conclude that Hybrid\_F\_S achieves a reasonable service quality while providing a commendable utility for workers.

\vspace{-0.1cm}
\subsubsection{Running time and interaction overhead}
\begin{figure}
	\vspace{-1.8cm}
\subfigtopskip=0pt
\subfigbottomskip=1pt
\subfigcapskip=-2.0cm
\setlength{\abovecaptionskip}{-1.6cm}
\subfigure[] {
	\label{fig:a}     
	\includegraphics[width=0.5\columnwidth]{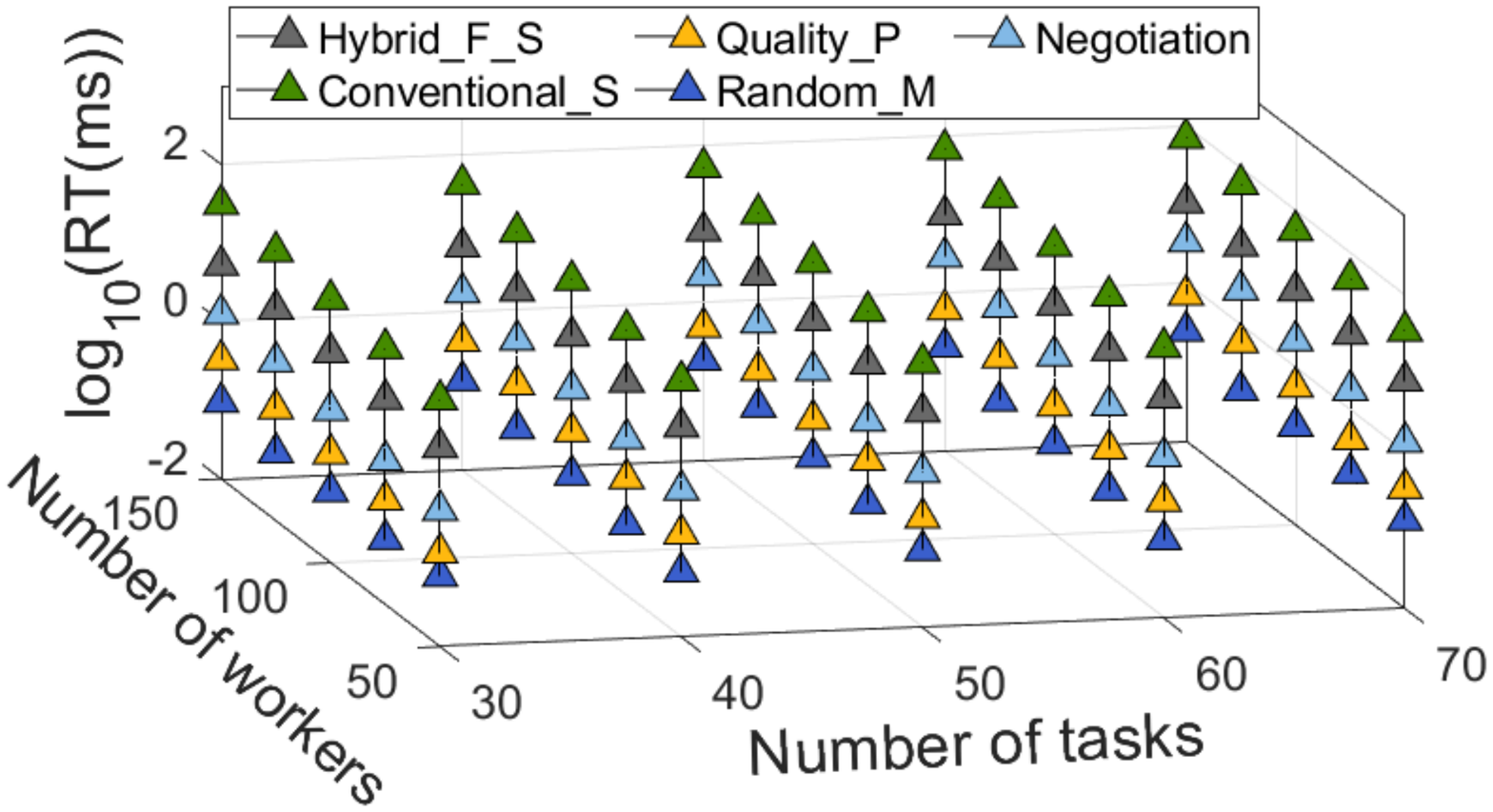}  
}    \hspace{-6mm} \vspace{-34mm}
\subfigure[] { 
	\label{fig:b}     
	\includegraphics[width=0.5\columnwidth]{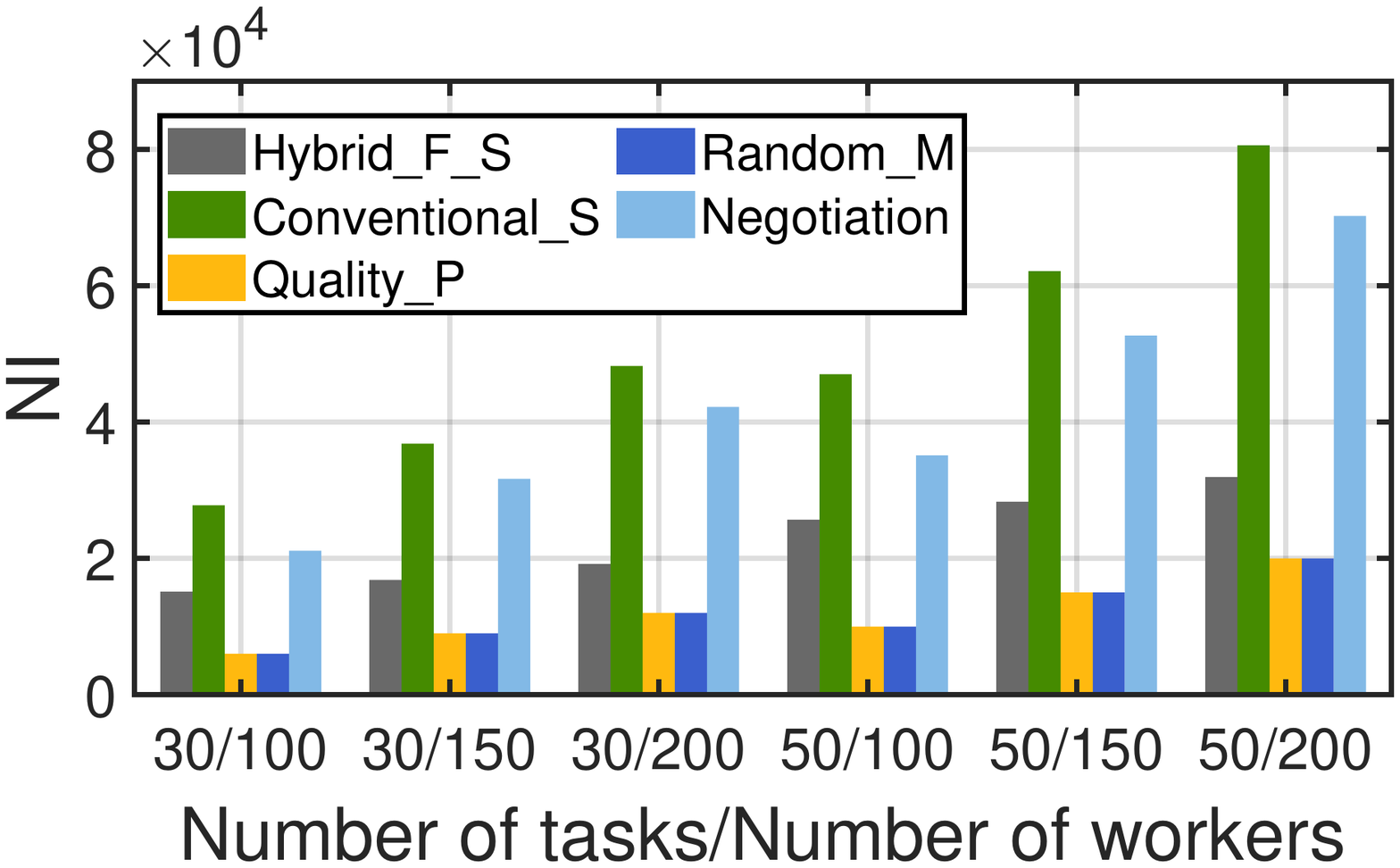}     
}     
\subfigure[] {
	\label{fig:c}     
	\includegraphics[width=0.5\columnwidth]{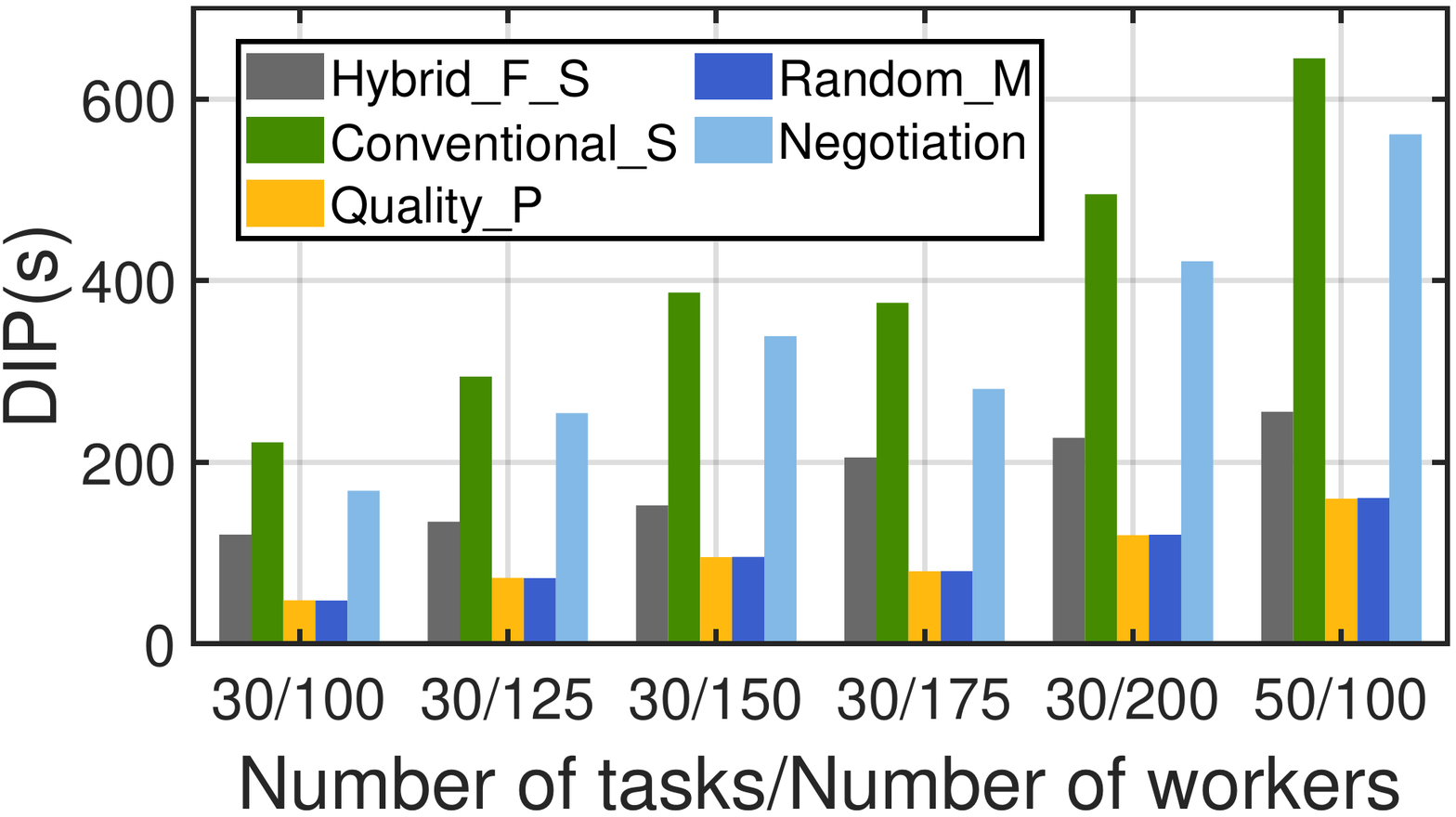}  
}    \hspace{-6mm} 
\subfigure[] { 
	\label{fig:d}     
	\includegraphics[width=0.5\columnwidth]{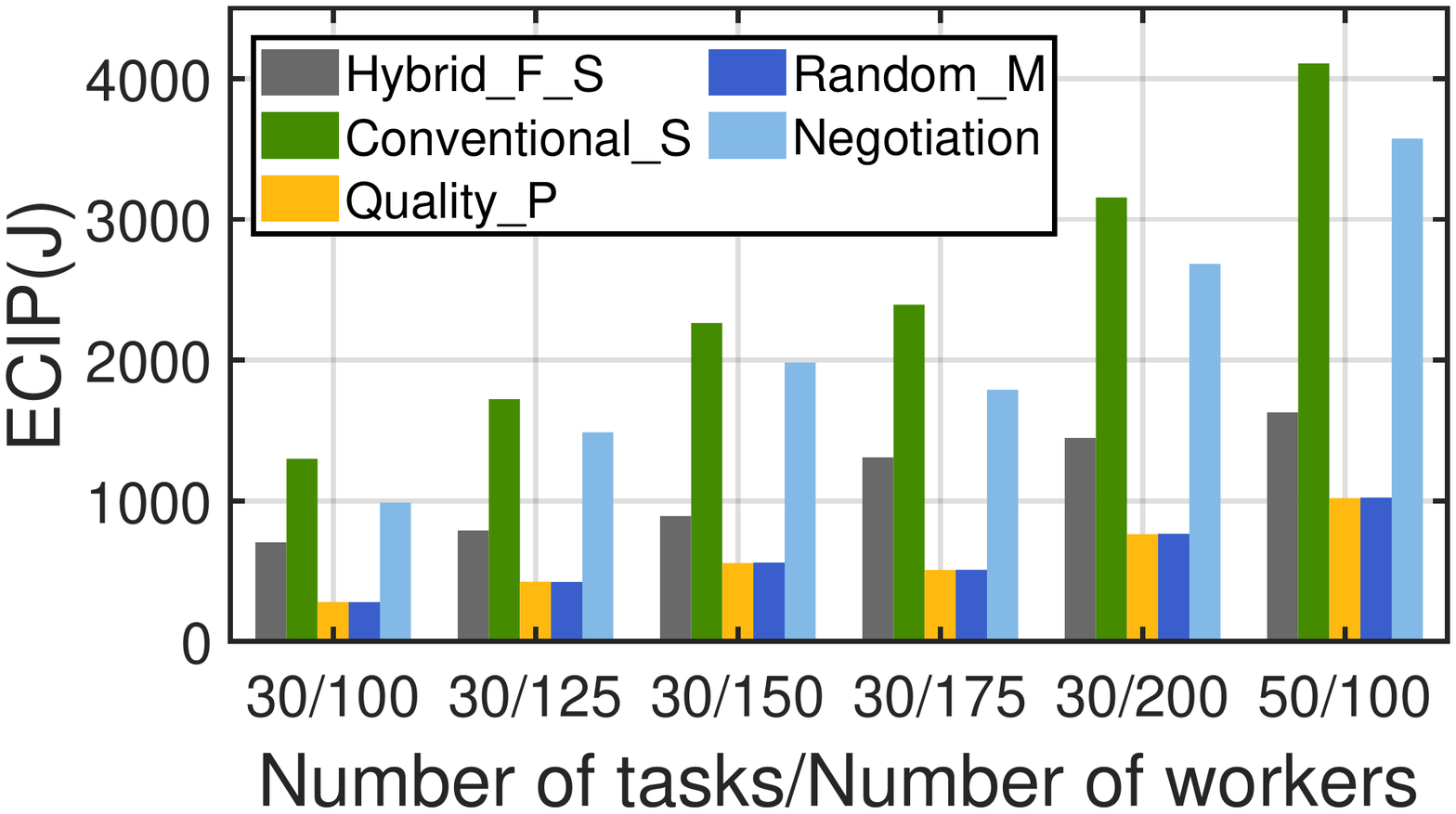}     
}      
	\caption{Performance comparisons in terms of the running time (a), number of interactions (NI) (b), delay of interactions among participants (c), and energy consumption of interactions (d), under different problem sizes. }     
	\label{fig} 
\vspace{-0.49cm}       
\end{figure}
One of the main advantages of a hybrid service trading market is a reduction in the overhead, which can lower latency and energy consumption of decision-making. To illustrate the performance in terms of time efficiency, we evaluate the running time and the total number of interactions of different methods in Fig. 4(a) and Fig. 4(b), respectively.
In Fig. 4(a), the logarithmic-representation is utilized in the y-axis for better illustrations. Also, the performance of Conventional\_F has been omitted here, since it does not consider spot trading, and thus incurs no latency in terms of onsite decision-making.

As can be seen from Fig. 4(a), as the numbers of workers and tasks increases, the running time of Hybrid\_F\_S remains to be better than that of Conventional\_S, since long-term workers in Hybrid\_F\_S do not engage in onsite decision-making. In addition, due to the simplicity of strategies pursued by Quality\_P and Random\_M, e.g., there is no bargain on service prices, the running time of these two methods is lower than that of the other methods. Note that Negotiation outperforms Conventional\_S and Hybrid\_F\_S since participants only have to discuss a unified asked payment. However, this advantage of Negotiation, Quality\_P, and Random\_M, comes with a lower performance in terms of service quality, as shown in Fig. 2.

Figs. 4(b)-(d) illustrate the overhead of matching decisions (e.g., DIP and ECIP). These figures show that Hybrid\_F\_S outperforms Conventional\_S and Negotiation owing to its overbooking-enabled futures trading mode, requiring only a relatively small portion of participants to communicate with each other during each practical transaction. Similar to Fig. 4(a), Quality\_P and Random\_M achieve a high overhead efficiency, which comes at the price of sacrificing the service quality. In conclusion, Hybrid\_F\_S balances a trade-off between latency and overhead.

\vspace{-0.1cm}
\subsubsection{Individual rationality}
\begin{figure*} \centering
	\vspace{-2.4cm}
	\subfigtopskip=1pt
	\subfigbottomskip=0pt
	\subfigcapskip=-2.1cm
	\setlength{\abovecaptionskip}{-1.6cm}
	\subfigure[] {
		\label{fig:a}     
		\includegraphics[width=0.53\columnwidth]{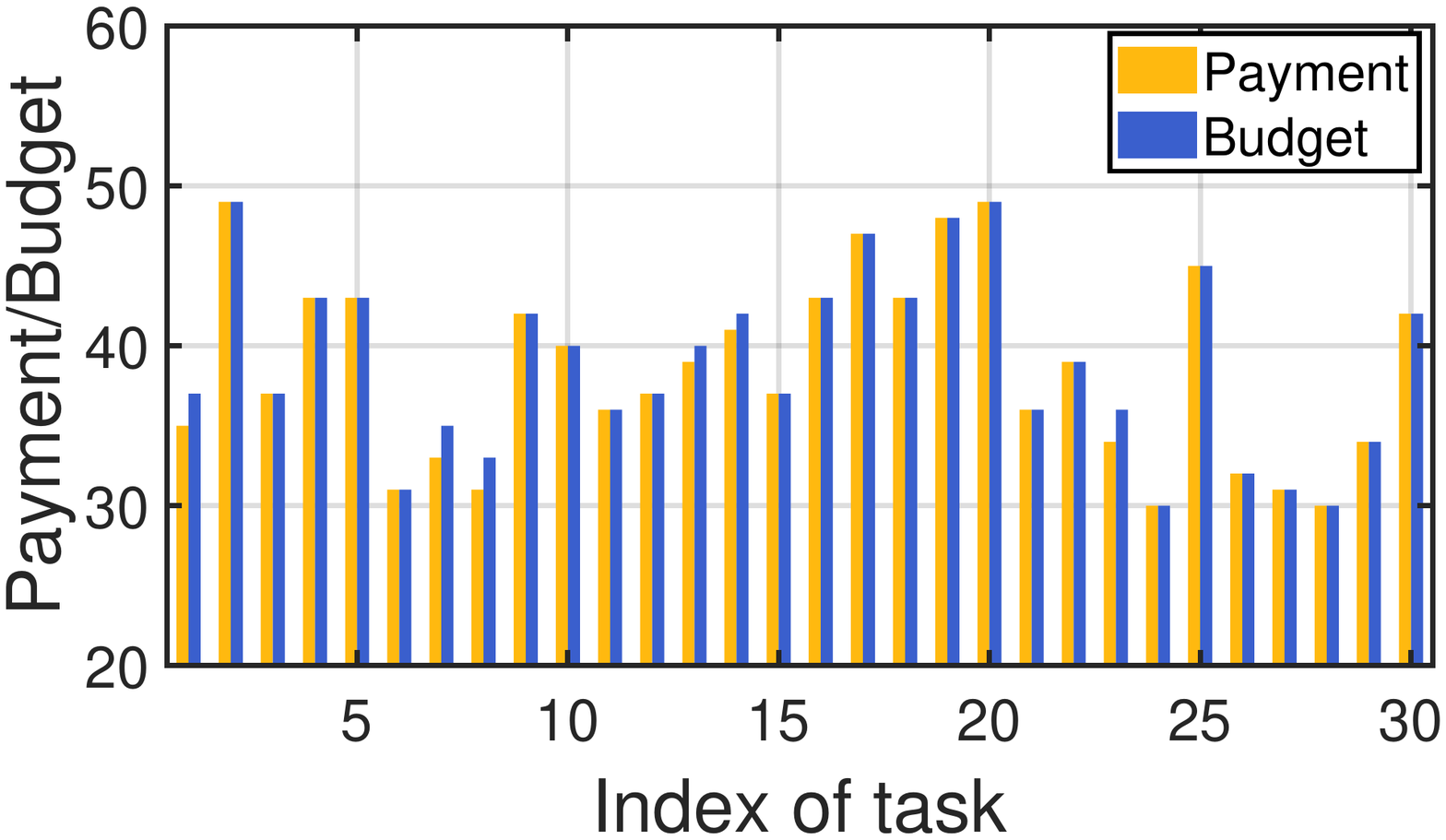}  
	}    \hspace{-7.5mm}
	\subfigure[] { 
		\label{fig:b}     
		\includegraphics[width=0.53\columnwidth]{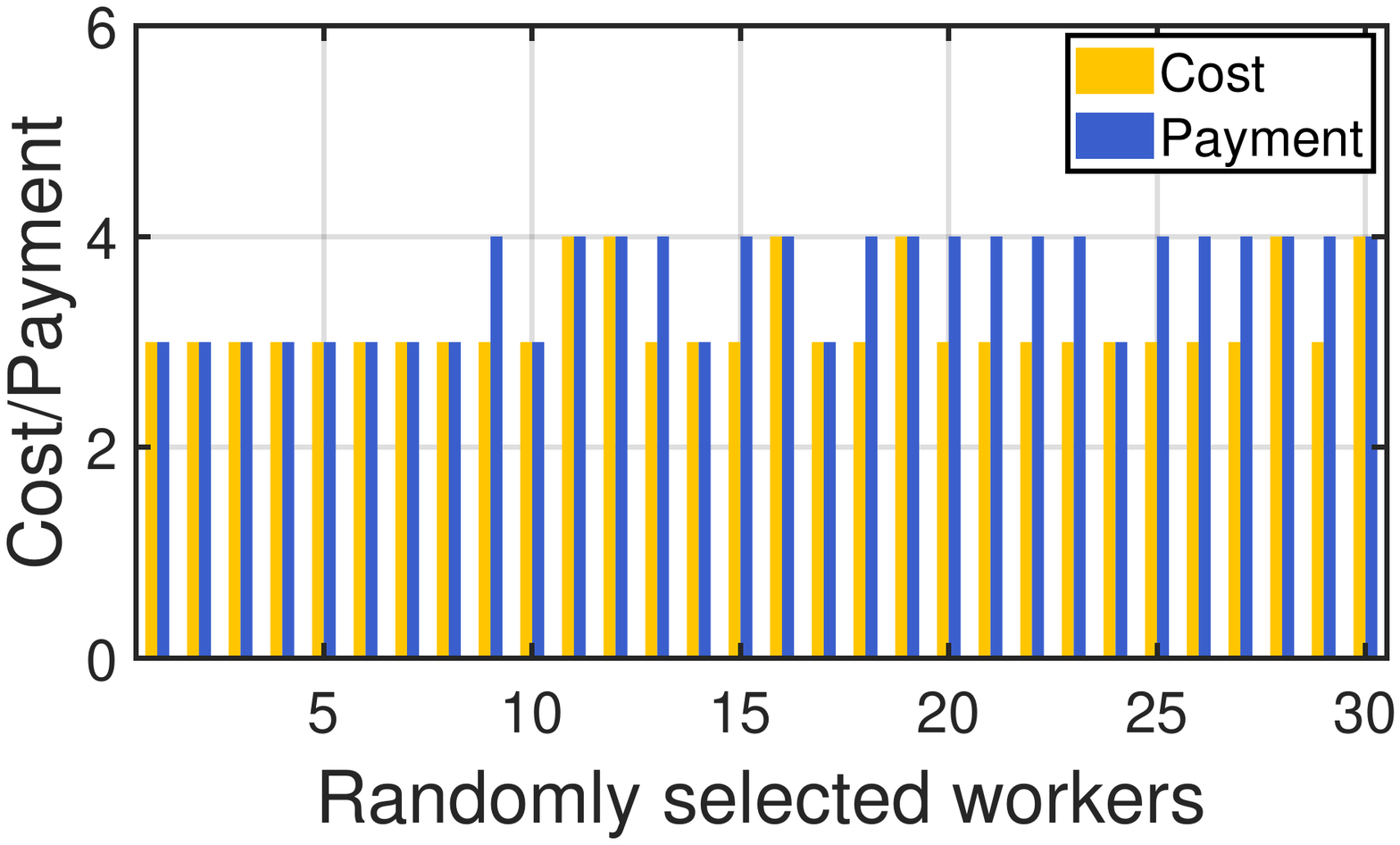}     
	}     \hspace{-6.8mm} 
	\subfigure[] {
		\label{fig:c}     
		\includegraphics[width=0.53\columnwidth]{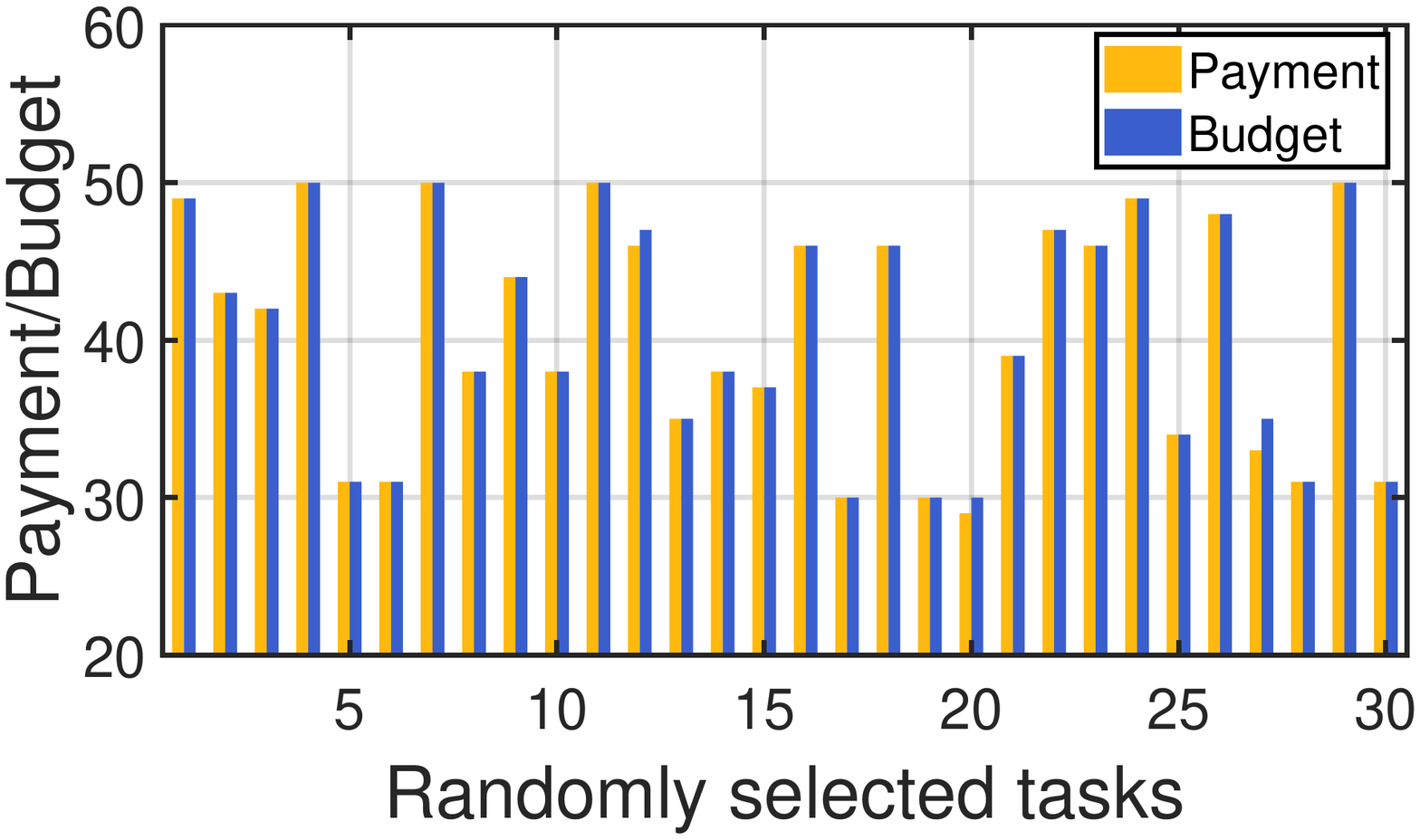}  
	}    \hspace{-7.5mm} 		
	\subfigure[] { 
		\label{fig:d}     
		\includegraphics[width=0.53\columnwidth]{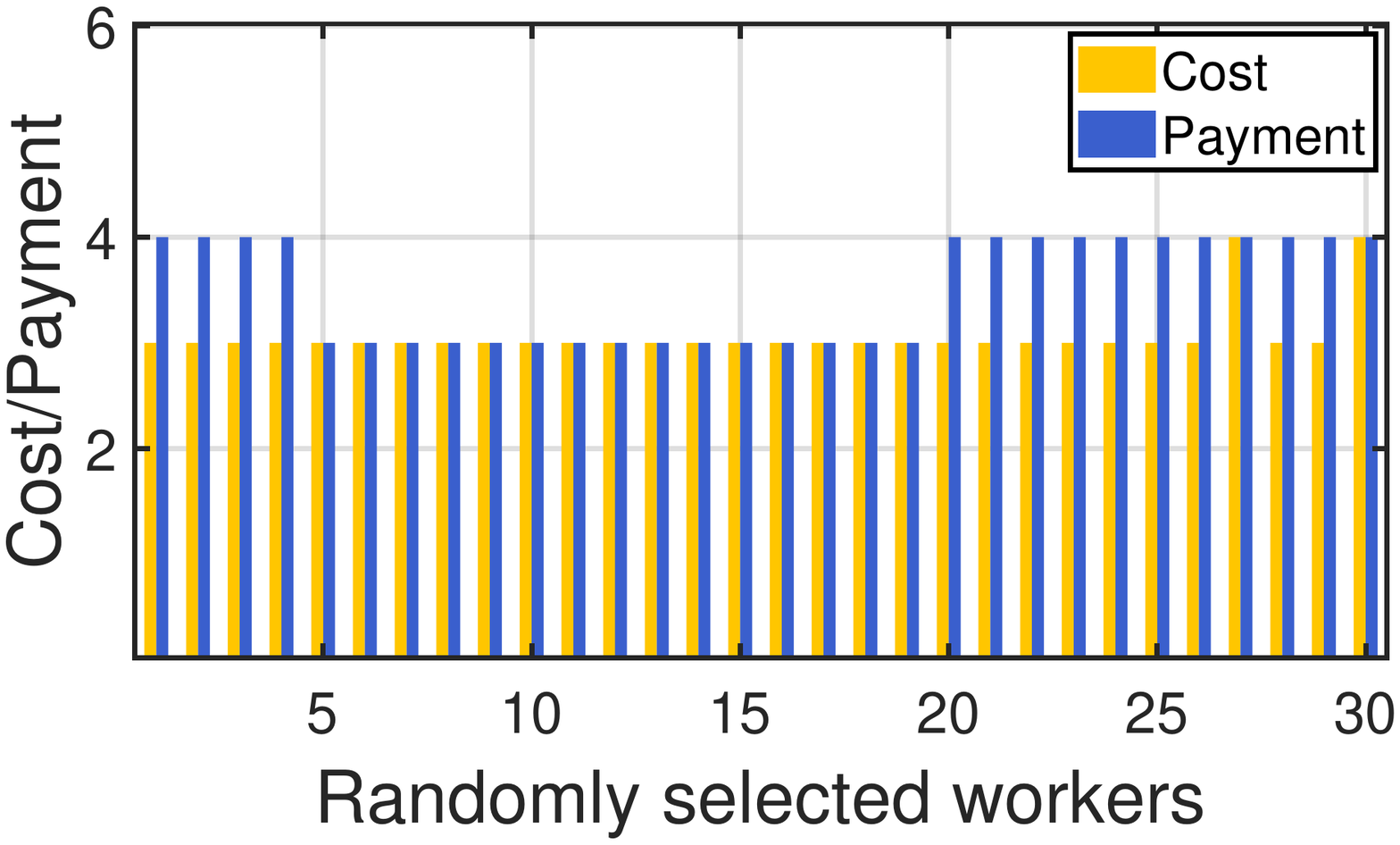}     
	}  
	
	\caption{Individual rationality of tasks and workers under different problem sizes, where (a)(b) consider 30 tasks and (c)(d) consider 50 tasks.}     
	\label{fig}     
	\vspace{-0.55cm}
\end{figure*}
To show the individual rationality achieved under Hybrid\_F\_S, we study the total payments and budgets of MCS tasks. We randomly selected 30 workers out of 100 and show their service costs as well as asked payments in Fig. 5. Fig. 5(a) and Fig. 5(b) consider 30 tasks, Fig. 5(c) and Fig. 5(d) consider 50 tasks.

Fig. 5(a) and Fig. 5(c) show that the total payments of tasks never exceed their budgets. This verifies that OMOM and O3M mechanisms guarantee the individual rationality of each task owner. 
As can be seen from Fig. 5(b) and Fig. 5(d), the total service cost of each worker never exceeds its total received payments, revealing that OIA3M, OMOM, and O3M ensure the individual rationality of workers.

\vspace{-0.1cm}
\subsubsection{Risk and overbooking}
\begin{figure} \centering
	\vspace{-1.9cm}
	\subfigtopskip=0pt
	\subfigbottomskip=1pt
	\subfigcapskip=-2.0cm
	\setlength{\abovecaptionskip}{-1.6cm}
	\subfigure[] {
		\label{fig:a}     
		\includegraphics[width=0.5\columnwidth]{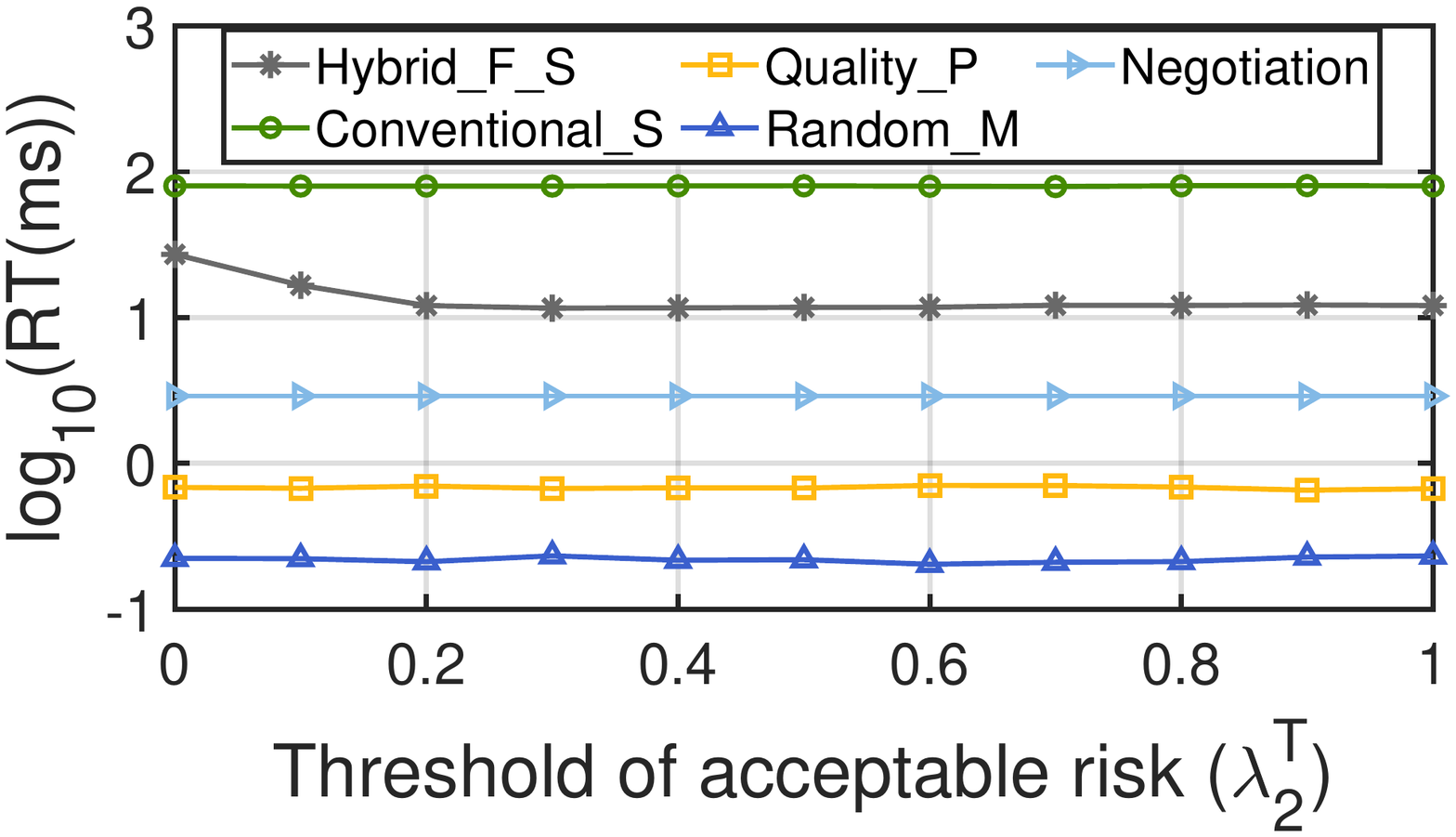}  
	}    \hspace{-6mm} \vspace{-34.8mm}
	\subfigure[] { 
		\label{fig:b}     
		\includegraphics[width=0.5\columnwidth]{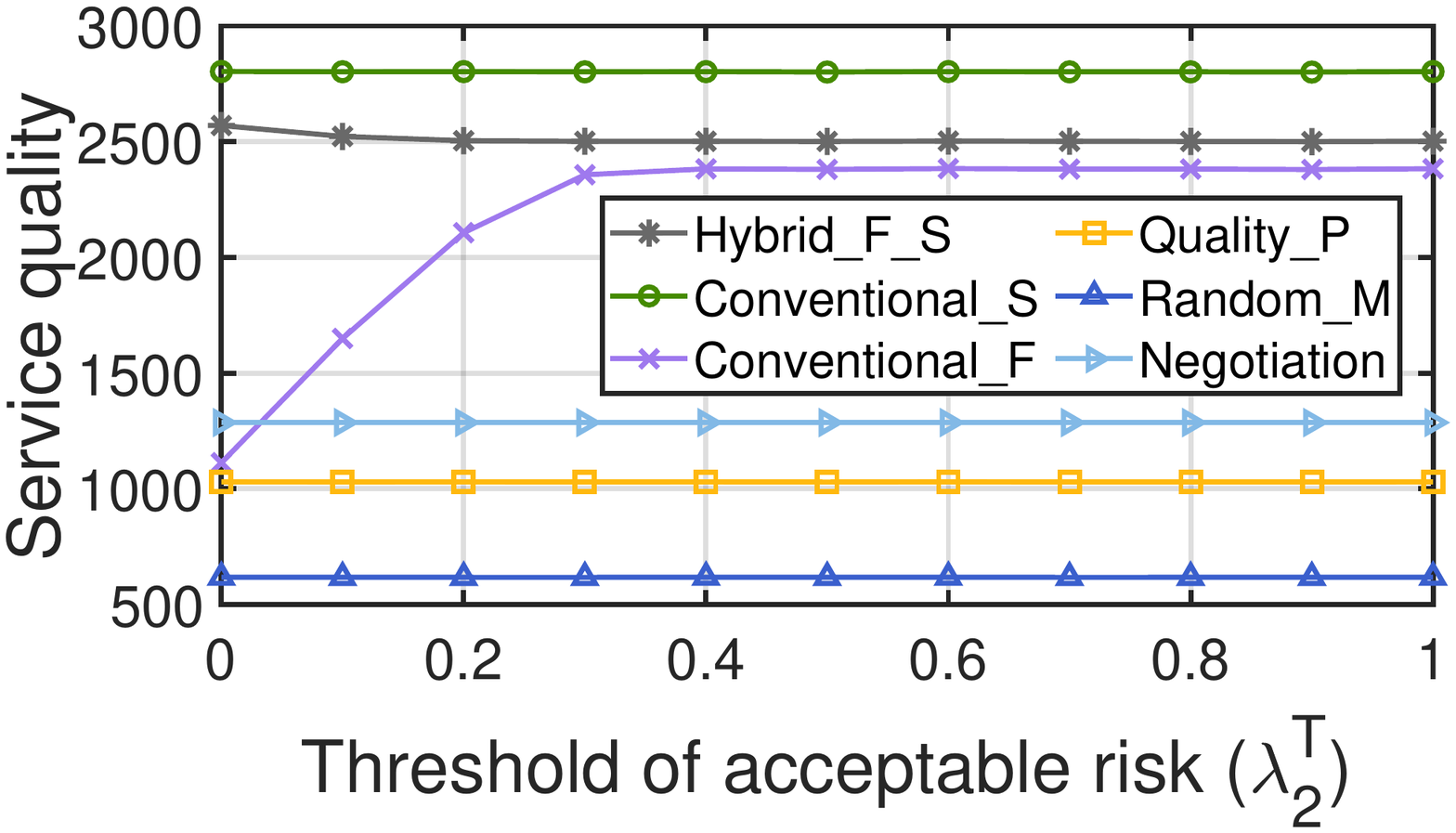}     
	}     
	\subfigure[] {
		\label{fig:c}     
		\includegraphics[width=0.505\columnwidth]{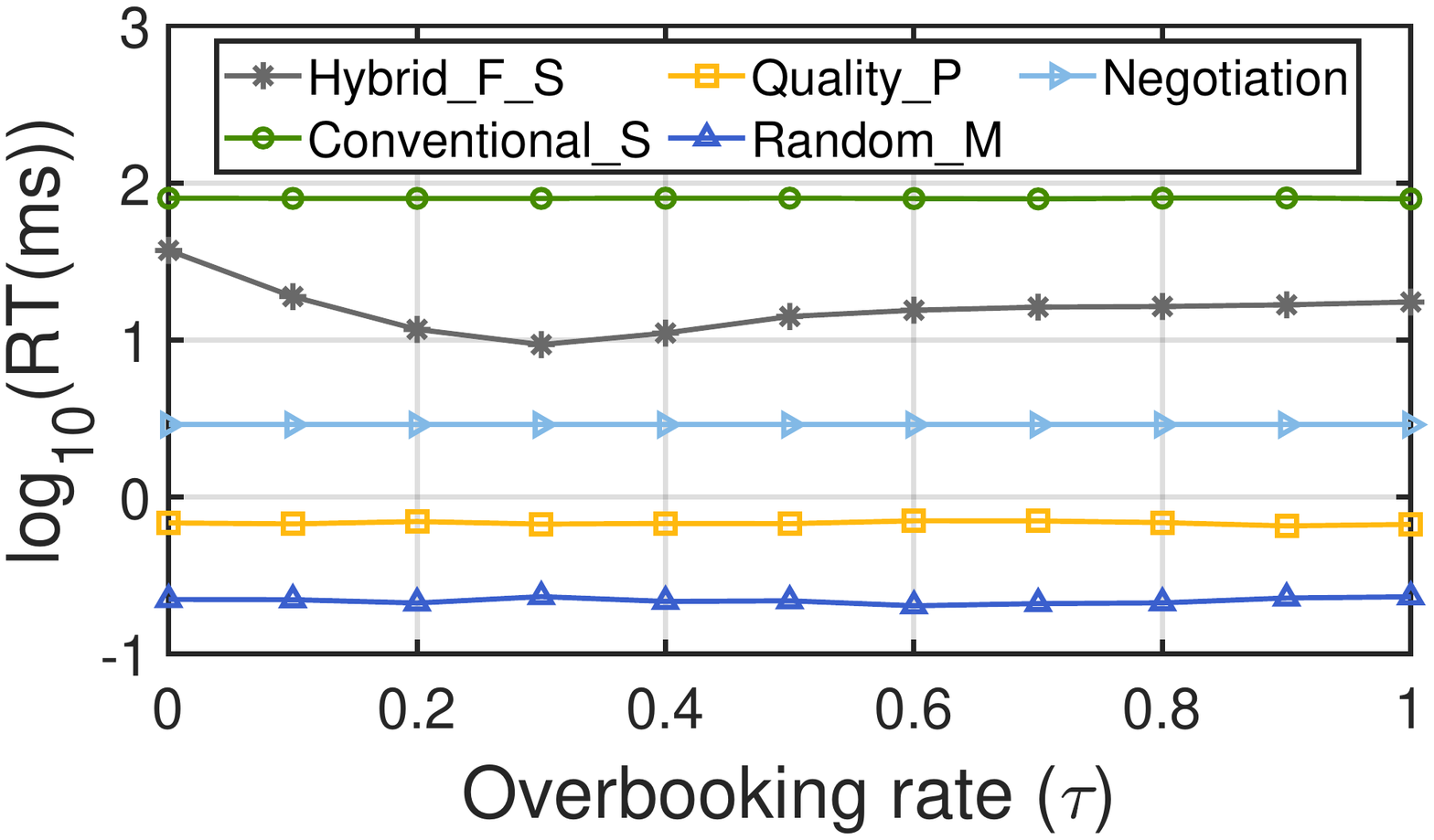}  
	}    \hspace{-7mm} 
	\subfigure[] { 
		\label{fig:d}     
		\includegraphics[width=0.51\columnwidth]{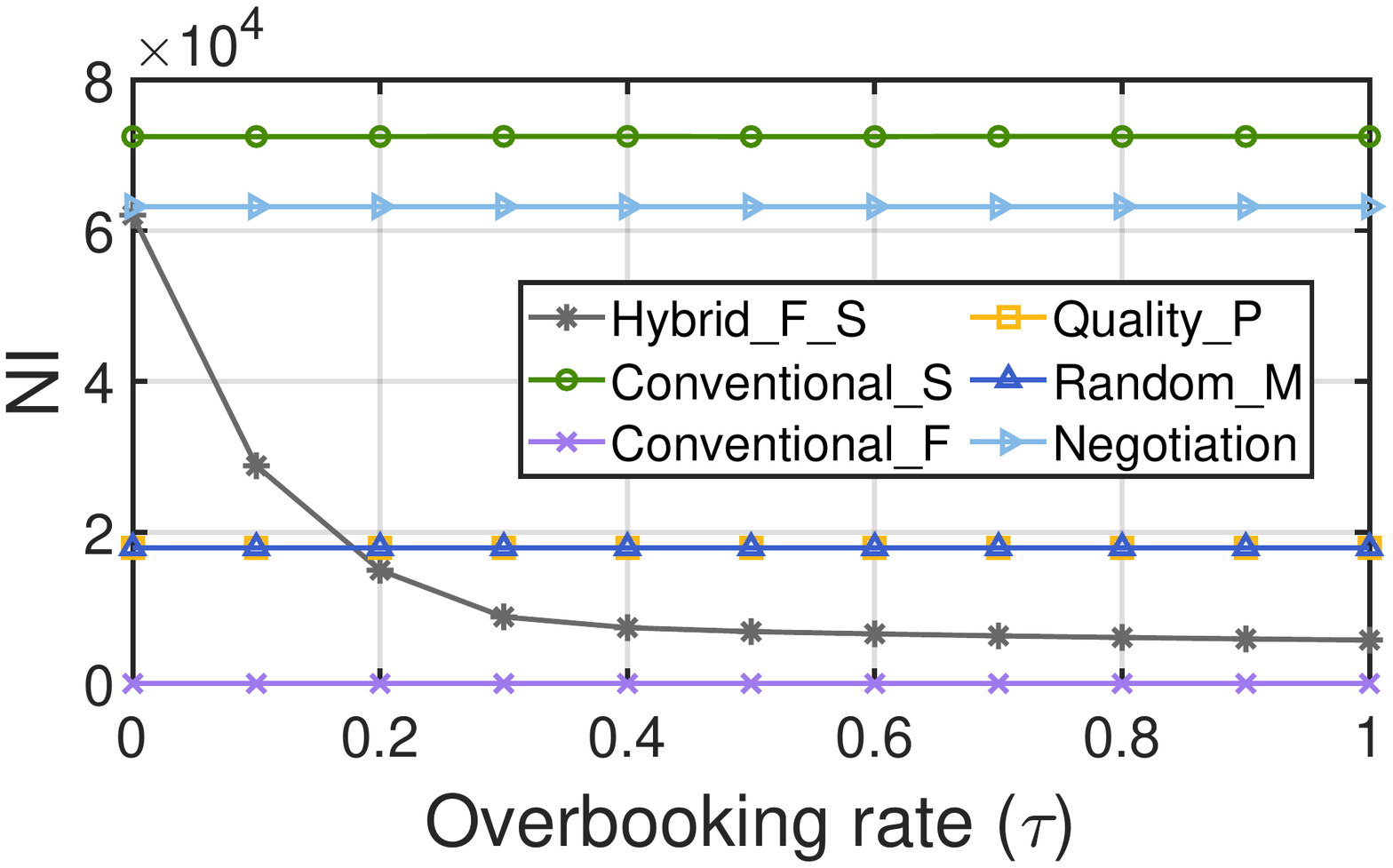}     
	}   
	\caption{(a)(b) Performance comparisons in terms of running time and service quality under different thresholds of acceptable risk ($ \lambda_{2}^{T} $). (c)(d) Performance comparisons in terms of running time and number of interactions (NI) under different overbooking rates ($ \tau $).}
	\label{fig}
\vspace{-0.5cm}     
\end{figure}
We next aim to illustrate the advantages of risk analysis and overbooking in terms of \textit{i)} service quality assurance; and \textit{ii)} coping with the network dynamics. We first evaluate the running time and service quality performance in Figs. 6(a)-6(b) under different acceptable risk thresholds (e.g., $ \lambda_{2}^T $). We then evaluate the advantages provided by overbooking in Figs. 6(c)-6(d) in terms of running time and the number of interactions under various overbooking rates ($\tau$). Figs. 6(a)-6(b) consider 45 MCS tasks and 200 workers, in which the desired service quality of tasks is set by $ Q_i \in [50,60] $ to better show the importance and impact of risk analysis. Figs. 6(c)-6(d) consider 45 tasks and 200 workers, where the threshold of acceptable risk is $ \lambda_{2}^{T}=0.3 $, and the desired service quality of tasks is $ Q_i \in [30,35] $.

As can be seen from Fig. 6(a), with increasing the threshold of acceptable risk ($ \lambda_{2}^T $), the running time of Hybrid\_F\_S initially decreases and then remains relatively stable/flat since a large value of $ \lambda_2^T $ implies that more tasks can sign long-term contracts with workers, which eliminates the need for bargaining during each transaction. This in turn leads to a reduction in the running time. Focusing on the curve after $ \lambda_2^T=0.3 $, since all tasks have signed contracts in the futures market, the running time remains stable. Overall, the running time of Hybrid\_F\_S outperforms that of Conventional\_S, as also shown in Fig. 4(a).

Fig. 6(b) shows that increasing $ \lambda_{2}^T $ leads to a slight decrease in service quality of Hybrid\_F\_S. This is because a large value of $ \lambda_2^T $ implies that tasks face higher chances of receiving unsatisfying service qualities. Also, the service quality of Conventional\_F initially improves with increasing the value of $ \lambda_2^T $, and then saturates after $ \lambda_2^T=0.4 $. This is because more tasks are willing to take part in the futures market since a larger risk can be accepted. Note that Hybrid\_F\_S outperforms Conventional\_F and the other three baselines in Fig. 6(b) thanks to considering the spot trading mode as a backup plan.

Fig. 6(c) shows that the running time of Hybrid\_F\_S decreases sharply as the overbooking rate $ \tau $ increases from 0 to 0.3. The reason behind this is that more workers are given the chance to be recruited in the designed futures market due to a large overbooking rate. Thus, fewer workers and tasks will take part in the spot market, reducing the latency of decision-making. However, after $ \tau=0.3 $, the running time of Hybrid\_F\_S increases slightly since a larger overbooked budget $ (1+\tau) B_i $ makes a task to recruit more long-term workers in the futures market. This further calls for the execution of our proposed OMOM algorithm, which further increases running time. 

Fig. 6(d) reveals that the number of interactions under Hybrid\_F\_S sharply decreases as $\tau$ increases, since more workers can take part in the futures market and no longer have to engage in onsite decision-making.
Note that because Conventional\_S, Negotiation, Quality\_P, and Random\_M do not consider risks and overbooking, their curves in Fig. 6 stay relatively stable, while Hybrid\_F\_S provides a commendable performance in terms of both running time and the number of interactions.

All in all, our proposed Hybrid\_F\_S can achieve commendable performance in terms of service quality in the dynamic MCS networks, while balancing the trade-off between the overhead on decision-making and utility of participants.

\vspace{-0.3cm}
\section{Conclusion}
\noindent
We proposed a set of stable matching mechanisms for a hybrid MCS service trading market that encompasses both futures and spot trading. We first determined a set of long-term workers for MCS tasks via OIA3M mechanism in the futures market. Since the required service quality of each task may not be satisfied due to fluctuations in the supplied resources of long-term workers, we further introduced a spot trading mode, where the task-worker mapping was conducted via O3M and OMOM mechanisms. We theoretically showed that our matching mechanisms satisfy stability, computation efficiency, individual rationality, and fairness. We then verified these theoretical findings through numerical evaluations, which also reveal that our methodology can achieve a commendable performance in terms of various metrics of interest, as compared to existing methods. Future directions include overbooking rate optimization and intelligent risk management for receiving an undesired service quality. Also, the impacts of the dynamics of service demand and time-varying wireless channel qualities on MCS service provisioning should be carefully investigated.
\vspace{-0.1cm}

\vspace{-1.0cm}
\begin{IEEEbiography}[{\includegraphics[width=1in,height=1.25in,clip,keepaspectratio]{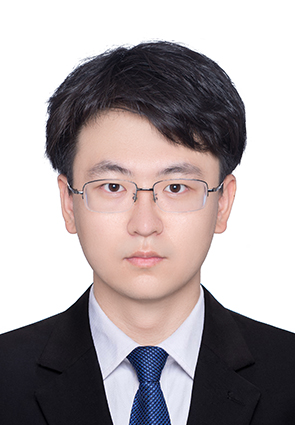}}]{Houyi Qi} received his B.S. degree in electronic information engineering from Zhengzhou University, China, in 2021. He is currently working toward the M.S. degree in School of Informatics, Xiamen University, China. His research interests include mobile crowdsensing networks, matching theory and cloud/edge/service computing.
\end{IEEEbiography}

\vspace{-0.79cm}
\begin{IEEEbiography}[{\includegraphics[width=1in,height=1.25in,clip,keepaspectratio]{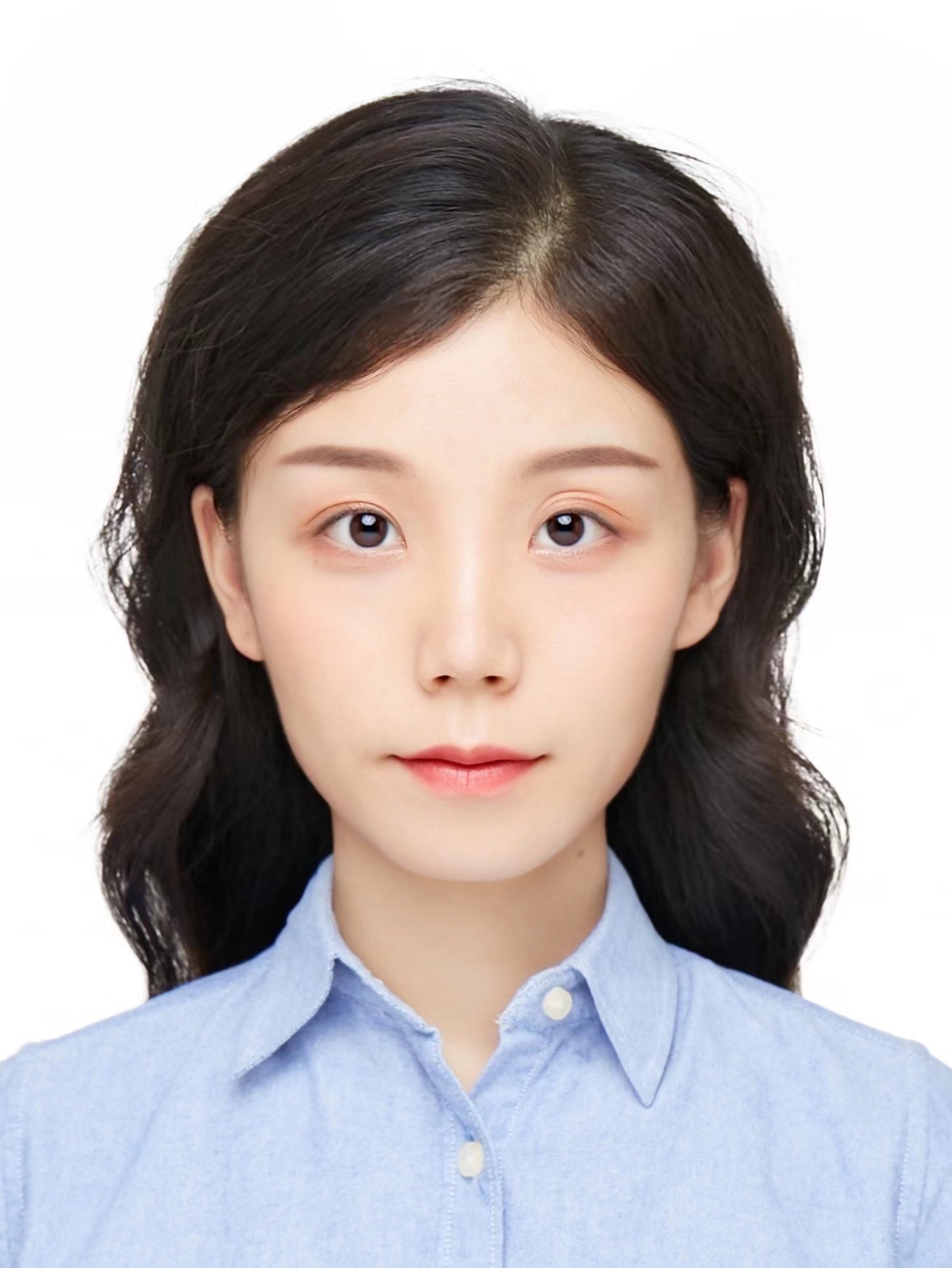}}]{Minghui Liwang [M'19]} received her Ph. D. degree in School of Informatics, Xiamen University, China, in 2019. She is currently an assistant professor in School of Informatics, Xiamen University, China. Her research interests include wireless communication systems, Internet of Things, cloud/edge/service computing, federated learning as well as economic models and applications in wireless communication networks.
\end{IEEEbiography}

\vspace{-0.79cm}
\begin{IEEEbiography}[{\includegraphics[width=1in,height=1.25in,clip,keepaspectratio]{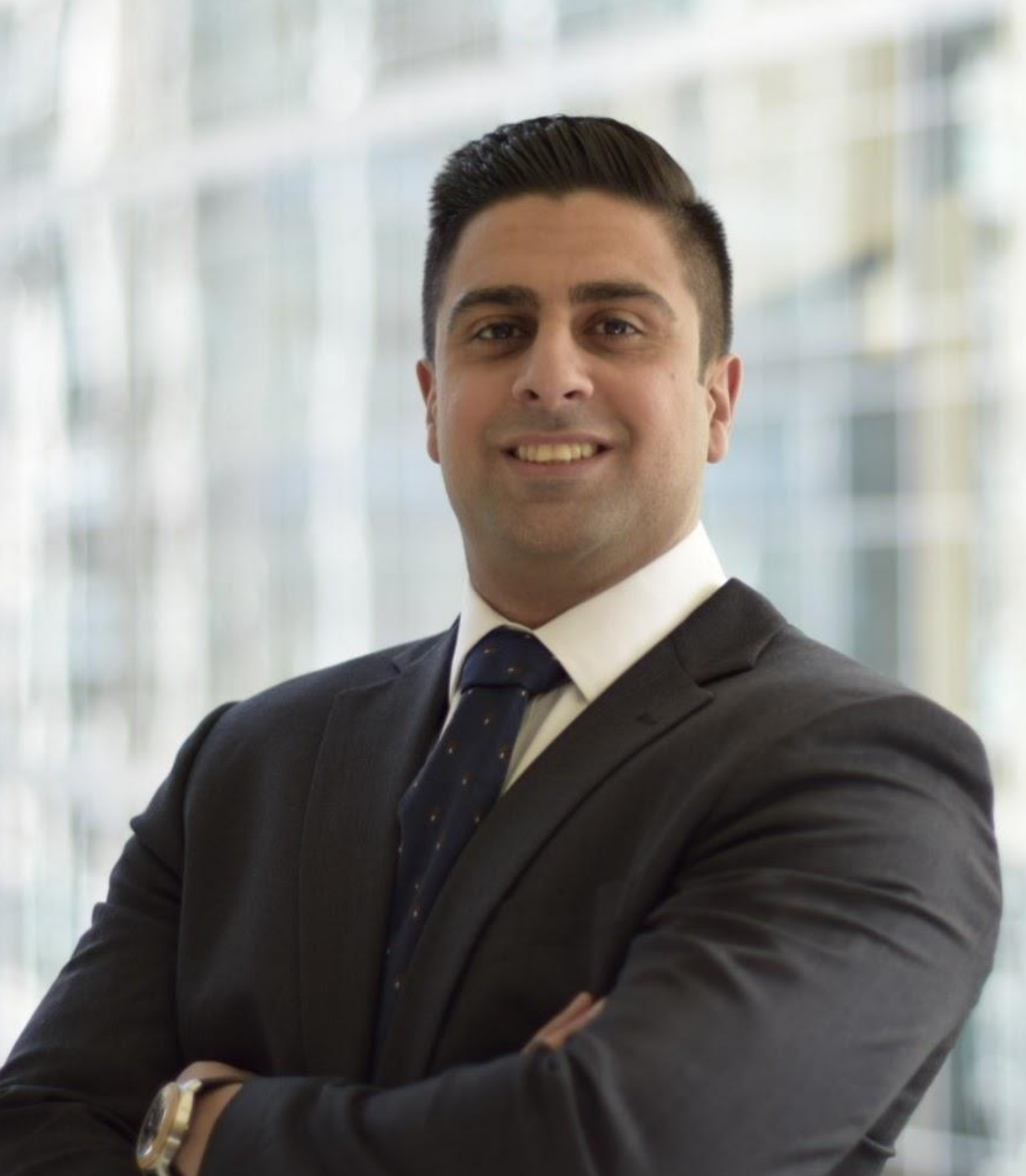}}]{Seyyedali Hosseinalipour [M'20]} received the Ph.D. degree in electrical engineering from NC State University in 2020. He is currently an assistant professor with the Department of Electrical Engineering, University at Buffalo, SUNY, Buffalo, NY, USA. His researcdh interests include 6G, machine learning, federated learning, fog and edge computing, and network optimization.
\end{IEEEbiography}

\vspace{-0.79cm}
\begin{IEEEbiography}[{\includegraphics[width=1in,height=1.25in,clip,keepaspectratio]{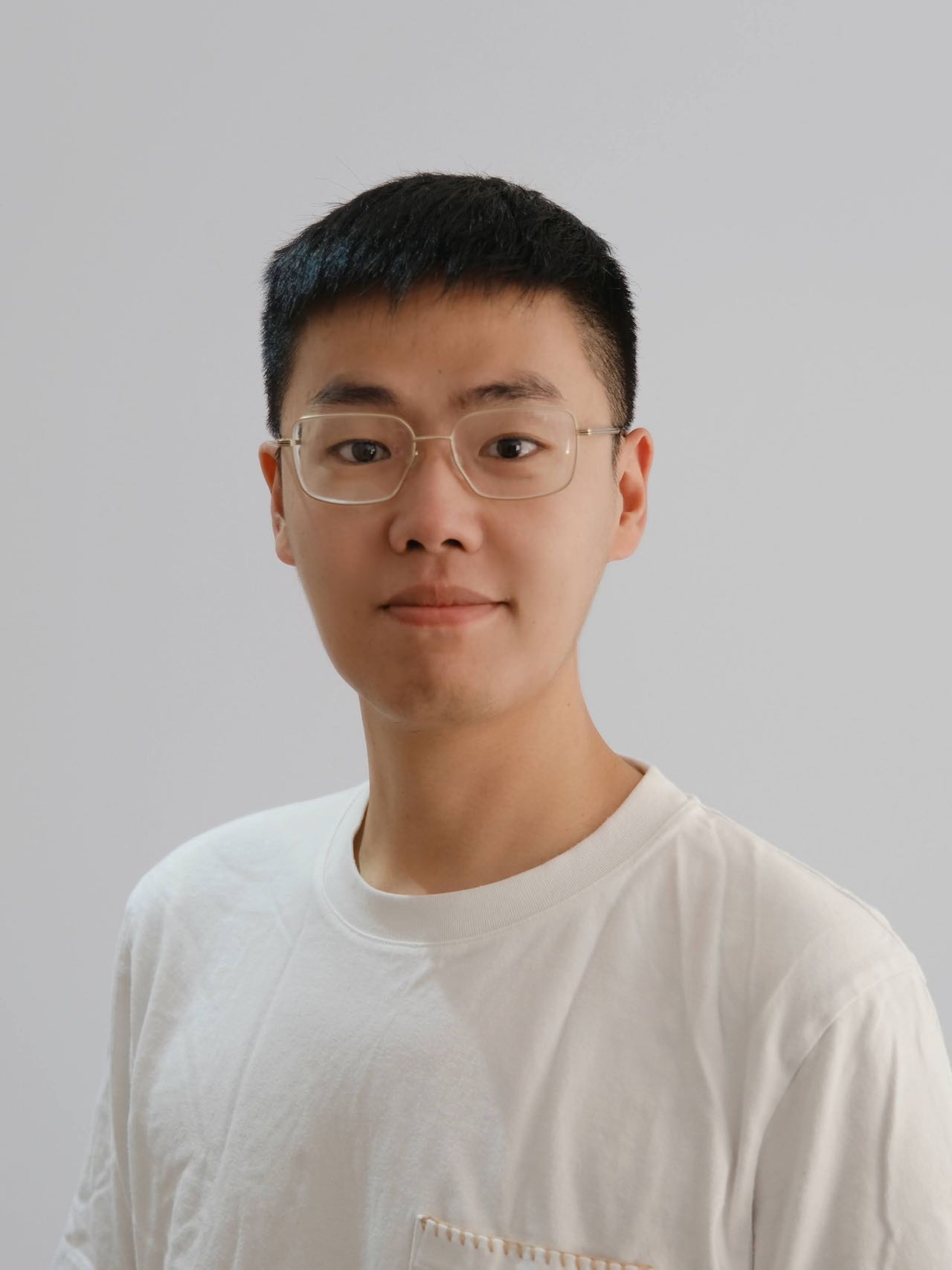}}]{Xiaoyu Xia [M'21]} received his Master degree from The University of Melbourne, Australia and his PhD degree from Deakin University, Australia. He is currently a lecturer in the School of Computing Technologies, RMIT University, Melbourne, Victoria, Australia. His research interests include edge computing, service computing, software engineering and cloud computing. 
\end{IEEEbiography}

\vspace{-0.79cm}
\begin{IEEEbiography}[{\includegraphics[width=1in,height=1.25in,clip,keepaspectratio]{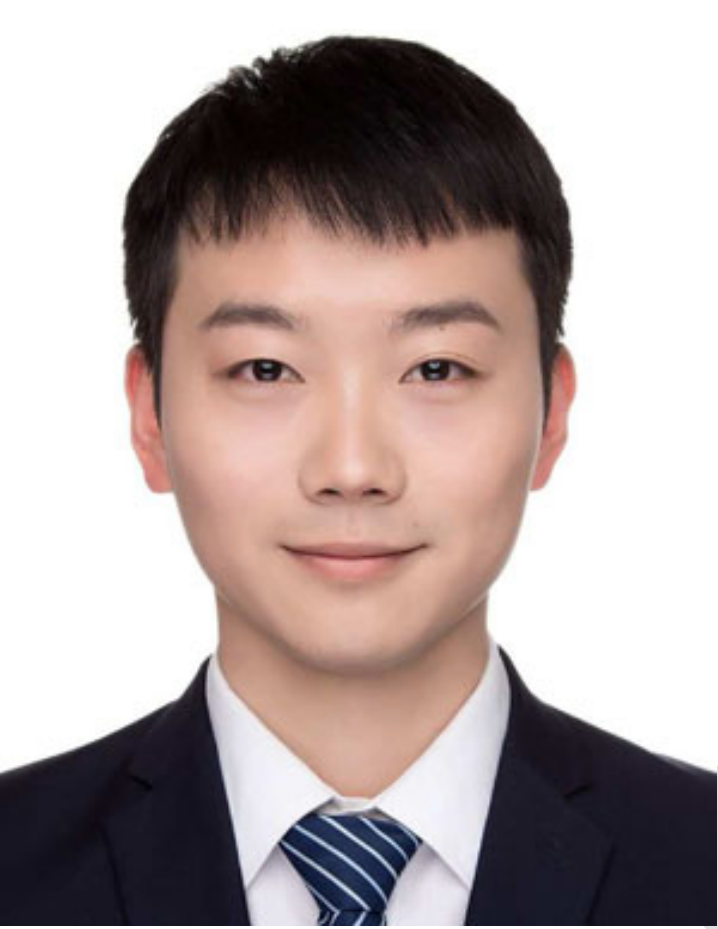}}]{Zhipeng Cheng [S'19]} received his B.S. degree in communication engineering from Jiangnan University, Wuxi, China, in 2017. He received the Ph.D. degree in communication engineering with Xiamen University, Xiamen, China, in 2023. He is now a lecturer in the School of Future Science and Engineering, Soochow University, Suzhou, China. His research interests include UAV communication networks, cloud/edge/service computing, reinforcement learning and distributed machine learning applications.
\end{IEEEbiography}

\vspace{-0.79cm}
\begin{IEEEbiography}[{\includegraphics[width=1in,height=1.25in,clip,keepaspectratio]{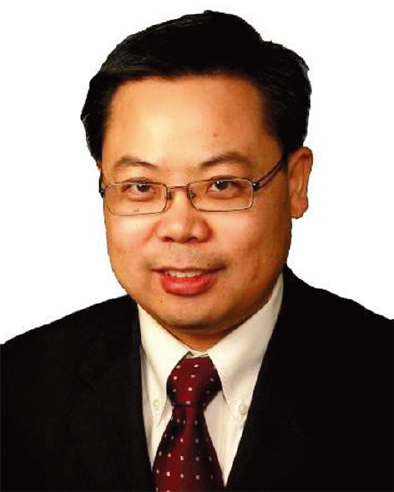}}]{Xianbin Wang [S'98-M'99-SM'06-F'17]} (Fellow, IEEE) is a professor and a Tier-1 Canada Research Chair in 5G and Wireless IoT Communications with Western University, Canada. His current research interests include 5G/6G technologies, Internet of Things, communications security, machine learning, and intelligent communications. He is a Fellow of the Canadian Academy of Engineering and a Fellow of the Engineering Institute of Canada. He has received many prestigious awards and recognitions, including the IEEE Canada R.A. Fessenden Award, Canada Research Chair, Engineering Research Excellence Award with Western University, Canadian Federal Government Public Service Award, Ontario Early Researcher Award, and nine Best Paper Awards. 
\end{IEEEbiography}

\vspace{-0.79cm}
\begin{IEEEbiography}[{\includegraphics[width=1in,height=1.25in,clip,keepaspectratio]{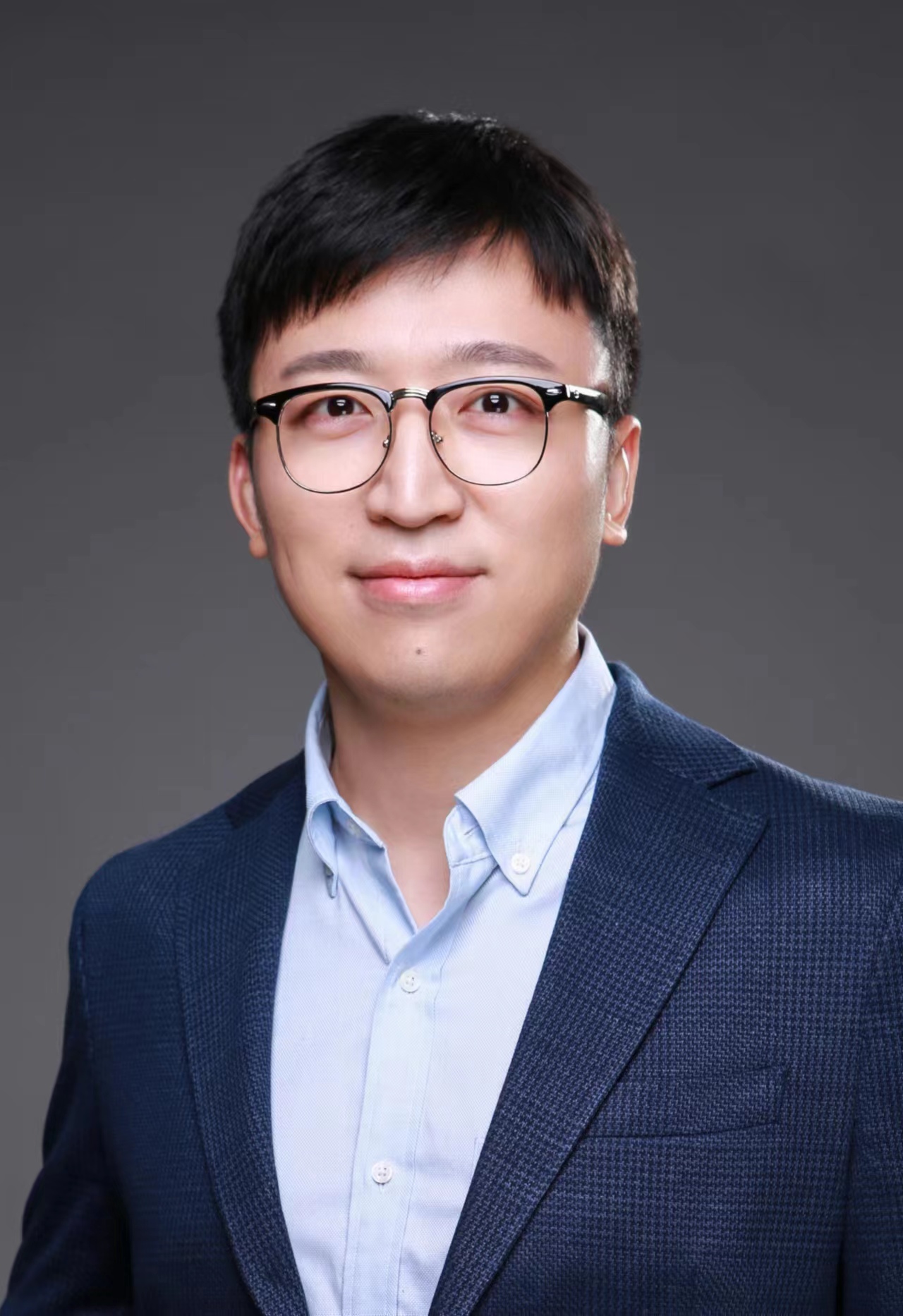}}]{Zhenzhen Jiao} received the Ph.D. degree from University of Chinese Academy of Sciences in computer science. He was an associate professor at the Institute of Computing Technology, Chinese Academy of Sciences and the director of a blockchain research laboratory. He is currently the director of the Teleinfo iF-Labs in China Academy of Information and Communications Technology. His research interests include blockchain, self-organized systems and networks, and computing power network.
\end{IEEEbiography}
\vfill

\newpage
\clearpage
\appendices

\section{Computational Complexity Analysis}
\textbf{Computational complexity of OIA3M:} The computational complexity of our proposed OIA3M depends on the number of rounds (denoted by $ k_{Max} $), the overbooked budget $ (1+\tau)B_i $, and the number of workers $ |\bm{W}| $, and the complexity of DP algorithm for the considered 0-1 knapsack problem, which is $ O(|\bm{W}|(1+\tau)B_i) $. Therefore, the complexity of OIA3M for each task $ t_i \in \bm{T} $ is $ O(k_{Max}|\bm{W}|(1+\tau)B_i) $.

\noindent
\textbf{Computational complexity of OMOM:} The corresponding computational complexity of OMOM depends on the overall rounds denoted by $ k_{Max}^{\prime} $, the practical budget $ B_i $, and the number of workers $ |\bm{W_i^{\prime}}| $. Since complexity of DP algorithm for our proposed 0-1 knapsack problem is $ O(|\bm{W_i^{\prime}}|B_i) $, the computational complexity of each task $ t_i \in \bm{T^{\prime}} $ is $ O(k_{Max}^{\prime}|\bm{W_i^{\prime}}|B_i) $

\noindent
\textbf{Computational complexity of O3M:} The computational complexity of O3M depends on the overall number of rounds denoted by $ k_{Max}^{\prime\prime} $, the remaining budget $ B_i^{\prime}$, and the number of workers $ |\bm{W^{\prime\prime}}| $, and the complexity of DP algorithm for our proposed 0-1 knapsack problem, which is $ O(|\bm{W^{\prime\prime}}|B_i^{\prime}) $. Thus, the computational complexity of each task $ t_i \in \bm{T^{\prime\prime}} $ is $ O(k_{Max}^{\prime\prime}|\bm{W^{\prime\prime}}|B_i^{\prime}) $.

\section{Property Analysis of OIA3M}
OIA3M satisfies the properties of convergence, individual rationality, fairness, non-wastefulness, and strong stability in the futures trading market. 

\noindent
\textbf{Lemma 1.} \textit{(Convergence of OIA3M) Algorithm 1 converges within finite rounds.}	\vspace{-0.1cm}
\begin{proof}
	Since DP algorithm is utilized to transform the problem into a two-dimensional 0-1 knapsack problem, after a finite number of rounds, each worker's payment can be either accepted or reaches its corresponding cost (e.g., $ c_{i,j} $).
\end{proof}

\noindent
\textbf{Lemma 2.} \textit{(Individual rationality of OIA3M) OIA3M satisfies the individual rationality of all tasks and workers in the futures market.}	\vspace{-0.1cm}
\begin{proof}
	The individual rationality of each task and worker is proved respectively, as the following: 
	
	\textbf{Individual rationality of tasks.} For each task $ t_i\in\bm{T} $, since the designed 0-1 knapsack problem regards $ (1+\tau)B_i $ as the corresponding capacity, the overall payment of $ t_i $ will thus not exceed $ (1+\tau)B_i $. Moreover, thanks to the factor of risk analysis and control of possible risk, e.g., constraint (10), each task $ t_i $ can decide whether to sign long-term contracts with the matched workers under an acceptable risk, which thus ensures that each task can receive a service quality as it desires, at a high probability. 
	
	\textbf{Individual rationality of workers.} Since we have 
	$ p_{i,j}^{F}\left\langle {k + 1} \right\rangle \leftarrow {\max}\left\{ p_{i,j}^{F}\left\langle k \right\rangle - \mathrm{\Delta}p_j~,{ c}_{i,j} \right\} $ (line 12, Algorithm 1), the final payment of each worker will definitely be larger than or at least equal to the corresponding cost, which guarantees a non-negative revenue of a worker.
	
	As a result, our proposed OIA3M in the futures market is individual rationality.
\end{proof}

\noindent
\textbf{Lemma 3.} \textit{(Fairness of OIA3M) OIA3M guarantees fairness in the futures market.} 	\vspace{-0.1cm}
\begin{proof}
	According to Definition 4, fairness indicates the case without type 1 blocking coalition, we offer the proof of Lemma 3 by contradiction. 
	
	Under a given matching $ \gamma $ and expected payment profile $ \mathbb{P}^E $, worker $ w_j $ and task set $ \mathbb{T} $ can form a type 1 blocking coalition $ (w_j; \mathbb{T}) $, if there exists a expected payment profile $ 
	\widetilde{\ P_j^E} $, as shown by (17) and (18).
	
	If task $ t_i $ does not sign a long-term contract with worker $ w_j $, the payment of worker $ w_j $ during the last round can only be the cost, as given by (50) and (51).
	\begin{equation}\label{key}
\setlength{\abovedisplayskip}{2pt} 
\setlength{\belowdisplayskip}{2pt}
{\small
	\begin{aligned}
		p_{i,j}^{F}\left\langle k \right\rangle = { c}_{i,j},
	\end{aligned}}
	\end{equation}
	\begin{equation}\label{key}
\setlength{\abovedisplayskip}{2pt} 
\setlength{\belowdisplayskip}{2pt}
{\small
	\begin{aligned}
			\overline{U^{T}}\left({ t_{i},\left\{\gamma\left( t_{i} \right)\backslash\gamma^{\prime}\left( t_{i} \right)\right\} \cup \left\{ w_{j} \right\} } \right) < \overline{U^{T}}\left( t_{i},\gamma\left( t_{i} \right) \right).
		\end{aligned}}
	\end{equation}
	If task $ t_i $ selects worker $ w_j $, we have $ p_{i,j}^{F}\left\langle k^{*} \right\rangle \geq p_{i,j}^{F}\left\langle k \right\rangle = {c}_{i,j} $ and the following (52)
	\begin{equation}\label{key}
\setlength{\abovedisplayskip}{2pt} 
\setlength{\belowdisplayskip}{2pt}
{\small
	\begin{aligned}
			&\overline{U^{T}}\left(t_{i},\left\{\gamma\left( t_{i} \right)\backslash\gamma^{\prime}\left( t_{i} \right)\right\} \cup \left\{ w_{j} \right\} \right) \geq\\& \overline{U^{T}}\left( t_{i},\left\{\gamma\left( t_{i} \right)\backslash\gamma^{\prime\prime}\left( t_{i} \right)\right\} \cup \left\{ w_{j} \right\}  \right),\\
		\end{aligned}}
	\end{equation}
	where $ 
	\gamma^{\prime\prime}\left( t_{i} \right) \subseteq \gamma^{\prime}\left( t_{i} \right) $. From (51) and (52), we can get
	\begin{equation}\label{key}
\setlength{\abovedisplayskip}{2pt} 
\setlength{\belowdisplayskip}{2pt}
{\small
	\begin{aligned}
			\overline{U^{T}}\left( {t_{i},\gamma\left( t_{i} \right) } \right) > \overline{U^{T}}\left(t_{i},\left\{\gamma\left( t_{i} \right)\backslash\gamma^{\prime\prime}\left( t_{i} \right)\right\} \cup \left\{ w_{j} \right\} \right),
		\end{aligned}}
	\end{equation}
	which is contrary to (18). Thus, our proposed OIA3M ensures the property of fairness. 
\end{proof}

\noindent
\textbf{Lemma 4.} \textit{(Non-wastefulness of OIA3M) Algorithm 1 satisfies the property of non-wastefulness.}	\vspace{-0.1cm}
\begin{proof}
	We conduct the proof of Lemma 4 by contradiction. Under a given matching $ \gamma $ and $ \mathbb{P}^E $, worker $ w_j $ and task set $ \mathbb{T} $ form a type 2 blocking coalition $ (w_j; \mathbb{T}) $ if there exists a expected payment profile $ 
	\widetilde{\ P_j^E} $, as shown by (19) and (20).
	
	If task $ t_i $ rejects $ w_j $, the payment of $ w_j $ during the last round can only be $ p_{i,j}^{F}\left\langle k \right\rangle = {c}_{i,j} $, where the only reason of the rejection between $ t_i $ and $ w_j $ is the overall payment exceeds the limited budget $ (1+\tau)B_i $. However, the coexistence of (19) and (20) shows that task $ t_i $ has an adequate budget to recruit workers, which contradicts the aforementioned assumption. Therefore, our proposed OIA3M in the futures market is non-wasteful.
\end{proof}

\noindent
\textbf{Theorem 1.} \textit{(Strong stability of OIA3M) OIA3M is strongly stable. }	\vspace{-0.1cm}
\begin{proof}
	Since the matching result of Algorithm 1 holds Lemma 2, Lemma 3, and Lemma 4, according to Definition 6, our proposed OIA3M in the futures market is strongly stable.
\end{proof}

\section{Algorithm and Property Analysis of OMOM}
The process of OMOM is similar to that of OIA3M, where the key differences are: \textit{i)} we conduct many-to-one-matching procedures for different tasks separately since the final determined workers of task $ t_i $ are selected from set $ \bm W_i^{\prime} $; \textit{ii)} $ B_i $ is considered as the practical budget. The pseudo-code of OMOM is given in Algorithm 2.

\begin{algorithm}[]
	{\small  
	\caption{\small{OMOM in the spot market}}
	\LinesNumbered 
	\textbf{Initialization: $ 
		k \leftarrow  1 $, $ p_{i,j}^{S}\left\langle 1 \right\rangle \leftarrow  P_{i,j}^{Desire} $, for $ \forall i,j $, $ {flag}_{j} \leftarrow  1 $}\; 
	\textbf{\% Action of each worker $ w_j\in\bm{W_i^{\prime}} $}
	
		\While{$ {flag}_{j} $}{
			\textbf{Calculate $ \mu^{k}\left( w_{j} \right) \leftarrow  \left\{ t_{i} \middle| {p_{i,j}^{S}\left\langle k \right\rangle \geq c_{i,j},t_{i} \in \bm{T^\prime}} \right\} $}\;
				\textbf{$ {flag}_{j} \leftarrow  0 $}\;
			\If{$ \forall\mu^{k}\left( w_{j} \right) \neq \varnothing $}{
				\For{$ 
					t_{i} \leftarrow \mu^{k}\left( w_{j} \right) $}{$ w_j $ sends a proposal including $ p_{i,j}^{S}\left\langle k \right\rangle $ and $ q_{i,j} $ to $ t_i $;}
				\textbf{Wait decisions from task $ t_i $}\;
				\For{
					$ \left\{ t_{i} \right\} \leftarrow  \mu^{k}\left( w_{j} \right) $
				}{
					\If{$ w_j $ is rejected by $ t_i $ and $ p_{i,j}^{S}\left\langle k \right\rangle > c_{i,j} $}{
						$ p_{i,j}^{S}\left\langle {k + 1} \right\rangle \leftarrow  \max\left\{ p_{i,j}^{S}\left\langle k \right\rangle - \mathrm{\Delta}p~,{~c}_{i,j} \right\} $;}
					\Else{$ p_{i,j}^{S}\left\langle {k + 1} \right\rangle \leftarrow  p_{i,j}^{S}\left\langle k \right\rangle $;}
				}
			
				\If{there exists $
					p_{i,j}^{S}\left\langle {k + 1} \right\rangle \neq p_{i,j}^{S}\left\langle k \right\rangle $, $ t_{i} \leftarrow  \mu^{k}\left( w_{j} \right) $}{
					$ {flag}_j\leftarrow 1 $\;
					$ k\leftarrow k+1 $;\
				}
			}
	}
	\textbf{\% Action of each task $ t_{i} \in \bm{T^\prime} $}\
	
		\While{
			$ \Sigma_{w_{j}\in\bm{W_i^{\prime}}}{flag}_{j} > 0 $}{
			Collect proposals from the workers in $ \bm{W_i^{\prime}} $, e.g., using $ {\widetilde{\mu}}^k\left(t_i\right) $ to include the workers that send proposals to $ t_i $;
			
			$ \mu^k(t_i)\leftarrow $ choose workers from $ {\widetilde{\mu}}^k\left(t_i\right) $ that can achieve the maximization of the utility under budget $ B_i $;
			
			use DP;
			
			$ t_i $ temporally accepts the workers in$ \ \mu^k(t_i) $, and rejects the others;

	}}
\end{algorithm}
Without losing generality, let $ p_{i,j}^{S}\left\langle k \right\rangle $ denote the payment asked by worker $ w_j $ for task $ t_i $ in the $ k^\text{th} $ round. Also, let $ \Delta{p}_j $ denote the reduction in the asked payment that worker $ w_j $ is willing to tolerate every time being rejected by a task, as long as the worker's utility (26) remains non-negative.

We next describe the steps of OMOM, the pseudo-code of which is given in Algorithm 2.

\noindent
\textbf{Step 1. Initialization:} At the beginning of each transaction, each worker $ w_j $'s asked payment is set to $ p_{i,j}^{S}\left\langle 1 \right\rangle \leftarrow  p^{Desire}_{i,j} $ (line 1, Algorithm 2). We also initialize $ \mu^k (w_j ) $ that contains the tasks that worker $ w_j $ is interested in and $ \mu^k(t_i) $ that contains the workers temporarily selected by task $ t_i $ in the $ k^\text{th} $ round.

\noindent
\textbf{Step 2. Proposal of workers:} At each round $ k $, each worker $ w_j \in \bm{W_i^{\prime}}$ first sends proposals to each task $ t_i $ in $ \mu^k(w_j) $, including its asked payments $ p_{i,j}^{S}\left\langle k \right\rangle $ and service quality $ q_{i,j} $ (lines 7-8, Algorithm 2).

\noindent
\textbf{Step 3. Worker selection from the task's side:} After collecting the information from workers in set $ \widetilde{\mu}^k(t_i) $, task $ t_i $ will solve a knapsack problem to determine a collection of temporarily accepted workers, e.g., set $ \mu^k(t_i)$, where $\mu^k(t_i)\subseteq \widetilde{\mu}^k(t_i)$ that maximize service quality under the practical budget $ B_i $, as given in (54).

\noindent
\textbf{Step 4. Solving the 0-1 knapsack problem:} Considering its practical budget $ B_i $, the task $ t_i $ obtains a set of workers $ \mu^k(t_i) $ to ensure the highest service quality, through solving a 0-1 knapsack problem,
\begin{equation}\label{key}
\setlength{\abovedisplayskip}{2pt} 
\setlength{\belowdisplayskip}{2pt}
{\small
	\begin{aligned}
	\underset{\mu^k{(t_{i})}}{\max} \sum_{w_j\in\mu^k\left(t_i\right)}(p_{i,j}^S-c_{i,j})
\end{aligned} }
\end{equation}
\vspace{-1.0mm}
{
	\setlength{\abovedisplayskip}{2pt} 
	\setlength{\belowdisplayskip}{2pt}
	\small
	\begin{flalign}
		&\ \text{s.t.}~~~~~~~~~~~~~~\sum_{w_j\in\mu^k\left(t_i\right)}{p_{i,j}^S} \leq B_i,~ \forall w_j\in {\widetilde{\mu}}^k\left(t_i\right),&
\end{flalign}}
which can generally be solved via DP \cite{22,47,48,49} (lines 20-23, Algorithm 2). Then, $ t_i $ reports the selected result of the current round to workers.

\noindent
\textbf{Step 5. Decision-making on workers' side:} After obtaining decisions from task $ t_i $, worker $ w_j $ inspects the following conditions:

	\textbf{Condition 1.} If $ w_j $ is accepted by a task $ t_i $ or its current asked payment $ p_{i,j}^{S}\left\langle k \right\rangle $ equals to its cost $ { c}_{i,j} $, its payment asked for $ t_i $ remains unchanged, i.e., $ p_{i,j}^{S}\left\langle {k + 1} \right\rangle \leftarrow p_{i,j}^{S}\left\langle k \right\rangle $ (line 14, Algorithm 2);

\textbf{Condition 2.} If $ w_j $ is rejected by a task $ t_i $ and its asked payment $ p_{i,j}^{S}\left\langle k \right\rangle $ is still above its cost $ { c}_{i,j} $, it decreases its asked payment for $ t_i $ in the next round, while avoiding a negative utility, as follows (line 12, Algorithm 2): 
\begin{equation}\label{key}	
\setlength{\abovedisplayskip}{2pt} 
\setlength{\belowdisplayskip}{2pt}
{\small
	\begin{aligned}
	p_{i,j}^{S}\left\langle {k + 1} \right\rangle =  \max\left\{ p_{i,j}^{S}\left\langle k \right\rangle - \mathrm{\Delta}p_j~,{~c}_{i,j} \right\},
\end{aligned} }
\end{equation}

\noindent
\textbf{Step 6. Repeat:} If all the asked payments stay unchanged from the $ (k-1)^{\text{th}} $ round to the $ k^{\text{th}} $ round, the matching will be terminated at this round. We use $ \Sigma_{w_j\in \bm{W_i^{\prime}}}{flag}_j=0 $ to denote this situation (line 5, Algorithm 2).
Otherwise, the algorithm repeats the above steps (e.g., lines 3-17, Algorithm 2) in the next round.

OMOM satisfies the properties of convergence, individual rationality, fairness, non-wastefulness, and strong stability. Relevant proofs are as follows.

\noindent
\textbf{Lemma 5.} \textit{(Convergence of OMOM) Algorithm 2 converges within finite rounds.}	\vspace{-0.1cm}

\begin{proof}
	Since DP algorithm is utilized to transform the problem into a 0-1 knapsack problem, after a finite number of rounds, each worker's payment can either be accepted or reaches its corresponding cost (e.g., $ c_{i,j} $).
\end{proof}

\noindent
\textbf{Lemma 6.} \textit{(Individual rationality of OMOM) OMOM satisfies the individual rationality of all tasks and workers in the spot market.}	\vspace{-0.1cm}
\begin{proof}
	\textbf{Individual rationality of tasks.} For task $ t_i\in\bm{T^\prime} $, since the designed 0-1 knapsack problem regards $ B_i $ as the corresponding capacity, the overall payment of $ t_i $ will thus not exceed $ B_i $. 
	
	\textbf{Individual rationality of workers.} Since we have 
	$ p_{i,j}^{S}\left\langle {k + 1} \right\rangle \leftarrow  \max\left\{ p_{i,j}^{S}\left\langle k \right\rangle - \mathrm{\Delta}p_j~,{~c}_{i,j} \right\} $ in Algorithm 2, the final payment of each worker will definitely be larger than or at least equal to the corresponding cost, which guarantees a non-negative revenue, and our proposed OMOM is thus individual rational.
\end{proof}

\noindent
\textbf{Lemma 7.} \textit{(Fairness of OMOM) OMOM guarantees fairness in the spot market.}	\vspace{-0.1cm}
\begin{proof}
	According to Definition 4, fairness indicates the case without the type 1 blocking pair, we offer the proof of Lemma 7 by contradiction.
	
	Under a given a matching $ \mu $ and payment profile $ \mathbb{P}^{S1} $, worker $ w_j $ and task $ t_i $ can form a blocking pair $ (w_j; t_i) $, if there exists a payment$ \widetilde{\ P_j^{S1}} $, as shown by (33).
	
	If task $ t_i $ does not recruit worker $ w_j $, the payment of worker $ w_j $ during the last round can only be the cost, as given by (57) and (58)
	\begin{equation}\label{key}
	\setlength{\abovedisplayskip}{2pt} 
	\setlength{\belowdisplayskip}{2pt}
		{\small
	\begin{aligned}
		p_{i,j}^{S}\left\langle k \right\rangle = { c}_{i,j},
		\end{aligned} }
	\end{equation}
	\begin{equation}\label{key}
	\setlength{\abovedisplayskip}{2pt} 
	\setlength{\belowdisplayskip}{2pt}
		{\small
		\begin{aligned} 
		 U ^{T}\left(  t_{i},\left\{\mu\left( t_{i} \right)\backslash\mu^{\prime}\left( t_{i} \right) \right\}\cup \left\{ w_{j} \right\} \right) < U ^{T}\left( t_{i},\mu\left( t_{i} \right) \right).\\
		\end{aligned} }
	\end{equation}
	If task $ t_i $ recruits worker $ w_j $, we have $ p_{i,j}^{S}\left\langle k^{*} \right\rangle \geq p_{i,j}^{S}\left\langle k \right\rangle = { c}_{i,j} $ and the following (59)
	\begin{equation}\label{key}
	\setlength{\abovedisplayskip}{2pt} 
	\setlength{\belowdisplayskip}{2pt}
		{\small
	\begin{aligned}
 &U ^{T}\left( t_{i},\left\{\mu\left( t_{i} \right)\backslash\mu^{\prime}\left( t_{i} \right)\right\} \cup \left\{ w_{j} \right\} \right) \geq \\&
 U ^{T}\left( t_{i},\left\{\mu\left( t_{i} \right)\backslash\mu^{\prime\prime}\left( t_{i} \right)\right\} \cup \left\{ w_{j} \right\} \right),\\ 
		\end{aligned}}
	\end{equation}
	where $ 
	\mu^{\prime\prime}\left( t_{i} \right) \subseteq \mu^{\prime}\left( t_{i} \right) $. From (58) and (59), we can get
	\begin{equation}\label{key}
	\setlength{\abovedisplayskip}{2pt} 
	\setlength{\belowdisplayskip}{2pt}
		{\small
	\begin{aligned} 
			&U ^{T}\left( t_{i},\mu\left( t_{i} \right) \right) > U ^{T}\left( t_{i},\left\{\mu\left( t_{i} \right)\backslash\mu^{\prime\prime}\left( t_{i} \right)\right\} \cup \left\{ w_{j} \right\} \right), 
		\end{aligned}}
	\end{equation}
	which is contrary to (33). Thus, our proposed OMOM ensures the property of fairness. 
\end{proof}

\noindent
\textbf{Lemma 8.} \textit{(Non-wastefulness of OMOM) Algorithm 2 satisfies the property of non-wastefulness.}	\vspace{-0.1cm}
\begin{proof}
	We conduct the proof of Lemma 8 by contradiction. Under a given matching $\mu$ and $ \mathbb{P}^{S1} $, worker $ w_j $ and task $ t_i $ form a type 2 blocking pair $ (w_j; t_i) $ under a payment profile $ \widetilde{~P_{j}^{S1}} $, as shown by (34).
	
	If task $ t_i $ rejects $ w_j $, the payment of $ w_j $ during the last round can only be $ p_{i,j}^{S}\left\langle k \right\rangle = { c}_{i,j} $, where the only reason of the rejection between $ t_i $ and $ w_j $ is the overall payment exceeds the limited budget $ B_i $. However, the coexistence of (34) shows that task $ t_i $ has an adequate budget to recruit workers, which contradicts the aforementioned assumption. Therefore, our proposed OMOM is non-wasteful.
\end{proof}

\noindent
\textbf{Theorem 2.} \textit{(Strong stability of OMOM) OMOM is strongly stable. } 	\vspace{-0.1cm}
\begin{proof}
	Since the matching result of Algorithm 2 holds Lemma 6, Lemma 7, and Lemma 8, according to Definition 6, our proposed OMOM is strongly stable.
\end{proof}

\section{Algorithm and Property Analysis of $ \rm \bf O3M $ }
\begin{algorithm}[] 
	{\small
		\caption{\small{O3M in the spot market}}
		\LinesNumbered 
		\textbf{Initialization: 	$ 
			k \leftarrow  1 $, $ p_{i,j}^{S}\left\langle 1 \right\rangle \leftarrow  p^{Desire}_{i,j} $, for $ \forall i,j$, ${flag}_{j} \leftarrow  1  $}\; 
		\textbf{\% Action of each worker $ w_j\in \bm{W^{\prime\prime}}\ $}
		
		\While{$ {flag}_{j} $}{
			\textbf{Calculate $ \varphi^{k}\left( w_{j} \right) \leftarrow  \left\{ t_{i} \middle| {p_{i,j}^{S}\left\langle k \right\rangle \geq c_{i,j},\forall t_{i} \in \bm{T^{\prime\prime}},t_{i} \notin \gamma\left( w_{j} \right)} \right\} $}\;
			\textbf{$ {flag}_{j} \leftarrow  0 $}\;
			\If{$ \forall\varphi^{k}\left( w_{j} \right) \neq \varnothing $}{
				\For{$ 
					\forall t_{i} \in \varphi^{k}\left( w_{j} \right) $}{$ w_j $ sends a proposal including $ p_{i,j}^{S}\left\langle k \right\rangle $ and $ q_{i,j} $ to $ t_i $;}
				\textbf{Wait decisions from tasks in $ \varphi^{k}\left( w_{j} \right) $}\;
				\For{
					$ \forall t_{i} \in \varphi^{k}\left( w_{j} \right) $
				}{
					\If{$ w_j $ is rejected by $ t_i $ and $ p_{i,j}^{S}\left\langle k \right\rangle > c_{i,j} $}{
						$ p_{i,j}^{S}\left\langle {k + 1} \right\rangle \leftarrow  \max\left\{ p_{i,j}^{S}\left\langle k \right\rangle - \mathrm{\Delta}p~,{~c}_{i,j} \right\} $;}
					\Else{$ p_{i,j}^{S}\left\langle {k + 1} \right\rangle \leftarrow  p_{i,j}^{S}\left\langle k \right\rangle $;}
				}
				
				\If{there exists $
					p_{i,j}^{S}\left\langle {k + 1} \right\rangle \neq p_{i,j}^{S}\left\langle k \right\rangle $, $ \forall t_{i} \in \varphi^{k}\left( w_{j} \right) $}{
					$ {flag}_j\leftarrow 1 $\;
					$ k\leftarrow k+1 $;\
				}
			}
		}
		\textbf{\% Action of each task $ t_i\in \bm{T^{\prime\prime}} $}\
		
		\While{
			$ \Sigma_{w_{j}\in \bm{W^{\prime\prime}}}{flag}_{j} > 0 $}{
			Collect proposals from the workers in $ \bm{W^{\prime\prime}} $, e.g., using $ {\widetilde{\varphi}}^k\left(t_i\right) $ to include the workers that send proposals to $ t_i $;
			
			$ \varphi^k(t_i)\leftarrow $ choose workers from $ {\widetilde{\varphi}}^k\left(t_i\right) $ that can achieve the maximization of the task utility under budget $ B_i^\prime $;
			
			use DP;
			
			$ t_i $ temporally accepts the workers in $ \varphi^k(t_i) $, and rejects the others;
		}
	}
\end{algorithm}
We next describe the steps of O3M, the pseudo-code of which is given in Algorithm 3.

\noindent
\textbf{Step 1. Initialization:} At the beginning of each transaction, each worker $ w_j $'s asked payment is set to $ p_{i,j}^{S}\left\langle 1 \right\rangle = p^{Desire}_{i,j} $ (line 1, Algorithm 3). We also initialize $ \varphi^k\left(w_j\right) $ that contains the tasks that the worker $ w_j $ is interested in and $ \varphi^k(t_i) $ that contains the workers temporarily selected by task $ t_i $ in the $ k^\text{th} $ round.

\noindent
\textbf{Step 2. Proposal of workers:} At each round $ k $, each worker $ w_j \in \bm{W^{\prime\prime}} $ first chooses tasks that generate non-negative revenue (37), and records them in $  \varphi^k\left(w_j\right) $, i.e., $ 
\varphi^{k}\left( w_{j} \right) \leftarrow  \left\{ t_{i} \middle| {p_{i,j}^{S}\left\langle k \right\rangle \geq c_{i,j},\forall t_{i} \in \bm{T^{\prime\prime}}} \right\} $. Then, $ w_j $ sends a proposal to each task $ t_i $ in $ \varphi^k(w_j) $, including its asked payment $ p_{i,j}^{S}\left\langle k \right\rangle $ and offered service quality $ q_{i,j} $ (lines 7-8, Algorithm 3).

\noindent
\textbf{Step 3. Worker selection on tasks' side:} After collecting the information from workers in set $ \widetilde{\varphi}^k\left(t_i\right) $, each task $ t_i $ solves a knapsack problem to determine a collection of temporary workers, e.g., set $ \varphi^k\left(t_i\right)$, where $\varphi^k\left(t_i\right)\subseteq \widetilde{\varphi}^k\left(t_i\right) $ that maximize its service quality under remaining budget $ B_i^{\prime} $, as given in (61).

\noindent
\textbf{Step 4. Solving the 0-1 knapsack problem:} Considering its remaining budget $ B_i^{\prime} $, each task $ t_i $ obtains a set of workers to ensure the highest service quality, through solving a 0-1 knapsack problem given below
\begin{equation}\label{key}
	\setlength{\abovedisplayskip}{2pt} 
	\setlength{\belowdisplayskip}{2pt}
	{\small
		\begin{aligned}
			\underset{\varphi^k{(t_{i})}}{\max} \sum_{w_j\in\varphi^k\left(t_i\right)}(p_{i,j}^S-c_{i,j})
	\end{aligned}}
\end{equation}
\vspace{-1.0mm}
{
	\setlength{\abovedisplayskip}{2pt} 
	\setlength{\belowdisplayskip}{2pt}
	\small
	\begin{flalign}
		&\ \text{s.t.}~~~~~~~~~~~~~~\sum_{w_j\in\varphi^k\left(t_i\right)}{p_{i,j}^S} \leq B_i^{\prime},~ \forall w_j\in {\widetilde{\varphi}}^k\left(t_i\right),&
\end{flalign} }
%
which generally be solved by via \cite{22,47,48,49} (lines 20-23, Algorithm 3). Then, $ t_i $ reports the selected result of the current round to workers.

\noindent
\textbf{Step 5. Decision-making on workers' side:} After obtaining the decisions from each task $ t_i\in\varphi^k\left(w_j\right) $, worker $ w_j $ inspects the following conditions:

\textbf{Condition 1.} If $ w_j $ is accepted by a task $ t_i $ or its current asked payment $ p_{i,j}^{S}\left\langle k \right\rangle $ equals to its cost $ {\ c}_{i,j} $, payment asked for $ t_i $ remains unchanged, i.e., $ p_{i,j}^{S}\left\langle {k + 1} \right\rangle \leftarrow p_{i,j}^{S}\left\langle k \right\rangle $ (line 14, Algorithm 3);

\textbf{Condition 2.} If $ w_j $ is rejected by a task $ t_i $ and its asked payment $ p_{i,j}^{S}\left\langle k \right\rangle $ is still above its cost $ {\ c}_{i,j} $, it decreases its asked payment for $ t_i $ in the next round, while avoiding a negative utility, as follows (line 12, Algorithm 3).

\noindent
\textbf{Step 6. Repeat:} If all the asked payments stay unchanged from the $ (k-1)^{\text{th}} $ round to the $ k^{\text{th}} $ round, the matching will be terminated at round $ k $. We use $ \Sigma_{w_j\in \bm{W^{\prime\prime}}}{flag}_j=0 $ to denote this situation (line 5, Algorithm 3).
Otherwise, the algorithm repeats the above steps (e.g., lines 3-17, Algorithm 3) in the next round.

Algorithm 3 is similar to Algorithm 1, where the key differences are: \textit{i)} the original $ \forall t_i\in \bm{T} $ is updated as $ \forall t_i\in \bm{T^{\prime\prime}} $ (line 18, Algorithm 3);
\textit{ii)} the expected utility of tasks and workers in OIA3M are changed to the practical utility of tasks and workers (line 21, Algorithm 3);
\textit{iii)} the overbooked budget $ (1+\tau)B_i $ is changed to the remaining budget $ B_i^{\prime} $ (line 21, Algorithm 3); 
\textit{iv)} O3M does not consider risk analysis. 

O3M satisfies the properties of convergence, individual rationality, fairness, non-wastefulness, and strong stability in the spot trading market.

\noindent
\textbf{Lemma 9.} \textit{(Convergence of O3M) Algorithm 3 converges within finite rounds.} 	\vspace{-0.1cm}

\begin{proof}
	Since DP algorithm is utilized to transform the problem into a two-dimensional 0-1 knapsack problem, after a finite number of rounds, each worker's payment can either be accepted or reaches its corresponding cost (e.g., $ c_{i,j} $).
\end{proof}

\noindent
\textbf{Lemma 10.} \textit{(Individual rationality of O3M) O3M satisfies the individual rationality of all tasks and workers in the futures market.} 	\vspace{-0.1cm}

\begin{proof}
	\textbf{Individual rationality of tasks.} For each task $ t_i\in\bm{T^{\prime\prime}} $, since the designed 0-1 knapsack problem regards $ B_i^\prime $ as the corresponding capacity, the overall payment of $ t_i $ will not exceed $ B_i^\prime $.
	
	\textbf{Individual rationality of workers.} Since we have 
	$ p_{i,j}^{S}\left\langle {k + 1} \right\rangle = \max\left\{ p_{i,j}^{S}\left\langle k \right\rangle - \mathrm{\Delta}p~,{~c}_{i,j} \right\} $ in Algorithm 3, the final payment of each worker will definitely be larger than or at least equal to the corresponding cost, which guarantees a non-negative revenue, and our proposed O3M is thus individual rational.
\end{proof}

\noindent
\textbf{Lemma 11.} \textit{(Fairness of O3M) O3M guarantees fairness in the futures market.} 	\vspace{-0.1cm}
\begin{proof}
	According to Definition 4, fairness indicates the case without type 1 blocking coalition, we offer the proof of Lemma 11 by contradiction.
	
	Under given a matching $ \varphi $ and payment profile $ \mathbb{P}^{S2} $, worker $ w_j $ and task set $ \mathbb{T}^{\prime\prime} $ can form a type 1 blocking coalition $ (w_j; \mathbb{T}^{\prime\prime})$, if there exists a payment$ \widetilde{\ P_j^{S2}} $, as shown by (45) and (46).

	If task $ t_i $ does not recruit worker $ w_j $, the payment of worker $ w_j $ during the last round can only be the cost, as given by (63) and (64).
	\begin{equation}\label{key}
	\setlength{\abovedisplayskip}{2pt} 
	\setlength{\belowdisplayskip}{2pt}
			{\small
		\begin{aligned}
		p_{i,j}^{S}\left\langle k \right\rangle = {~c}_{i,j}
			\end{aligned}}
	\end{equation}
	\begin{equation}\label{key}
	\setlength{\abovedisplayskip}{2pt} 
	\setlength{\belowdisplayskip}{2pt}
		{\small
	\begin{aligned} 
			&	U ^{T}\left( t_{i},\left\{\varphi\left( t_{i} \right)\backslash\varphi^{\prime}\left( t_{i} \right)\right\} \cup \left\{ w_{j} \right\} \right) < U ^{T}\left( t_{i},\varphi\left( t_{i} \right) \right). 
		\end{aligned} }
	\end{equation}  
	If task $ t_i $ recruits worker $ w_j $, we have $ p_{i,j}^{S}\left\langle k^{*} \right\rangle \geq p_{i,j}^{S}\left\langle k \right\rangle = {~c}_{i,j} $ and the following (65)
	\begin{equation}\label{key}
	\setlength{\abovedisplayskip}{2pt} 
	\setlength{\belowdisplayskip}{2pt}
		{\small
	\begin{aligned} 
			&	U ^{T}\left( t_{i},\left\{\varphi\left( t_{i} \right)\backslash\varphi^{\prime}\left( t_{i} \right)\right\} \cup \left\{ w_{j} \right\} \right) \geq\\ 		&U ^{T}\left( t_{i},\left\{\varphi\left( t_{i} \right)\backslash\varphi^{\prime\prime}\left( t_{i} \right)\right\} \cup \left\{ w_{j} \right\} \right),\\
		\end{aligned}}
	\end{equation}
	where $ 
	\varphi^{\prime\prime}\left( t_{i} \right) \subseteq \varphi^{\prime}\left( t_{i} \right) $. From (64) and (65), we can get
	\begin{equation}\label{key}
	\setlength{\abovedisplayskip}{2pt} 
	\setlength{\belowdisplayskip}{2pt}
		{\small
	\begin{aligned} 
 U ^{T}\left( t_{i},\varphi\left( t_{i} \right)  \right) > U ^{T}\left( t_{i},\left\{\varphi\left( t_{i} \right)\backslash\varphi^{\prime\prime}\left( t_{i} \right)\right\} \cup \left\{ w_{j} \right\}  \right), 
		\end{aligned}}
	\end{equation}
	which is contrary to (46). Thus, our proposed O3M ensures the property of fairness is fair.
\end{proof}

\noindent
\textbf{Lemma 12.} \textit{(Non-wastefulness of O3M) Algorithm 3 satisfies the property of non-wastefulness.} 	\vspace{-0.1cm}
\begin{proof}
	We conduct the proof of Lemma 12 by contradiction. Under a given matching $ \varphi $ and $ \mathbb{P}^{S2} $, worker $ w_j $ and task set $ \mathbb{T}^{\prime\prime} $ form a type 2 blocking coalition ($ w_j; \mathbb{T}^{\prime\prime} $), if there exists a payment$ \widetilde{\ P_j^{S2}} $, shown by (47) and (48).

	If task $ t_i $ rejects $ w_j $, the payment of $ w_j $ during the last round can only be $ p_{i,j}^{S}\left\langle k \right\rangle = {c}_{i,j} $, where the only reason of the rejection between $ t_i $ and $ w_j $ is the overall payment exceeds the limited budget $ B_i^{'} $. However, the coexistence of (47) and (48) shows that task $ t_i $ has an adequate budget to recruit workers, which contradicts the aforementioned assumption, Therefore, our proposed O3M is non-wasteful.
\end{proof}

\noindent
\textbf{Theorem 3.} \textit{(Strong stability of O3M) O3M is strongly stable.}	\vspace{-0.1cm}
\begin{proof}
	Since the matching result of Algorithm 3 holds Lemma 10, Lemma 11, and Lemma 12, according to Definition 6, our proposed O3M is strongly stable.
\end{proof}
\end{document}